\documentclass[american,english,aps,pra,superscriptaddress,showkeys,showpacs]{revtex4-1}
\usepackage[T1]{fontenc}
\usepackage[latin9]{inputenc}
\usepackage{units}
\usepackage{amsmath}
\usepackage{amssymb}
\usepackage{graphicx}

\makeatletter
\usepackage{hyperref}

\makeatother

\usepackage{babel}
\begin{document}

\title{Diagrammatic treatment of few-photon scattering\\from a Rydberg
blockaded atomic ensemble in a cavity}

\author{A. Grankin}

\affiliation{Institute for Theoretical Physics, University of Innsbruck, and Institute
for Quantum Optics and Quantum Information, Austrian Academy of Sciences,
Innsbruck, Austria.}

\affiliation{Laboratoire Charles Fabry, Institut d'Optique Graduate School, CNRS,
Université Paris-Saclay, 91127 Palaiseau, France.}

\affiliation{Laboratoire Aimé Cotton, Université Paris-Sud, ENS Cachan, CNRS,
Université Paris-Saclay, 91405 Orsay Cedex, France.}

\author{P. Grangier}

\affiliation{Laboratoire Charles Fabry, Institut d'Optique Graduate School, CNRS,
Université Paris-Saclay, 91127 Palaiseau, France.}

\author{E. Brion}

\affiliation{Laboratoire Aimé Cotton, Université Paris-Sud, ENS Cachan, CNRS,
Université Paris-Saclay, 91405 Orsay Cedex, France.}

\date{\today}
\begin{abstract}
In a previous letter \citep{GBB16} we studied the giant optical nonlinearities
of a Rydberg atomic medium within an optical cavity, in the Schwinger-Keldysh
formalism. In particular, we calculated the non-linear contributions
to the spectrum of the light transmitted through the cavity. In this
article we spell out the essential details of this calculation, and
we show how it can be extended to higher input photon numbers, and
higher order correlation functions. As a relevant example, we calculate
and discuss the three-photon correlation function of the transmitted
light, and discuss its physical significance in terms of the polariton
energy levels of the Rydberg medium within the optical cavity. 
\end{abstract}
\maketitle

\section{Introduction}

Optical quantum information processing requires photonic gates, that
may be implemented either deterministically or non-deterministically
\citep{NC10}. For the sake of efficiency and scalability it is preferable
to implement them in a deterministic way, which requires photon-photon
interactions. Though impossible to achieve directly, such interactions
can be effectively emulated by coupling photons to a medium with a
``giant'' optical non-linearity, i.e. large enough to allow photonic
qubits to interact. In this article we study an example of such a
medium, consisting in an atomic ensemble driven in a configuration
of electromagnetically induced transparency (EIT), involving a highly
excited Rydberg level. Following this approach, few-photon non-linearities
were achieved in free-space configuration setups: antibunching of
photons was observed in dispersive \citep{FPL13} and absorptive regimes
\citep{PFL12}, photon switches/transistors were implemented \citep{GTS14,TBS14}
and photon blockade was demonstrated \citep{DK12,MSB13}. By placing
such a medium in an optical cavity, strong nonlinearities for classical
light were predicted and demonstrated \citep{GBB14,GBB15}, as well
as quantum effects \citep{BUB16,PBS12,SPB13} recently observed \citep{JSG17}.
Such a non-linear medium is actually a strongly correlated many-body
system, and its full dynamics as well as its effects on the incoming
photons cannot be computed exactly. So far, analytic expressions of
dynamical variables like, the correlation functions of the transmitted
field could be derived either using \emph{ad hoc} models \textendash{}
such as the Rydberg bubble picture \citep{POF11,GNP13}, or resorting
to the perturbation theory restricted to the lowest non-vanishing
order in the number of incoming photons \citep{GBB15,GOF11}. 

In this article, we employ the Schwinger-Keldysh contour formalism
\citep{S61,R07,SL13} to derive analytic expressions for field correlation
functions for a Rydberg-EIT medium within an optical cavity, beyond
the lowest non-vanishing order in the excitation number \citep{GBB15}.
By opening a systematic and manageable way to deal with higher-order
terms, our approach breaks new ground for solving the outstanding
problem set by the many-body dynamics of Rydberg-blockaded ensembles
interacting with quantized light. It also allows us to unveil nontrivial
physical features of the transmitted light spectrum that we explain
by a simple polaritonic picture. Finally, it is important to notice
that parameters used for simulations correspond to experimental setups
such as the one used in \citep{JSG17}, or in \citep{PBS12,BUB16}
with an upgraded cavity. Therefore, the effects predicted by our model
can be, in principle, experimentally observed. 

The purpose of this article is to present the calculation of photon-photon
correlation functions using the formalism quoted above. For all physical
quantities of interest, we will perform the expansion and full resummation,
for the first few orders in the cavity feeding rate. In Sec. II we
introduce the model and notations, and in Sec. III we present the
elements of the Schwinger-Keldysh formalism that are useful for our
purpose. In Sec. \ref{sec:CTP_First-order} we derive the first-order
averages for cavity and atomic variables, and in Sec. \ref{sec:CTP_Intensity-correlation}
we analytically derive the the photonic pair correlation in the lowest
non-vanishing order. In Sec. \ref{sec:CTP_Transmission-spectrum}
we go beyond the lowest order and derive the analytic expression of
the transmission spectrum of the cavity, distinguishing its elastic
and inelastic parts. We give a physical explanation to the inelastic
part using a simple polaritonic picture. In the last section, we derive
the third-order correlation function of the transmitted light by adopting
the approach developed by L. D. Faddeev in application to the quantum-mechanical
three-body scattering problem. We get thus new results about three-photon
correlation functions, that are discussed from a physical point of
view. 

\section{Model and notations}

\paragraph*{Coupled atom-cavity system}

\begin{figure}
\begin{centering}
\includegraphics[width=0.3\paperwidth]{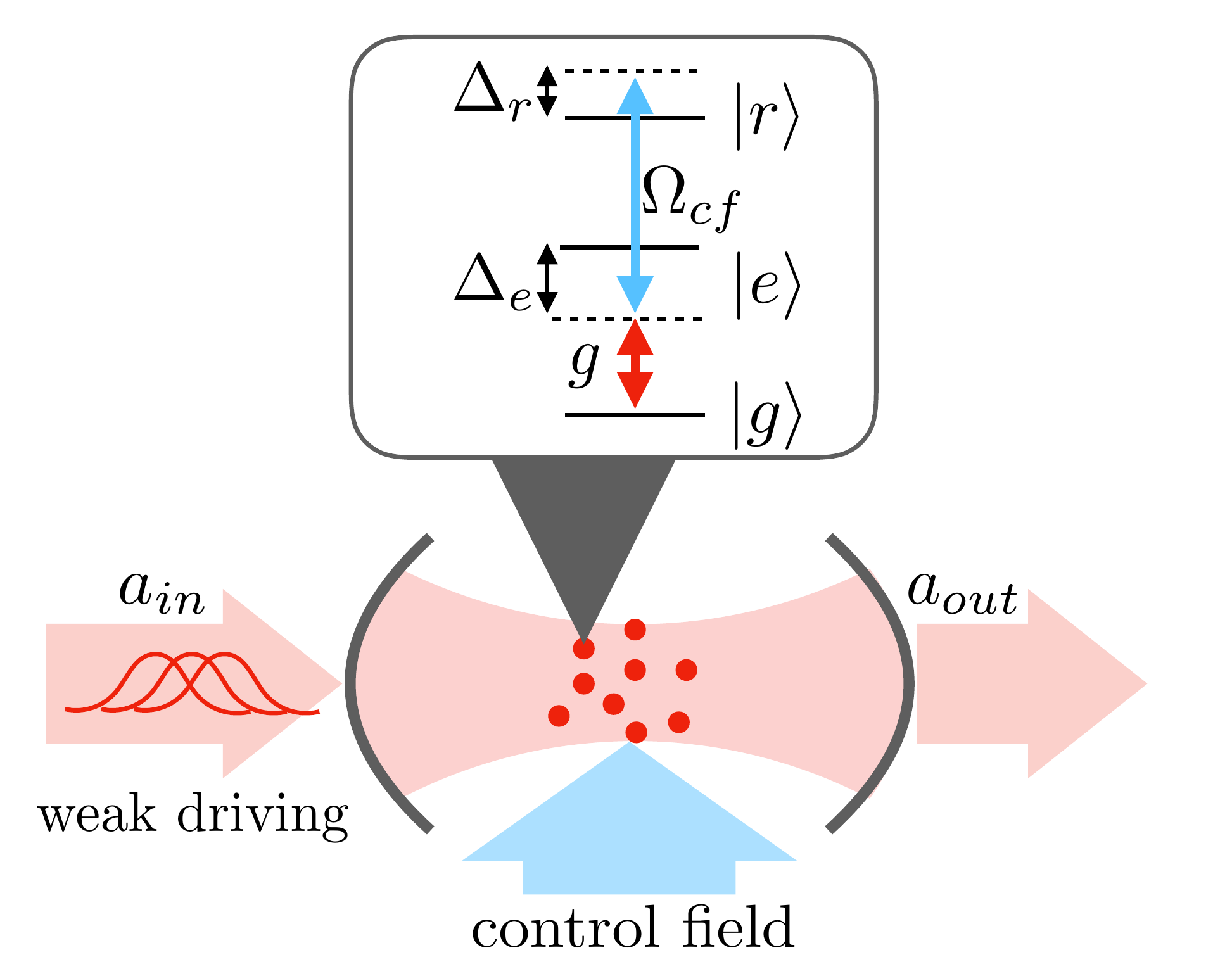}
\par\end{centering}
\caption{Atomic level scheme and the setup.}

\label{L_Scheme}
\end{figure}

We consider an ensemble of $N$ atoms with a ground, intermediate
and Rydberg states, denoted by $\left\vert g\right\rangle $, $\left\vert e\right\rangle $
and $\left\vert r\right\rangle $, respectively, loaded in an optical
cavity \citep{GBB14} (see Fig. \ref{L_Scheme}). The transitions
$g\leftrightarrow e$ and $e\leftrightarrow r$ are respectively driven
by the cavity mode, of frequency $\omega_{c}$ and annihilation operator
$a$, and the strong control field, with the coupling strength $g$
and the Rabi frequency $\Omega_{cf}$, respectively. The cavity is
fed through an input mirror with decay rate $\gamma_{c}^{\left(f\right)}$
by a weak probe laser of frequency $\omega_{p}$, while the field
transmitted by the cavity can be detected through an output mirror
with decay rate $\gamma_{c}^{\left(d\right)}$; we moreover set $\gamma_{c}\equiv\gamma_{c}^{\left(f\right)}+\gamma_{c}^{\left(d\right)}$.
We define detunings for the cavity $\Delta_{c}=\left(\omega_{p}-\omega_{c}\right)$,
single-photon $\Delta_{e}=\left(\omega_{p}-\omega_{eg}\right)$ and
two-photon $\Delta_{r}=\left(\omega_{p}+\omega_{cf}-\omega_{rg}\right)$,
with respect to the frequencies $\omega_{eg}$ and $\omega_{rg}$
of the $g\leftrightarrow e$ and $g\leftrightarrow r$ transitions.
We denote by $\gamma_{e}$ and $\gamma_{r}$ the decay rates from
the intermediate $\left\vert e\right\rangle $ and Rydberg $\left\vert r\right\rangle $
states, respectively.

If there were no atomic interactions, the cloud driven under perfect
EIT conditions $\left(\gamma_{r}\approx\Delta_{r}\approx0\right)$,
would be transparent for the probe light \citep{FL00}. The dipole-dipole-interaction-induced
blockade phenomenon \citep{SWM10,LFC01} actually prevents most of
the atoms in the sample from being Rydberg excited. If $\Delta_{e}\approx0$,
spontaneous emission from the intermediate state is strongly enhanced
which significantly modifies the shape of the transmitted light spectrum.
This effect can be characterized by the steady state correlation function
of the intracavity light $\left\langle a^{\dagger}\left(t\right)a\left(0\right)\right\rangle $
\citep{Walls}. At the lowest non-vanishing order in the feeding rate
$\left|\alpha\right|\equiv\sqrt{2\gamma_{c}^{\left(f\right)}I_{in}}$,
where $I_{in}$ is the incident photon flux fed into the cavity, the
correlation function was shown to factorize, \emph{i.e.} $\left\langle a^{\dagger}\left(t\right)a\left(0\right)\right\rangle ^{\left(2\right)}=\left\langle a^{\dagger}\left(t\right)\right\rangle ^{\left(1\right)}\left\langle a\left(0\right)\right\rangle ^{\left(1\right)}$
\citep{GBB15}, where the superscript denotes the order in $\alpha$.
To reveal nonlinear features, one has to investigate orders higher
than four \textendash{} by conservation of excitation number the third
order vanishes. Usual techniques are not suited to this task. In particular,
the standard fourth-order perturbative expansion would already lead
to a cumbersome hierarchy of Heisenberg equations which could hardly
be generalized further. Here, we show that the Schwinger-Keldysh contour
formalism \citep{S61,R07,SL13} allows one to compute dynamical variables
of the system up to \emph{a priori} arbitrary order in the feeding
strength, in a systematic and handy way. Besides bringing physical
insight into the specific problem considered here, our calculation
demonstrates how powerful this approach is to deal with non-equilibrium
dynamics of atomic systems as already stressed in \citep{FY99}.

\paragraph*{Dynamical equations of the system. }

According to Holstein-Primakoff approximation, the atomic lowering
operators $\sigma_{ge}^{\left(n\right)}$ and $\sigma_{gr}^{\left(n\right)}$
can be treated as bosons $b_{n}$ and $c_{n}$, respectively, in the
low excitation regime \citep{GBB15,BCF14}. The intrinsic (saturation)
nonlinearity of the EIT ladder scheme (for the probe beam) is neglected
from our consideration as it is much smaller than the nonlinear effects
induced by Rydberg interactions, for the chosen regime of parameters.
The Hamiltonian of the full system writes $H=H_{0}+H_{int}$ where\textbf{ }

\begin{align*}
H_{0} & =H_{at}+V_{a-c}+H_{bath}+V_{cav-bath}+V_{at-bath}\\
H_{int} & =H_{dd}+H_{f}\\
H_{dd} & =\frac{1}{2}\sum_{m,n}^{N}\kappa_{mn}c_{m}^{\dagger}c_{n}^{\dagger}c_{m}c_{n}\\
H_{f} & =\alpha\left(a+a^{\dagger}\right)\\
H_{at}= & \sum_{n=1}^{N}\left\{ -\Delta_{e}b_{n}^{\dagger}b_{n}-\Delta_{r}c_{n}^{\dagger}c_{n}+\frac{\Omega_{cf}}{2}\left(b_{n}^{\dagger}c_{n}+b_{n}c_{n}^{\dagger}\right)\right\} \\
V_{a-c}= & \sum_{n=1}^{N}g\left(ab_{n}^{\dagger}+a^{\dagger}b_{n}\right)-\Delta_{c}a^{\dagger}a
\end{align*}
where $\kappa_{mn}\equiv C_{6}/\left|\vec{r}_{m}-\vec{r}_{n}\right|^{6}$
denotes the van der Waals interaction potential. Performing the rotating
wave and Markov approximations, the relevant Heisenberg-Langevin equations
are 

\begin{eqnarray}
\frac{d}{dt}a & = & -\Gamma_{c}a-i\alpha-igb_{n}+\sqrt{2\gamma_{c}^{\left(f\right)}}a_{in}^{\left(f\right)}+\sqrt{2\gamma_{c}^{\left(d\right)}}a_{in}^{\left(d\right)}\label{HL1}\\
\frac{d}{dt}b_{n} & = & -\Gamma_{e}b_{n}-iga-i\frac{\Omega_{cf}}{2}c_{n}+b_{in,n}\label{HL2}\\
\frac{d}{dt}c_{n} & = & -\Gamma_{r}c_{n}-i\frac{\Omega_{cf}}{2}b_{n}-i\sum_{m}\kappa_{m,n}c_{m}^{\dagger}c_{m}c_{n}+c_{in,n}\label{HL3}
\end{eqnarray}
where $\left\{ a_{in}^{\left(f\right)},\;a_{in}^{\left(d\right)},\;b_{in,n},\;c_{in,n}\right\} $
denote the respective Langevin forces associated to the incoming fields
from the feeding and detection sides, and to the atomic operators
$b_{n}$ and $c_{n}$.\textbf{ }We use complex decay rates $\Gamma_{\nu}\equiv\gamma_{\nu}+i\Delta_{\nu}$
where $\nu=c,e,r$ for simplicity. 

\section{Schwinger-Keldysh formalism}

\subsection{Contour-ordered representation of correlation functions \label{subsec:CTP_ contour}}

Throughout this paper, we will focus on evaluating correlation functions
of the light transmitted through the cavity, which can be experimentally
obtained via multitime measurements of the light outgoing from the
setup. Input-output theory shows that, under Markov approximation,
these functions simply relate to the intracavity field correlation
functions, themselves coupled to the atomic correlation functions
via Heisenberg-Langevin equations. The generic form for such correlation
functions is

\begin{equation}
\left\langle \widetilde{\mathcal{T}}\left\{ \prod_{i=1}^{r}{\cal O}_{H,i}^{\dagger}\left(t_{i}\right)\right\} \mathcal{T}\left\{ \prod_{j=r+1}^{r+s}{\cal O}_{H,j}\left(t_{j}\right)\right\} \right\rangle \label{eq:CTP_corr}
\end{equation}
where ${\cal O}_{H,i}\left(t\right)\equiv e^{{\rm i}H\left(t-t_{0}\right)}{\cal O}_{i}e^{-{\rm i}H\left(t-t_{0}\right)}$
is an arbitrary operator of our system, expressed in the Heisenberg
picture with respect to the Hamiltonian $H$ given in the previous
subsection. In (\ref{eq:CTP_corr}) $\mathcal{T}$ and $\mathcal{\tilde{T}}$
stand for the usual chronological and anti-chronological time-ordering
operators, respectively. We also notice that averaging in Eq. (\ref{eq:CTP_corr})
is performed over the initial state of the system (i.e.\emph{ }at
$t=t_{0}$), that we assume to be the vacuum $\rho_{0}=\left\vert \textrm{Ø}\right\rangle \left\langle \textrm{Ø}\right\vert $
( i.e. $\left\langle \cdots\right\rangle \equiv{\rm Tr}\left[\rho_{0}\cdots\right]$
). 

Using the relation 
\begin{eqnarray}
\mathcal{O}_{H}\left(t\right) & = & \mathcal{\widetilde{T}}\left\{ e^{-i\,\int_{t}^{t_{0}}dsH_{int}\left(s\right)}\right\} \mathcal{O}_{H_{0}}\left(t\right){\cal T}\left\{ e^{-i\,\int_{t_{0}}^{t}dsH_{int}\left(s\right)}\right\} \label{eq:CTP_O_H}
\end{eqnarray}
where $H_{int}\left(t\right)\equiv e^{i\,H_{0}\left(t-t_{0}\right)}H_{int}e^{-i\,H_{0}\left(t-t_{0}\right)}$
we get:

\begin{eqnarray}
 &  & \left\langle \mathcal{\widetilde{T}}\left\{ \prod_{i=1}^{r}{\cal O}_{H,i}^{\dagger}\left(t_{i}\right)\right\} \mathcal{T}\left\{ \prod_{j=r+1}^{r+s}{\cal O}_{H,j}\left(t_{j}\right)\right\} \right\rangle \label{eq:CTP_corr-1-1}\\
 & = & {\rm Tr}\left[\begin{array}{c}
\rho_{0}\widetilde{\mathcal{T}}\left\{ {\cal O}_{H_{0},1}^{\dagger}\left(t_{1}\right)\ldots{\cal O}_{H_{0},r}^{\dagger}\left(t_{r}\right)e^{-i\,\int_{+\infty}^{-\infty}dsH_{int}\left(s\right)}\right\} \\
\times\mathcal{T}\left\{ {\cal O}_{H_{0},r+1}\left(t_{r+1}\right)\ldots{\cal O}_{H_{0},r+s}\left(t_{r+s}\right)e^{-i\,\int_{-\infty}^{+\infty}dsH_{int}\left(s\right)}\right\} 
\end{array}\right]\nonumber 
\end{eqnarray}

The form of Eq. (\ref{eq:CTP_corr-1-1}) suggests to introduce a new
variable, which does not merely follow the real axis $\left(-\infty,\infty\right)$
but rather a contour $\mathcal{C}$ made of two branches $\mathcal{C}_{+}=\left(-\infty,+\infty\right)$
and $\mathcal{C}_{-}=\left(+\infty,-\infty\right)$ (Fig. \ref{L_Scheme}).
A contour-ordering operator $\mathcal{T_{C}}$ can be defined, accordingly,
by 

\[
\mathcal{T}_{{\cal C}}\left\{ A\left(z_{1}\right)B\left(z_{2}\right)\right\} =\begin{cases}
A\left(z_{1}\right)B\left(z_{2}\right); & \text{if}\,\,z_{1}\epsilon\mathcal{C}_{-},z_{2}\epsilon\mathcal{C}_{+}\\
B\left(z_{2}\right)A\left(z_{1}\right); & \text{if}\,\,z_{1}\epsilon\mathcal{C}_{+},z_{2}\epsilon\mathcal{C}_{-}\\
\mathcal{T}\left\{ \,A\left(z_{1}\right)B\left(z_{2}\right)\,\right\} ; & \text{if}\,\,z_{1}\epsilon\mathcal{C}_{+},z_{2}\epsilon\mathcal{C}_{+}\\
\tilde{\mathcal{T}}\left\{ \,A\left(z_{1}\right)B\left(z_{2}\right)\,\right\} ; & \text{if}\,\,z_{1}\epsilon\mathcal{C}_{-},z_{2}\epsilon\mathcal{C}_{-}
\end{cases}
\]
Finally, introducing the notation $\mathcal{O}_{\pm}\left(t\right)\equiv\mathcal{O}_{H_{0}}$$\left(t\,\epsilon\,\mathcal{C}_{\pm}\right)$,
we may rewrite Eq. (\ref{eq:CTP_corr-1-1}) under the form:

\begin{eqnarray}
\left\langle \mathcal{\tilde{T}}\left\{ \prod_{i=1}^{r}{\cal O}_{H,i}^{\dagger}\left(t_{i}\right)\right\} \mathcal{T}\left\{ \prod_{j=r+1}^{r+s}{\cal O}_{H,j}\left(t_{j}\right)\right\} \right\rangle  & = & \left\langle \mathcal{T_{C}}\left\{ \prod_{i=1}^{r}\prod_{j=r+1}^{r+s}{\cal O}_{-,i}^{\dagger}\left(t_{i}\right){\cal O}_{+,j}\left(t_{j}\right)e^{-{\rm i}\int_{{\cal C}}dsH_{int}\left(s\right)}\right\} \right\rangle \label{eq:CTP_cont_ordered}
\end{eqnarray}

For future reference, we expand Eq. (\ref{eq:CTP_cont_ordered}) with
respect to $H_{f}$ and introduce the operator 
\begin{equation}
A_{q}\equiv\frac{1}{\sqrt{2\pi}}\int_{-\infty}^{\infty}\left(a_{+}-a_{-}\right)ds\label{eq:A_q}
\end{equation}
 in Eq. (\ref{eq:CTP_cont_ordered}), where the $q$ subscript stands
for the so-called ``quantum'' variable \citep{K11}:

\begin{eqnarray}
 &  & \left\langle {\cal T}_{{\cal C}}\left\{ e^{-\mathrm{i}\left(\int_{{\cal C}}H_{dd}\right)-\mathrm{i}\sqrt{2\pi}\alpha\left(A_{q}+A_{q}^{\dagger}\right)}\prod_{i=1}^{r}{\cal O}_{-,i}^{\dagger}\left(t_{i}\right)\prod_{j=r+1}^{r+s}{\cal O}_{+,j}\left(t_{j}\right)\right\} \right\rangle \nonumber \\
 & = & \sum_{n,k,p}\frac{\left(-\mathrm{i}\sqrt{2\pi}\alpha\right)^{n}}{\left(n-k\right)!k!}\left\langle {\cal T}_{{\cal C}}\left\{ \frac{\left(-i\int_{{\cal C}}H_{dd}\right)^{p}}{p!}A_{q}^{n-k}A_{q}^{\dagger k}\prod_{i=1}^{r}\prod_{j=r+1}^{r+s}{\cal O}_{-,i}^{\dagger}\left(t_{i}\right){\cal O}_{+,j}\left(t_{j}\right)\right\} \right\rangle \label{eq:CTP_General_bis}
\end{eqnarray}
where we used the fact that all operators behave as c-numbers inside
a ${\cal T}_{{\cal C}}$-ordered product. 

\begin{figure}
\begin{centering}
\includegraphics[width=0.3\paperwidth]{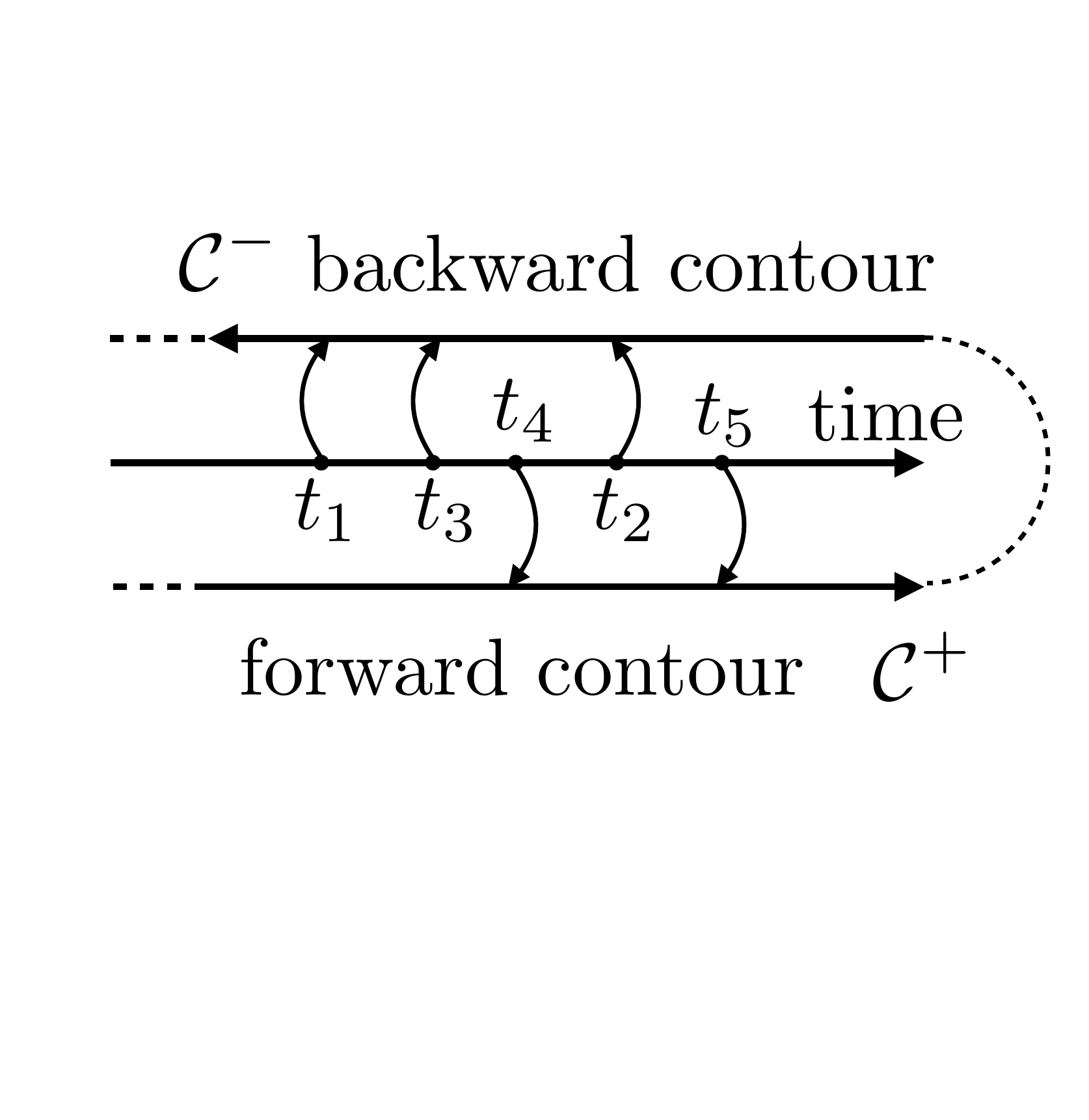}
\par\end{centering}
\caption{Representation of contour-ordering for the multitime correlation function
$\left\langle \tilde{\mathcal{T}}\left\{ \mathcal{O}^{1}\left(t_{1}\right)\mathcal{O}^{2}\left(t_{2}\right)\mathcal{O}^{3}\left(t_{3}\right)\right\} \mathcal{T}\left\{ \mathcal{O}^{4}\left(t_{4}\right)\mathcal{O}^{5}\left(t_{5}\right)\right\} \right\rangle $.}

\label{FIGCONT}
\end{figure}

The generic term of the double perturbative expansion in Eq. (\ref{eq:CTP_General_bis})
is an expectation value in the vacuum state $\rho_{0}$ of a contour-ordered
string of creation and annihilation operators 

\begin{equation}
{\cal S}=\left\langle {\cal T}_{{\cal C}}\left\{ e_{p+q}^{\dagger}\left(z_{p+q}\right)\ldots e_{p+1}^{\dagger}\left(z_{p+1}\right)e_{p}\left(z_{p}\right)\ldots e_{1}\left(z_{1}\right)\right\} \right\rangle \label{eq:CTP_wick_string}
\end{equation}
where $e_{1},e_{2},\cdots,e_{p+q}$ are bosonic annihilation operators
in the interaction picture with respect to $H_{0}$. Applied to our
system, Wick's theorem \citep{SL13,AGD} states that such a contour-ordered
string can be decomposed into a sum over all possible pairwise products
of creation and annihilation operators in the string in Eq. (\ref{eq:CTP_wick_string})

\begin{equation}
{\cal S}=\sum_{a.p.p}\prod_{k,l}\left\langle {\cal T}_{{\cal C}}\left\{ e_{l}\left(z_{l}\right)e_{k}^{\dagger}\left(z_{k}\right)\right\} \right\rangle \label{eq:CTP_wicks}
\end{equation}
The quantity $G_{e_{k}e_{l}}^{\left({\cal C}\right)}\left(z_{l},z_{k}\right)=-{\rm i}\left\langle {\cal T_{C}}\left\{ e_{l}\left(z_{l}\right)e_{k}^{\dagger}\left(z_{k}\right)\right\} \right\rangle $
is called the unperturbed contour-ordered Green's function for the
operators $e_{k}$ and $e_{l}$. 

Before evaluating the unperturbed Green's functions, it is important
to notice that an implicit part of the theorem's statement is that
the number of creation and annihilation operators should be equal.
In the general formula Eq. (\ref{eq:CTP_General_bis}) there are $k+r+2p$
creation operators (recalling that $H_{dd}=\frac{1}{2}\sum_{ij}\kappa_{ij}c_{i}^{\dagger}c_{j}^{\dagger}c_{i}c_{j}$)
and $n-k+s+2p$ annihilation operators: the series Eq. (\ref{eq:CTP_General_bis})
should be restricted to the terms which satisfy $k+r+2p=n-k+s+2p$,
or equivalently $k=\frac{n+s-r}{2}$. Defining $D=s-r$ we finally
have 

\begin{align}
 & \left\langle \mathcal{T}_{{\cal C}}\left\{ \prod_{i=1}^{r}\prod_{j=r+1}^{r+s}{\cal O}_{H,i}^{\dagger}\left(t_{i}\right){\cal O}_{H,j}\left(t_{j}\right)\right\} \right\rangle \label{eq:CTP_General_bis_s}\\
 & =\sum_{n:\frac{n+D}{2}\epsilon\mathbb{Z},p}\frac{\left(-\mathrm{i}\sqrt{2\pi}\alpha\right)^{n}}{\left(\frac{n+D}{2}\right)!\left(\frac{n-D}{2}\right)!}\left\langle {\cal T}_{{\cal C}}\left\{ \frac{\left(-\mathrm{i}\int_{{\cal C}}H_{dd}\right)^{p}}{p!}A_{q}^{\frac{n-D}{2}}A_{q}^{\dagger\frac{n+D}{2}}\prod_{i=1}^{r}\prod_{j=r+1}^{r+s}{\cal O}_{-,i}^{\dagger}\left(t_{i}\right){\cal O}_{+,j}\left(t_{j}\right)\right\} \right\rangle \nonumber 
\end{align}
where in the summation over $n$ we specified that $k$ should be
an integer. For future reference and for the sake of conciseness we
shall use the formally resummed version of this formula with respect
to $p$ 

\begin{equation}
\sum_{n:\frac{n+D}{2}\epsilon\mathbb{Z}}\frac{\left(-\mathrm{i}\sqrt{2\pi}\alpha\right)^{n}}{\left(\frac{n+D}{2}\right)!\left(\frac{n-D}{2}\right)!}\left\langle {\cal T}_{{\cal C}}\left\{ e^{-\mathrm{i}\left(\int_{{\cal C}}H_{dd}\right)}A_{q}^{\frac{n-D}{2}}A_{q}^{\dagger\frac{n+D}{2}}\prod_{i=1}^{r}\prod_{j=r+1}^{r+s}{\cal O}_{-,i}^{\dagger}\left(t_{i}\right){\cal O}_{+,j}\left(t_{j}\right)\right\} \right\rangle \label{eq:CTP_the_most_general}
\end{equation}

\subsection{Green's functions\label{subsec:CTP_Green}}

The contour-ordered Green's function $G_{e_{k}e_{l}}^{\left({\cal C}\right)}\left(z_{l},z_{k}\right)=-{\rm i}\left\langle {\cal T_{C}}\left\{ e_{l}\left(z_{l}\right)e_{k}^{\dagger}\left(z_{k}\right)\right\} \right\rangle $
physically characterizes the system's response at some time $z_{l}$
to the creation of a single excitation at time $z_{k}$. Depending
on the respective positions of the arguments $z_{k}$ and $z_{l}$
on the contour, $G_{e_{k}e_{l}}^{\left({\cal C}\right)}\left[z_{l},z_{k}\right]$
coincides with one of the four following real-time Green's functions:

\[
\begin{cases}
G_{e_{k}e_{l}}^{T}\left(z_{l}-z_{k}\right) & =G_{e_{k}e_{l}}^{\left({\cal C}\right)}\left[z_{l},z_{k}\right];\,\text{when}\,\,z_{l},z_{k}\,\epsilon\,\mathcal{C}_{+}\\
G_{e_{k}e_{l}}^{\tilde{T}}\left(z_{l}-z_{k}\right) & =G_{e_{k}e_{l}}^{\left({\cal C}\right)}\left[z_{l},z_{k}\right];\,\text{when}\,\,z_{l},z_{k}\,\epsilon\,\mathcal{C}_{-}\\
G_{e_{k}e_{l}}^{>}\left(z_{l}-z_{k}\right) & =G_{e_{k}e_{l}}^{\left({\cal C}\right)}\left[z_{l},z_{k}\right];\,\text{when}\,\,z_{k}\,\epsilon\,\mathcal{C}_{+},z_{l}\,\epsilon\,\mathcal{C}_{-}\\
G_{e_{k}e_{l}}^{<}\left(z_{l}-z_{k}\right) & =G_{e_{k}e_{l}}^{\left({\cal C}\right)}\left[z_{l},z_{k}\right];\,\text{when}\,\,z_{k}\,\epsilon\,\mathcal{C}_{-},z_{l}\,\epsilon\,\mathcal{C}_{+}
\end{cases}
\]
where we implicitly assumed the time invariance of $G^{T,\tilde{T},>,<}$
(resulting from the fact that $H_{0}$ is time-independent). Note
that, while $\left(z_{k},z_{l}\right)$ are contour arguments in $G_{e_{k}e_{l}}^{\left({\cal C}\right)}\left(z_{l},z_{k}\right)$,
they must be understood as ``real'' time arguments in the functions
$G^{T}$,$G^{\tilde{T}}$,$G^{>}$,$G^{<}$. To avoid any ambiguity,
here and below we will implicitly use the convention that same-time
Green's function is equal to a normally ordered product of the corresponding
operators, and therefore vanishes. It can be shown \citep{SL13} that
all four Green's functions are not independent. For any pair of operators
$\left(x,y\right)$ $G_{xy}^{\tilde{T}}\left[t^{\prime}-t\right]=-\left[G_{yx}^{T}\left[t-t^{\prime}\right]\right]^{*}=-\left[G_{xy}^{T}\left[t-t^{\prime}\right]\right]^{*}$,
or equivalently in the temporal Fourier space $G_{xy}^{\tilde{T}}\left[\omega\right]=-\left[G_{xy}^{T}\left[\omega\right]\right]^{*}$.
Moreover, the different Green's functions are related by $G_{xy}^{>}\left[\omega\right]+G_{xy}^{<}\left[\omega\right]=G_{xy}^{T}\left[\omega\right]+G_{xy}^{\tilde{T}}\left[\omega\right]$.
This can be further simplified by noticing that, since $H_{0}$ preserves
the excitation number and the state we average on is the vacuum $\rho_{0}$,
then $G_{xy}^{<}\left[\omega\right]=0$; therefore $G_{xy}^{>}\left[\omega\right]=2{\rm i}\,{\rm Im}\left[G_{xy}^{T}\left[\omega\right]\right]$.
As a consequence defining the so-called ''quantum'' variable $x_{q}\left(t\right)\equiv\frac{1}{\sqrt{2}}\left(x_{+}\left(t\right)-x_{-}\left(t\right)\right)$
(where as usual $x_{\pm}\left(t\right)=x_{H_{0}}\left(t\epsilon\mathcal{C}_{\pm}\right)$)
we get for any pair of two operators $\left(x,y\right)$: $\left\langle \mathcal{T}_{\mathcal{C}}\left\{ x_{q}\left(t\right)y_{q}^{\dagger}\left(t^{\prime}\right)\right\} \right\rangle =0$. 

Time-ordered unperturbed Green's functions $G_{e_{k}e_{l}}^{T}$$\left(t_{k}-t_{l}\right)$
can be deduced from the Heisenberg-Langevin equations, generated by
$H_{0}$ alone, i.e. from Eqs. (\ref{HL1}-\ref{HL3}) in which $\alpha$
and $\kappa_{m,n}$ are set to zero. For the sake of convenience we
introduce the collective spinwaves $b_{\vec{k}}\equiv\frac{1}{\sqrt{N}}\sum_{j}e^{i\vec{k}\vec{r_{j}}}b_{j}$,
$c_{\vec{k}}\equiv\frac{1}{\sqrt{N}}\sum_{j}e^{i\vec{k}\vec{r_{j}}}c_{j}$
defined in App. \ref{sec:A_bos_spinwave} which allow us to split
Eqs. (\ref{HL1}-\ref{HL3}) into a set of independent subsystems,
\emph{i.e.}

\begin{eqnarray}
\frac{d}{dt}a & = & -\Gamma_{c}a-\mathrm{i}g\sqrt{N}b_{0}+\sqrt{2\gamma_{c}}a_{in}\label{eq:CTP_bosonic1_k=00003D0}\\
\frac{d}{dt}b_{\vec{k}} & = & -\Gamma_{e}b_{\vec{k}}-\mathrm{i}\frac{\Omega_{cf}}{2}c_{\vec{k}}-{\rm i}g\sqrt{N}\delta_{\vec{k},0}a+\sqrt{2\gamma_{e}}b_{\vec{k},in}\\
\frac{d}{dt}c_{\vec{k}} & = & -\Gamma_{r}c_{\vec{k}}-\mathrm{i}\frac{\Omega_{cf}}{2}c_{\vec{k}}+\sqrt{2\gamma_{r}}c_{\vec{k},in}\label{eq:CTP_bosonic3_k=00003D0}
\end{eqnarray}
We define the matrix

\begin{eqnarray}
\hat{G}^{T}\left[t,t^{\prime}\right] & \equiv & -{\rm i}\left\langle {\cal T}\left(\vec{X}\left(t\right)\times\vec{X}^{\dagger}\left(t^{\prime}\right)\right)\right\rangle \label{eq:CTP_G_time}
\end{eqnarray}
where 
\[
\vec{X}\left(t\right)\equiv\left(\begin{array}{c}
a\left(t\right)\\
b_{0}\left(t\right)\\
c_{0}\left(t\right)\\
\left\{ \begin{array}{c}
b_{k}\left(t\right)\\
c_{k}\left(t\right)
\end{array}\right\} _{\vec{k}\neq0}
\end{array}\right)
\]
and $\vec{X}^{\dagger}\left(t\right)$ is the transconjugated vector
$\left(a^{\dagger}\left(t\right),b_{0}^{\dagger}\left(t\right),c_{0}^{\dagger}\left(t\right),\left\{ b_{k}^{\dagger}\left(t\right),c_{k}^{\dagger}\left(t\right)\right\} \right)$.
From Eqs. (\ref{eq:CTP_bosonic1_k=00003D0}-\ref{eq:CTP_bosonic3_k=00003D0})
we deduce the matrix equation \citep{R07} $\partial_{t}\hat{G}^{T}\left[t,t^{\prime}\right]=\hat{M}\times\hat{G}\left[t,t^{\prime}\right]-{\rm i}\delta\left(t-t^{\prime}\right)\mathbb{I}$,
where $\hat{M}$ the coefficient matrix of the system Eqs. (\ref{eq:CTP_bosonic1_k=00003D0}-\ref{eq:CTP_bosonic3_k=00003D0}).
Switching to the temporal Fourier space $\left(\hat{G}^{T}\left[\omega\right]\equiv\right.\left.-{\rm i}\int d\omega\,e^{{\rm i}\omega t}\left\langle {\cal T}\left(\vec{X}\left(t\right)\times\vec{X}^{\dagger}\left(0\right)\right)\right\rangle \right)$
we get: $\hat{G}^{T}\left[\omega\right]=\left(\omega-{\rm i}\hat{M}\right)^{-1}$
and finally find $\hat{G}^{T}\left[\omega\right]$ to be block-diagonal:

\[
\hat{G}^{T}\left[\omega\right]=\left[\begin{array}{cc}
\hat{G}_{0}^{T}\left[\omega\right] & 0\\
0 & \left\{ \hat{G}_{\vec{k}}^{T}\left[\omega\right]\right\} 
\end{array}\right]
\]
where 

\begin{equation}
\hat{G}_{0}^{T}\left[\omega\right]=\left(\begin{array}{ccc}
\omega+{\rm i}\Gamma_{c} & -g\sqrt{N} & 0\\
-g\sqrt{N} & \omega+{\rm i}\Gamma_{e} & -\frac{\Omega_{cf}}{2}\\
0 & -\frac{\Omega_{cf}}{2} & \omega+{\rm i}\Gamma_{r}
\end{array}\right)^{-1};\,\,\hat{G}_{\vec{k}}^{T}\left[\omega\right]=\left(\begin{array}{cc}
\omega+{\rm i}\Gamma_{e} & -\frac{\Omega_{cf}}{2}\\
-\frac{\Omega_{cf}}{2} & \omega+{\rm i}\Gamma_{r}
\end{array}\right)^{-1}\label{eq:CTP_G_k}
\end{equation}

Recalling the properties of the Green's functions $\hat{G}^{\tilde{T}},\hat{G}^{>}$
specified in the introduction to this subsection we may straightforwardly
deduce that they all exhibit the same block-diagonal structure as
$\hat{G}^{T}$. 

In the following sections we present the calculation of correlation
functions using the formalism presented above. For all physical quantities
of interest, we will perform the expansion and full resummation of
Eq. (\ref{eq:CTP_the_most_general}) with respect to $H_{dd}$, for
the first few orders in the feeding rate $\alpha$: therefore, unless
specified, the term ``order'' will refer to the order in power of
$\alpha$. In the next section we derive the first-order averages
for cavity and atomic variables, then the photonic pair correlation
function and the transmission spectrum of the cavity, and finally
we calculate the third-order correlation function of the transmitted
light using the Faddeev approach, and discuss some of its properties. 

\section{Linear EIT cavity response recovered\label{sec:CTP_First-order}}

We first briefly show how the contour formalism allows us to recover
well-known linear EIT response of the cavity. Setting ${\cal O}_{+}=a$,
$r=0$, $s=1$, $D=1$ in Eq. (\ref{eq:CTP_the_most_general}) we
get:

\begin{equation}
\left\langle a\left(t\right)\right\rangle ^{\left(1\right)}=\left(-\mathrm{i}\sqrt{2\pi}\alpha\right)\left\langle {\cal T}_{{\cal C}}\left\{ e^{-\mathrm{i}\left(\int_{{\cal C}}H_{dd}\right)}A_{q}^{\dagger}a_{+}\left(t\right)\right\} \right\rangle \label{eq:CTP_a^1}
\end{equation}

We split the contour integral into its forward and backward parts
$\int_{{\cal C}}H_{dd}=\int_{{\cal C}_{+}}H_{dd}+\int_{{\cal C}_{-}}H_{dd}$
and expand Eq. (\ref{eq:CTP_a^1}) with respect to each of them separately
to get:

\begin{equation}
\left\langle a\left(t\right)\right\rangle ^{\left(1\right)}=\left(-\mathrm{i}\sqrt{2\pi}\alpha\right)\sum_{p,q}\left\langle {\cal T}_{{\cal C}}\left\{ \frac{\left(-i\,\int_{{\cal C}_{+}}H_{dd}\right)^{p}}{p!}\frac{\left(-i\,\int_{{\cal C}_{-}}H_{dd}\right)^{q}}{q!}A_{q}^{\dagger}a_{+}\left(t\right)\right\} \right\rangle \label{eq:CTP_a^2_detailed}
\end{equation}
Applying Wick's theorem to this expression, we find that only terms
with $p=0,q=0$ yield non-vanishing contributions. As a consequence
of the fact that $H_{dd}$ acts in the doubly-excited subspace only,
all the other terms in the sum inevitably contain contractions, equivalent
to vanishing normally-ordered product of operators (e.g. $G^{<}$
Green's functions). Eq. (\ref{eq:CTP_a^2_detailed}) therefore simplifies
into:

\begin{equation}
\left\langle a\left(t\right)\right\rangle ^{\left(1\right)}=\left(-\mathrm{i}\sqrt{2\pi}\alpha\right)\left\langle {\cal T}_{{\cal C}}\left\{ A_{q}^{\dagger}a_{+}\left(t\right)\right\} \right\rangle =\left(-i\alpha\right)\left\langle {\cal T}_{{\cal C}}\left\{ \int dsa_{+}^{\dagger}a_{+}\left(t\right)\right\} \right\rangle \label{eq:CTP_a^1_t}
\end{equation}
where we used the definition (Eq. \ref{eq:A_q}) of $A_{q}$ and omitted
the vanishing vacuum average of a normally ordered product of operators
$\left\langle {\cal T}_{{\cal C}}\left\{ \int dsa_{-}^{\dagger}\left(s\right)a_{+}\left(t\right)\right\} \right\rangle $. 

In Fourier space we get $\frac{1}{\sqrt{2\pi}}\int e^{{\rm i}\omega t}dt\left\langle a\left(t\right)\right\rangle ^{\left(1\right)}=\left(-\mathrm{i}\alpha\sqrt{2\pi}\right){\rm i}G_{aa}\left[\omega\right]\delta\left(\omega\right)$.
The delta function in this expression results from the system being
in the steady state (we assume that the evolution starts at $t_{0}=-\infty$).
Using Eq. (\ref{eq:CTP_G_k}) we finally recover the standard cavity-EIT
response formula:

\begin{eqnarray*}
\left\langle a\left(t\right)\right\rangle ^{\left(1\right)} & = & \left(-\mathrm{i}\alpha\right){\rm i}G_{aa}^{T}\left[0\right]=\frac{\left(-\mathrm{i}\alpha\right)}{\Gamma_{c}+\frac{g^{2}N}{\left(\Gamma_{e}+\frac{\Omega_{cf}^{2}}{4\Gamma_{r}}\right)}}
\end{eqnarray*}

\section{Pair correlation function\label{sec:CTP_Intensity-correlation}}

In Apps. \ref{sec:Pair-correlation-function},\ref{subsec:CTP_T-matrix}
we explicitly re-derive the results presented in \citep{GBB15} for
the $\left\langle {\cal T}\left\{ a\left(t\right)a\left(t^{\prime}\right)\right\} \right\rangle $
correlation function. The calculations are only briefly sketched in
this section, which allows us to introduce various tools that we will
use to compute quantities beyond the lowest order in $\alpha$. 

Setting with $r=0$, $s=2$ and ${\cal O}_{+,1,2}=a$ in Eq. (\ref{eq:CTP_the_most_general})
we get the photon pair correlation function $\left\langle {\cal T}\left(a\left(t\right)a\left(t^{\prime}\right)\right)\right\rangle ^{\left(2\right)}$
to the second order in the feeding rate $\alpha$:
\begin{equation}
\left\langle \mathcal{T}\left\{ a\left(t\right)a\left(t^{\prime}\right)\right\} \right\rangle ^{\left(2\right)}=\frac{\left(-\mathrm{i}\sqrt{2\pi}\alpha\right)^{2}}{2!}\left\langle {\cal T}_{{\cal C}}\left\{ e^{-\mathrm{i}\left(\int_{{\cal C}_{+}}H_{dd}\right)}\left(A_{q}^{\dagger}\right)^{2}a_{+}\left(t\right)a_{+}\left(t^{\prime}\right)\right\} \right\rangle \label{eq:CTP_aa}
\end{equation}
Note that here we omitted the $e^{-\mathrm{i}\left(\int_{{\cal C}_{-}}H_{dd}\right)}$
factor under the contour ordering as only its 0th order in expansion
contributes to the average. We also notice that Eq. (\ref{eq:CTP_aa})
contains only ``+'' operators and therefore its Wick's expansion
comprises only time-ordered Green's functions. 

\begin{figure}
\begin{centering}
\emph{a}) \includegraphics[height=2cm]{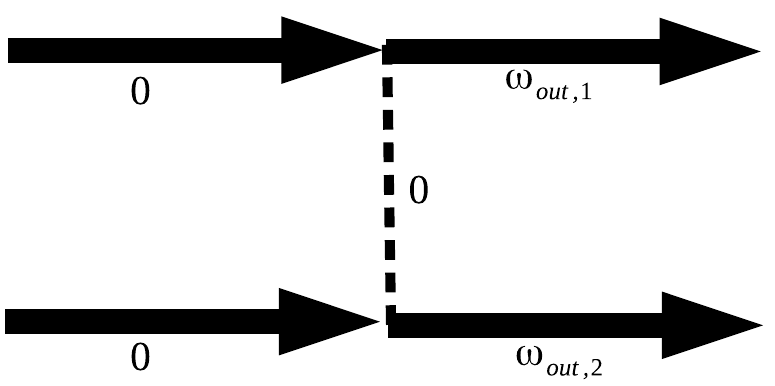}\emph{b}) \includegraphics[height=2cm]{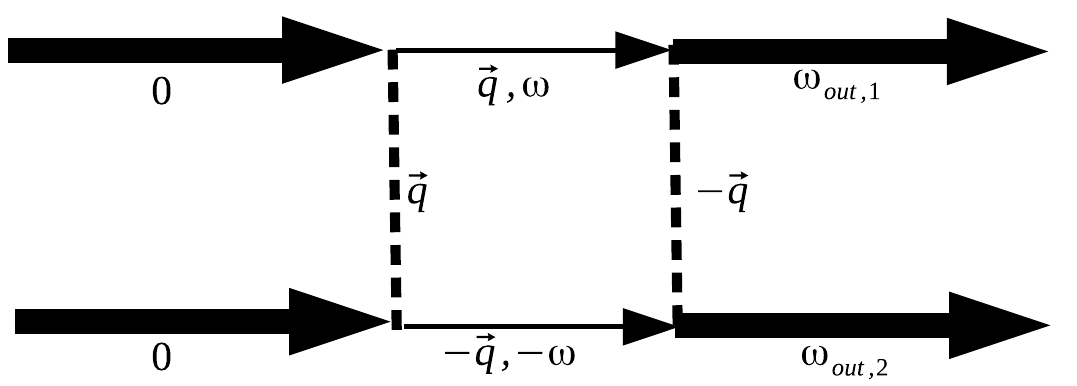}\emph{c})\includegraphics[height=2cm]{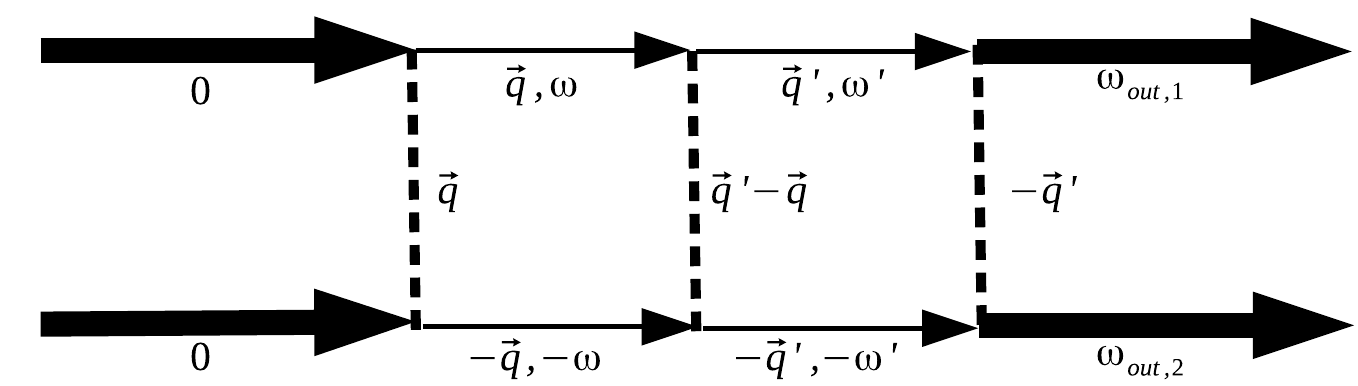}
\par\end{centering}
\begin{centering}
\emph{d})\includegraphics[height=2cm]{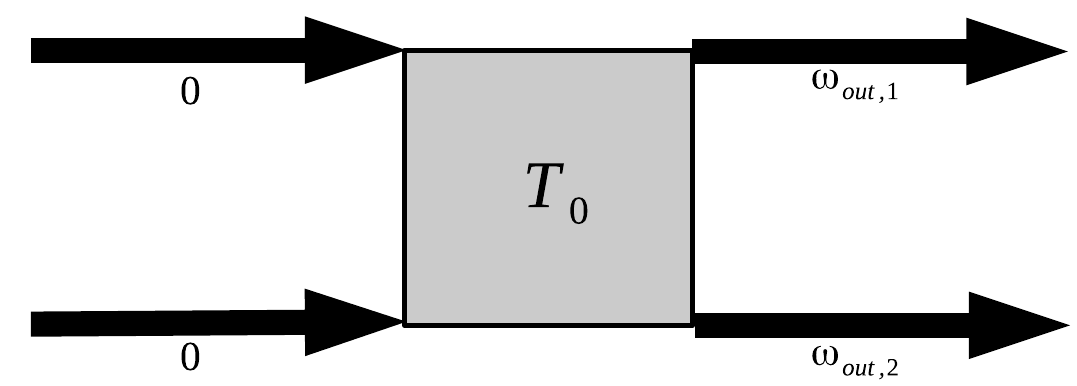}\emph{e})\includegraphics[height=2cm]{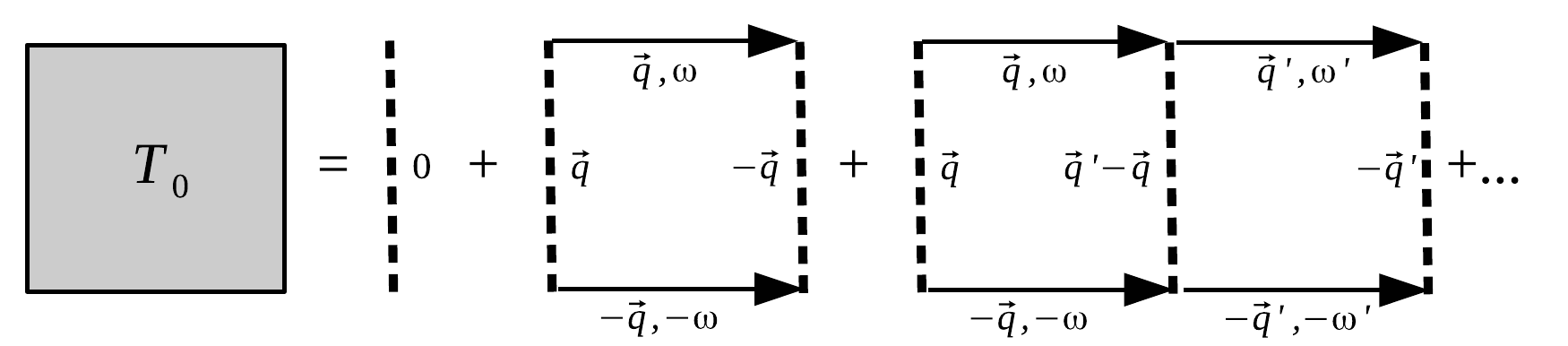}
\par\end{centering}
\caption{Feynman diagrams corresponding to\emph{ a}) first order, \emph{b})
second order, \emph{c}) third order contributions in $H_{dd}$. \emph{d})
Schematic representation of the sum of all orders starting with the
first\emph{. e}) diagrammatic representation of the perturbative expansion
of $T_{0}$. Here the thick arrows represent the Green's functions
$G_{ac},G_{ca}$, thin arrows stand for the polaritonic propagator
$G_{cc}$ and, finally the dashed vertical lines represent the Fourier
transform of the interaction potential $U$.}

\label{Fig_feyn}
\end{figure}

We now perform a perturbative expansion of Eq. (\ref{eq:CTP_aa})
with respect to the Hamiltonian of dipole-dipole interactions $H_{dd}$.
As shown in App. \ref{sec:Pair-correlation-function}, each term of
this expansion can be represented by a diagram (see Fig. \ref{Fig_feyn}).
More explicitly, denoting by the subscript $\left(i,j\right)$ a perturbation
order of the corresponding correlation function in $\alpha$ and $H_{dd}$
respectively, we get: 
\[
\left\langle {\cal T}\left\{ a\left(\omega_{out,1}\right)a\left(\omega_{out,2}\right)\right\} \right\rangle ^{\left(2\right)}=\left\langle a\left(\omega_{out,1}\right)\right\rangle ^{\left(1\right)}\left\langle a\left(\omega_{out,2}\right)\right\rangle ^{\left(1\right)}+\sum_{p>0}\left\langle {\cal T}\left\{ a\left(\omega_{out,1}\right)a\left(\omega_{out,2}\right)\right\} \right\rangle ^{\left(2,p\right)},
\]
where 

\begin{eqnarray}
\sum_{p>0}\left\langle {\cal T}\left\{ a\left(\omega_{out,1}\right)a\left(\omega_{out,2}\right)\right\} \right\rangle ^{\left(2,p\right)} & = & {\rm i}\alpha^{2}\delta\left(\omega_{out,1}+\omega_{out,2}\right)G_{ac_{0}}^{T}\left[\omega_{out,1}\right]G_{ac_{0}}^{T}\left[\omega_{out,2}\right]T_{0}\left(G_{c_{0}a}^{T}\left[0\right]\right)^{2}\label{eq:CTP_aa_p>1}
\end{eqnarray}
and by direct translation of Fig. \ref{Fig_feyn} (e) 

\begin{eqnarray}
T_{0} & = & U_{0}+{\rm i}\sum_{\vec{q}}U_{-\vec{q}}S_{\vec{q}}U_{\vec{q}}+{\rm i}^{2}\sum_{\vec{q}}U_{-\vec{q}}S_{\vec{q}}\sum_{\vec{q}^{\prime}}U_{\vec{q}-\vec{q}^{\prime}}S_{\vec{q}^{\prime}}U_{q^{\prime}}+\cdots\label{eq:CTP_t_0_vague}
\end{eqnarray}
with 
\begin{equation}
S_{\vec{q}}\equiv\frac{1}{2\pi}\int d\omega G_{c_{\vec{q}},c_{\vec{q}}}^{T}\left[\omega\right]G_{c_{-\vec{q}},c_{-\vec{q}}}^{T}\left[-\omega\right].\label{eq:S_q}
\end{equation}

In Eq. (\ref{eq:CTP_aa_p>1}) the term $\left(G_{c_{0}a}^{T}\left[0\right]\right)^{2}$
stands for the conversion of two incoming photons into symmetric Rydberg
polaritons. Resulting from the resummation of diagrams of all perturbative
orders in $H_{dd}$, the term $\frac{\left(-{\rm i}T_{0}\right)}{2\pi}$
represents the action of the Rydberg dipole-dipole-interaction-induced
non-linearity on the two symmetric polaritons, provided they return
to the symmetric subspace. Finally the term $G_{ac_{0}}^{T}\left[\omega_{out,1}\right]G_{ac_{0}}^{T}\left[\omega_{out,2}\right]$
represents the conversion of two symmetric polaritons back to the
cavity mode photons. In conclusion, we note that $T_{0}$ can be derived
analytically as we show in App. \ref{subsec:CTP_T-matrix}.

\section{$G^{\left(1\right)}$ correlation function\label{sec:CTP_Transmission-spectrum}}

In this section we use Schwinger-Keldysh contour formalism in order
to compute the correlation function $G_{out}^{\left(1\right)}\left(t,t^{\prime}\right)=\left\langle a_{out}^{\left(R\right)\dagger}\left(t\right)a_{out}^{\left(R\right)}\left(t^{\prime}\right)\right\rangle $
of the light transmitted through the cavity at fourth order in the
feeding rate $\alpha$ as presented in \citep{GBB16}. By virtue of
input-output relations this quantity is proportional to the correlation
function of the intracavity fields, i.e. $G_{out}^{\left(1\right)}\left(t,t^{\prime}\right)=2\gamma_{c}^{\left(R\right)}\left\langle a^{\dagger}\left(t\right)a\left(t^{\prime}\right)\right\rangle $.

At second order in the feeding rate $\alpha$ Eq. (\ref{eq:CTP_the_most_general})
straightforwardly yields $\left\langle a^{\dagger}\left(t\right)a\left(t^{\prime}\right)\right\rangle ^{\left(2\right)}=\left\langle a^{\dagger}\left(t\right)\right\rangle ^{\left(1\right)}\left\langle a\left(t^{\prime}\right)\right\rangle ^{\left(1\right)}$
which agrees with the factorization property shown in \citep{GBB15}.
At fourth order in $\alpha$ this property does not hold any longer
and Eq. (\ref{eq:CTP_the_most_general}) writes:

\begin{equation}
\left\langle a^{\dagger}\left(t\right)a\left(t^{\prime}\right)\right\rangle ^{\left(4\right)}=\frac{\left(-\mathrm{i}\sqrt{2\pi}\alpha\right)^{4}}{4}\left\langle {\cal T}_{{\cal C}}\left\{ e^{-\mathrm{i}\left(\int_{{\cal C}}H_{dd}\right)}A_{q}^{2}A_{q}^{\dagger2}a_{-}^{\dagger}\left(t\right)a_{+}\left(t^{\prime}\right)\right\} \right\rangle \label{eq:CTP_aa^4}
\end{equation}
Note that we placed the operators $a_{-}^{\dagger}\left(t\right)$
and $a_{+}\left(t\right)$ on the $\mathcal{C}_{-}$ and $\mathcal{C}_{+}$
branches, respectively, in order to impose the normal ordering $a^{\dagger}\left(t\right)a\left(t^{\prime}\right)$
whatever $t$ and $t^{\prime}$ are. We now expand the correlation
function Eq. (\ref{eq:CTP_aa^4}) with respect to the dipole-dipole
interactions, separating the forward and backward branches of the
contour $\mathcal{C}$ as follows: 

\begin{eqnarray}
 &  & \left\langle a^{\dagger}\left(t\right)a\left(t^{\prime}\right)\right\rangle ^{\left(4\right)}\label{eq:CTP_pq_expansion}\\
 & = & \frac{\left(-\mathrm{i}\sqrt{2\pi}\alpha\right)^{4}}{4}\sum_{p,q}\frac{\left(-{\rm i}\right)^{p+q}}{p!q!}\left\langle {\cal T}_{{\cal C}}\left\{ \left(\int_{{\cal C}_{+}}dsH_{dd}\left(s\right)\right)^{p}\left(\int_{{\cal C}_{-}}dsH_{dd}\left(s\right)\right)^{q}A_{q}^{2}A_{q}^{\dagger2}a_{-}^{\dagger}\left(t\right)a_{+}\left(t^{\prime}\right)\right\} \right\rangle \nonumber 
\end{eqnarray}

We first consider the partial resummation ${\cal E}\left(t,t^{\prime}\right)=\sum_{p=0,q>0}+\sum_{p>0,q=0}$
of Eq. (\ref{eq:CTP_pq_expansion}). Omitting the technical details
of the derivation, which are provided in App. \ref{sec:Transmission-Spectrum-technical},
the Fourier transform of this contribution can be put under the form 

\begin{eqnarray}
{\cal E}\left(\omega,\omega^{\prime}\right) & = & -2\pi\alpha^{4}\delta\left(\omega^{\prime}\right)\delta\left(\omega\right)G_{aa}^{*}\left[0\right]\left(G_{ac_{0}}^{T}\left[0\right]G_{ac_{0}}^{\tilde{T}}\left[0\right]\right)T_{0}\left(G_{c_{0}a}^{T}\left[0\right]\right)^{2}+{\rm c.c.}\label{eq:Eww}
\end{eqnarray}
where $T_{0}$ was defined in Eq. (\ref{eq:CTP_t_0_vague}). Since
$\left\langle a^{\dagger}\left(\omega\right)\right\rangle ^{\left(3\right)}\varpropto\delta\left(\omega\right),$
and $\left\langle a\left(\omega^{\prime}\right)\right\rangle ^{\left(1\right)}\varpropto\delta\left(\omega^{\prime}\right)$,
we see that ${\cal E}\left(\omega,\omega^{\prime}\right)$ corresponds
to the elastic part of the $G_{out}^{\left(1\right)}$ function at
fourth order in feeding, i.e. $\left\langle a_{out}^{\left(R\right)\dagger}\left(\omega\right)a_{out}^{\left(R\right)}\left(\omega^{\prime}\right)\right\rangle _{el}\varpropto\delta\left(\omega\right)\delta\left(\omega^{\prime}\right)$,
which is due to photons propagating through cavity without changing
their frequency. Note that this fourth-order elastic contribution
is actually a correction of the (necessarily elastic) second order
spectrum. 

\subsection{Inelastic contribution \label{subsec:CTP_inelastic}}

As shown in App. \ref{sec:Transmission-Spectrum-technical}, the remaining
contribution to Eq. (\ref{eq:CTP_pq_expansion}) takes the following
form

\begin{eqnarray}
{\cal I}\left(\omega,\omega^{\prime}\right) & = & -\alpha^{4}\delta\left(\omega-\omega^{\prime}\right)\left|T_{0}\right|^{2}{\rm i}G_{c_{0}c_{0}}^{>}\left[-\omega\right]G_{c_{0}a}^{\tilde{T}}\left[\omega^{\prime}\right]G_{ac_{0}}^{T}\left[\omega\right]\left(G_{c_{0}a}^{T}\left[0\right]\right)^{2}\left(G_{ac_{0}}^{\tilde{T}}\left[0\right]\right)^{2}\label{eq:CTP_aa_inelast}
\end{eqnarray}
and brings nonlinearity-induced inelastic features which were absent
at lower orders.

The results derived above allows us to investigate the spectrum of
the transmitted light

\[
\mathcal{S}^{out}\left(\omega\right)\equiv\int d\omega^{\prime}G_{out}^{\left(1\right)}\left(\omega,\omega^{\prime}\right)
\]

\begin{figure}
\begin{raggedright}
\emph{\Large{}a}{\Large{})}\includegraphics[scale=0.6]{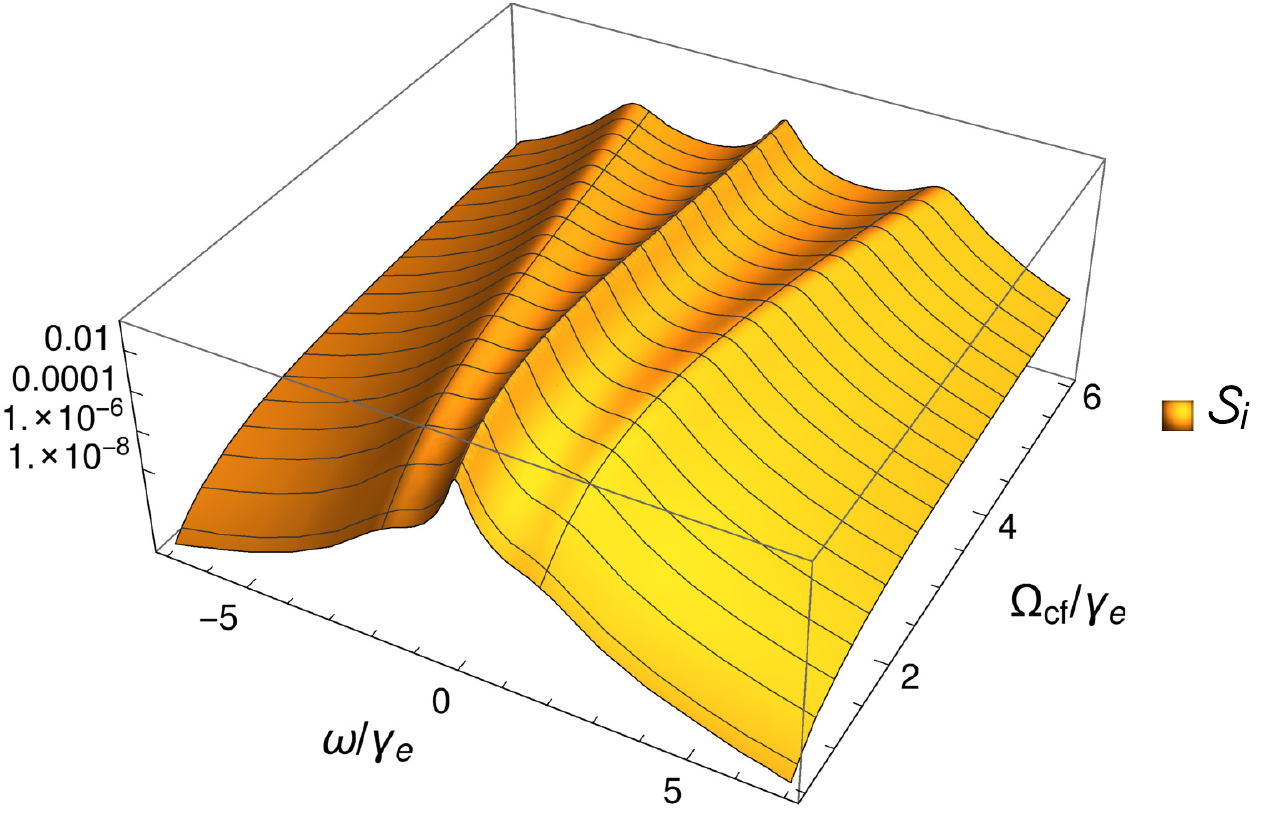}\emph{\Large{}b}{\Large{})}\includegraphics[scale=0.6]{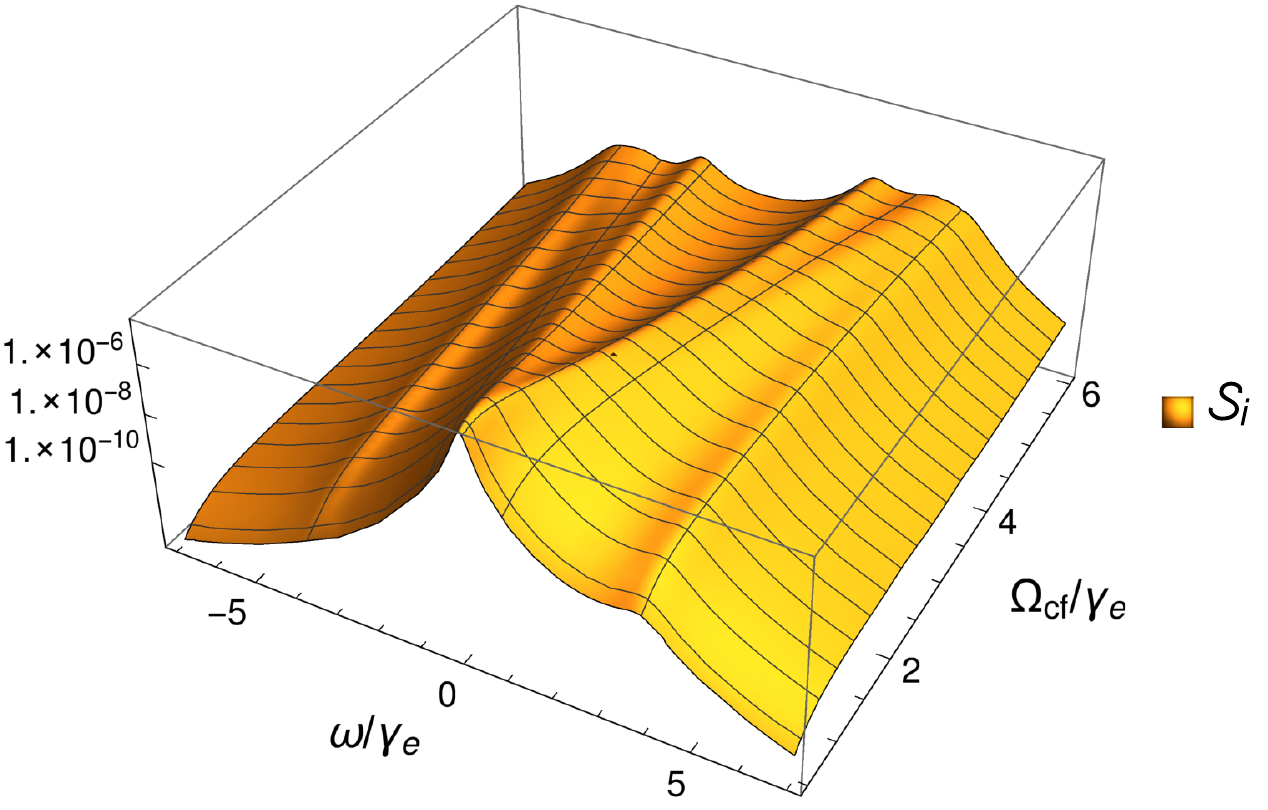}
\par\end{raggedright}
\caption{Inelastic component of the cavity transmission spectrum ${\cal S}_{i}\equiv2\gamma_{c}^{R}\int d\nu\left\langle a^{\dagger}\left(\omega\right),a\left(\nu\right)\right\rangle $
in logarithmic scale as a function of $\Omega_{cf}$ and the frequency
(in the frame rotating at $\omega_{p}$) for: a) the resonant case
$\Delta_{c}=\Delta_{e}=\Delta_{r}=0$, b) the detuned case. The transverse
curves give $\left(\pm\epsilon_{1},\pm\epsilon_{2},\pm\epsilon_{3}\right)$
as functions of $\Omega_{cf}$ (see main text).}

\label{FIG1}
\end{figure}

We are particularly interested in the inelastic part, \emph{i.e.}
${\cal S}_{i}^{out}\equiv2\gamma_{c}^{R}\int_{-\infty}^{\infty}d\nu\mathcal{I}\left(\omega,\nu\right)$
(see Eq. (\ref{eq:CTP_aa_inelast})), which is represented in Fig.
\ref{FIG1} in resonant ($\Delta_{c}=\Delta_{e}=\Delta_{r}=0$) as
well as detuned ($\Delta_{c}=-3\gamma_{e},\Delta_{e}=0,\Delta_{r}=0$)
configurations. For both regimes we assume a cloud cooperativity $C=5$,
and $\gamma_{c}^{R}=0.3\gamma_{e}\gg\gamma_{c}^{L}$ and $\gamma_{r}=0.15\gamma_{e}$
for the cavity and Rydberg decays respectively. All parameters are
expressed in units of the intermediate state decay rate $\gamma_{e}=2\pi\times3{\rm MHz}$. 

As can be seen on Fig. \ref{FIG1} the spectrum exhibits several resonances
which depend on the control field Rabi frequency. The resonance structure
shown in Fig. \ref{FIG1} (a) (resonant case) resembles the level
pattern of the Hamiltonian in the single excitation subspace

\[
\left(\begin{array}{ccc}
0 & g\sqrt{N} & 0\\
g\sqrt{N} & 0 & \frac{\Omega}{2}\\
0 & \frac{\Omega}{2} & 0
\end{array}\right)
\]

\begin{figure}
\begin{centering}
\includegraphics[width=0.4\paperwidth]{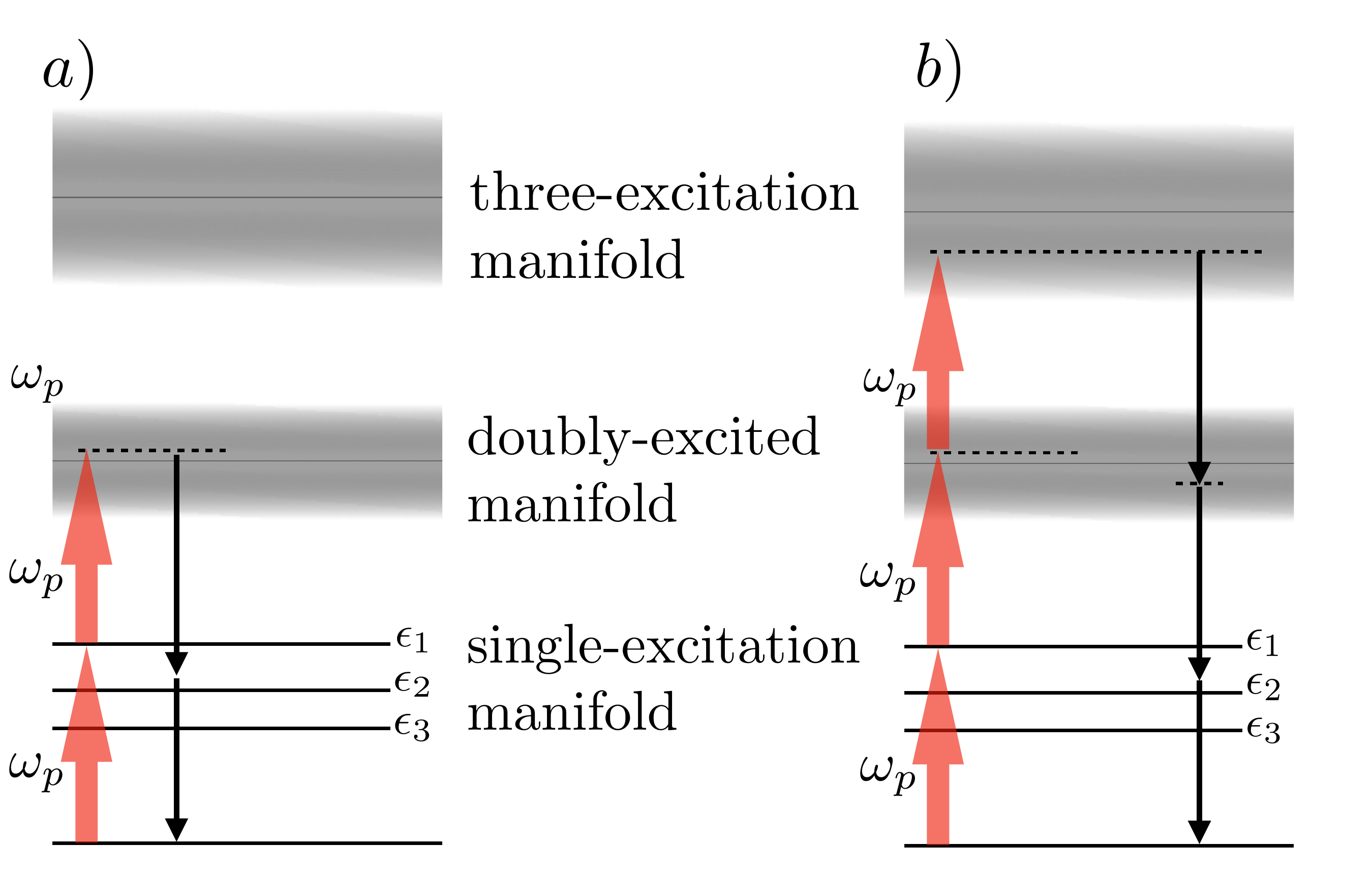}
\par\end{centering}
\caption{Schematic structure of the cavity-atom coupled system restricted to
at most three excitations. The two-photon (a) and three-photon (b)
cascades allow to understand the features observed in the cavity transmission
spectrum (see Sec. \ref{sec:CTP_Transmission-spectrum}) and the three-photon
correlation function (Sec. \ref{sec:Three-body-effects})}

\label{FIG2}
\end{figure}

In the detuned case, the structure shown on Fig. \ref{FIG1} (b) is
more complicated; resonances can still be identified as the eigenvalues
$\epsilon_{1},\epsilon_{2},\epsilon_{3}$ of the Hamiltonian 

\begin{equation}
\left(\begin{array}{ccc}
-\Delta_{c} & g\sqrt{N} & 0\\
g\sqrt{N} & -\Delta_{e} & \frac{\Omega}{2}\\
0 & \frac{\Omega}{2} & -\Delta_{r}
\end{array}\right)\label{eq:Single_Ham}
\end{equation}
but taken with positive and negative signs. This effect can be understood
by inspecting the level structure of the considered system, shown
in Fig. (a) \ref{FIG2} \citep{OKK11}. The system can be excited
by two photons of the probe laser frequency $\omega_{p}$. The strength
of dipole-dipole interactions does not affect the $\omega$-dependence
of the inelastic component at fourth order since $H_{dd}$ enters
Eq. (\ref{eq:CTP_aa_inelast}) only via the overall frequency-independent
factor $\left|T_{0}\right|^{2}$. Doubly excited states decay via
dissipative terms shown in the Heisenberg-Langevin equations Eqs.
(\ref{eq:CTP_bosonic1_k=00003D0}-\ref{eq:CTP_bosonic3_k=00003D0})
to three symmetric polaritons of energies $\epsilon_{1},\epsilon_{2}$
and $\epsilon_{3}$ respectively. The resonance frequencies of the
emitted photon pairs are therefore $\omega_{p}\pm\epsilon_{1}$, $\omega_{p}\pm\epsilon_{2}$
and $\omega_{p}\pm\epsilon_{3}$, respectively, or, in the frame rotating
at the probe frequency $\omega_{p}$, $\pm\epsilon_{1}$, $\pm\epsilon_{2}$
and $\pm\epsilon_{3}$. 

\section{Three-body effects\label{sec:Three-body-effects}}

Until now, we dealt with quantities whose perturbative expansion involved
at most two excitations. In this section we compute the three-photon
wavefunction of the transmitted light $\left\langle a_{out}\left(\omega_{1}\right)a_{out}\left(\omega_{2}\right)a_{out}\left(\omega_{3}\right)\right\rangle ^{\left(3\right)}$
whose calculation involves three-body terms. As in previous sections,
we use input-output theory to relate $a_{out}$ to the intracavity
field operator, and get:

\begin{equation}
\left\langle a_{out}\left(\omega_{1}\right)a_{out}\left(\omega_{2}\right)a_{out}\left(\omega_{3}\right)\right\rangle ^{\left(3\right)}=\left(\sqrt{2\gamma_{c}}\right)^{3}\left\langle {\cal T}\left\{ a\left(\omega_{1}\right)a\left(\omega_{2}\right)a\left(\omega_{3}\right)\right\} \right\rangle ^{\left(3\right)}\label{eq:a_out^3}
\end{equation}
Since we are focusing only on three-body effects, we discard one-
and two-body terms by considering the quantity $\left\langle {\cal T}\left\{ \delta a\left(\omega_{1}\right)\delta a\left(\omega_{2}\right)\delta a\left(\omega_{3}\right)\right\} \right\rangle ^{\left(3\right)}$
where $\delta a\left(\omega\right)\equiv a\left(\omega\right)-\left\langle a\left(\omega\right)\right\rangle $:

\begin{align}
 & \left\langle {\cal T}\left\{ \delta a\left(\omega_{1}\right)\delta a\left(\omega_{2}\right)\delta a\left(\omega_{3}\right)\right\} \right\rangle ^{\left(3\right)}\label{eq:aaa}\\
= & \left\langle {\cal T}\left\{ a\left(\omega_{1}\right)a\left(\omega_{2}\right)a\left(\omega_{3}\right)\right\} \right\rangle ^{\left(3\right)}-\left\langle {\cal T}\left\{ a\left(\omega_{1}\right)a\left(\omega_{2}\right)\right\} \right\rangle ^{\left(2\right)}\left\langle a\left(\omega_{3}\right)\right\rangle ^{\left(1\right)}-\left\langle {\cal T}\left\{ a\left(\omega_{1}\right)a\left(\omega_{3}\right)\right\} \right\rangle ^{\left(2\right)}\left\langle a\left(\omega_{2}\right)\right\rangle ^{\left(1\right)}\nonumber \\
 & -\left\langle {\cal T}\left\{ a\left(\omega_{2}\right)a\left(\omega_{3}\right)\right\} \right\rangle ^{\left(2\right)}\left\langle a\left(\omega_{1}\right)\right\rangle ^{\left(1\right)}+2\left\langle a\left(\omega_{1}\right)\right\rangle ^{\left(1\right)}\left\langle a\left(\omega_{2}\right)\right\rangle ^{\left(1\right)}\left\langle a\left(\omega_{3}\right)\right\rangle ^{\left(1\right)}\nonumber 
\end{align}
This term corresponds to the contribution of connected diagrams only
in the full perturbative expansion of $\left\langle {\cal T}\left\{ a\left(\omega_{1}\right)a\left(\omega_{2}\right)a\left(\omega_{3}\right)\right\} \right\rangle ^{\left(3\right)}$.
We now represent Eq. \eqref{eq:aaa} in the contour-ordered form;
noticing that only the forward part of the contour gives a non-vanishing
contribution (by the same kind of arguments as in Sec. \ref{sec:CTP_Intensity-correlation})
we get

\begin{equation}
\left\langle {\cal T}\left\{ \delta a\left(\omega_{1}\right)\delta a\left(\omega_{2}\right)\delta a\left(\omega_{3}\right)\right\} \right\rangle ^{\left(3\right)}=\frac{\left(-i\alpha\sqrt{2\pi}\right)^{3}}{3!}\left\langle {\cal T}\left\{ a\left(\omega_{1}\right)a\left(\omega_{2}\right)a\left(\omega_{3}\right)e^{-i\int\left(H_{dd}\right)}\left(a^{\dagger}\left(0\right)\right)^{3}\right\} \right\rangle _{conn}\label{eq:aaa_int}
\end{equation}
where the ``conn'' subscript stands for the summation over connected
diagrams only. We can now expand the righthand side of this expression
using Wick's theorem.  The resummation of diagrams is much simpler
if we combine them in groups, following Faddeev's original approach
to three-body scattering \citep{F60}. 

\subsection{Partial resummation}

\begin{figure}
\begin{centering}
\includegraphics[scale=0.6]{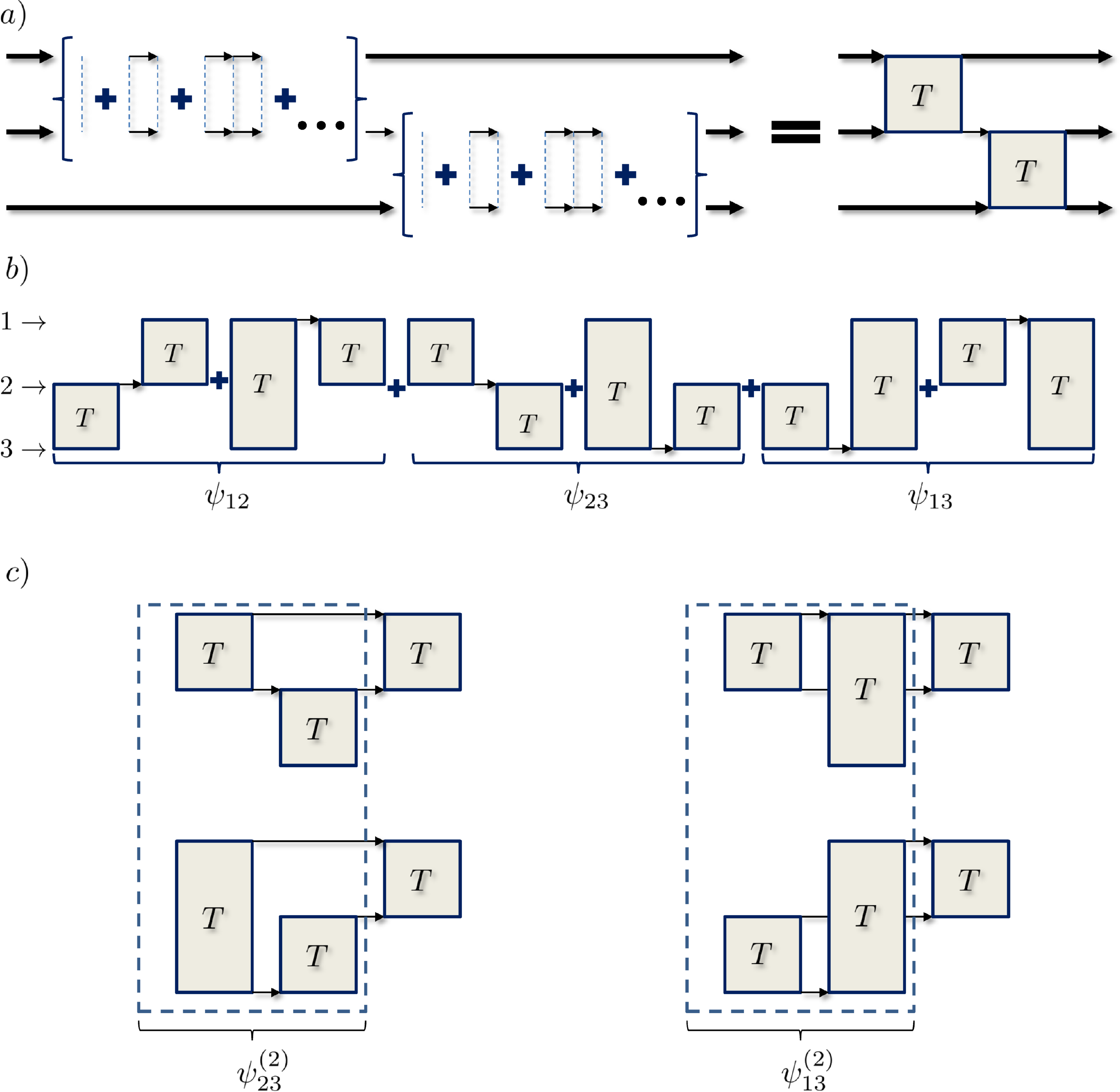}
\par\end{centering}
\caption{Diagrammatic expansion of the three-body correlation function: a)
combination of ladder diagrams into two-body $T$-matrices, b) effective
perturbative expansion with respect to $T$-matrices, c) expansion
of $\psi_{12}^{\left(3\right)}$ to the third order with respect to
$T$-matrices.}

\label{Fig_combination}
\end{figure}

In this subsection we perform the partial resummation of the perturbative
expansion of Eq. (\ref{eq:aaa_int}). To this end, we first group
diagrams so as to form elements of the two-body $T$-matrix expansion,
as shown in Fig. \ref{Fig_combination} a) on an example. Such a combination
of diagrams greatly facilitates the resummation: now the real interaction
potential $\kappa_{ij}$ is merely replaced by $T$ matrices. The
latter can effectively be considered as a new perturbation parameter,
with respect to which the original correlation function Eq. (\ref{eq:aaa_int})
is expanded. The second-order terms in $T$ matrices of perturbation
series are shown on Fig. \ref{Fig_combination} b) (the external lines
are omitted analogously to App. \ref{subsec:CTP_T-matrix}). It is
natural to collect diagrams into three groups, depending on which
lines are connected by the rightmost $T$-matrix. The corresponding
partial resummations of the three sets, which can be called Faddeev
components \citep{F60}, are denoted by $\psi_{kl}\left(k,l=1,2,3\right)$
where $\left(k,l\right)$ are line indexes (see fig. \ref{Fig_combination}
c). 

The first approximation we employ is to neglect hopping between atoms
through the cavity, that is $T$ coincides with $\mathring{T}$ when
$g$ is set to zero (see App. \ref{subsec:CTP_T-matrix}). This constitutes
a valid assumption in the regime when the number of blockaded atoms
is much smaller than the total number of atoms. Under this approximation,
it is convenient to work with real atomic positions instead of the
reciprocal space. Expressed in the temporal Fourier space Eq. (\ref{eq:aaa_int})
therefore takes the form:

\begin{align}
 & \left\langle {\cal T}\left\{ \delta a\left(\omega_{1}\right)\delta a\left(\omega_{2}\right)\delta a\left(\omega_{3}\right)\right\} \right\rangle ^{\left(3\right)}\label{eq:aaa_psi}\\
= & \sum_{r_{i},r_{j},r_{k}}\left(\alpha\sqrt{2\pi}\right)^{3}\delta\left(\sum\omega_{i}\right)G_{c_{i}a}\left[0\right]G_{c_{j}a}\left[0\right]G_{c_{k}a}\left[0\right]\left(\psi_{12}\left[\omega_{3}\right]+\psi_{23}\left[\omega_{2}\right]+\psi_{13}\left[\omega_{1}\right]\right)G_{ac_{i}}\left[\omega_{1}\right]G_{ac_{j}}\left[\omega_{2}\right]G_{ac_{k}}\left[\omega_{3}\right]\nonumber \\
= & \left(\alpha\sqrt{2\pi}G_{c_{0}a}\left[0\right]\right)^{3}\frac{\delta\left(\sum\omega_{i}\right)}{N^{3}}\sum_{r_{1},r_{2},r_{3}}\left(\psi_{12}\left[\omega_{3}\right]+\psi_{23}\left[\omega_{2}\right]+\psi_{13}\left[\omega_{1}\right]\right)G_{ac_{0}}\left[\omega_{1}\right]G_{ac_{0}}\left[\omega_{2}\right]G_{ac_{0}}\left[\omega_{3}\right]\nonumber 
\end{align}
where we used the following relation between Green's functions $G_{ac_{0}}\left[\omega\right]\equiv\frac{1}{\sqrt{N}}\sum_{i}G_{ac_{i}}\left[\omega\right]=\sqrt{N}G_{ac_{i,k,j}}\left[\omega\right]$.
According to these remarks, the second-order term in the expansion
can be written in the following algebraic form:

\begin{align}
\vec{\psi}^{\left(2\right)} & \equiv\left(\begin{array}{c}
\psi_{12}^{\left(2\right)}\left[\omega\right]\\
\psi_{23}^{\left(2\right)}\left[\omega\right]\\
\psi_{13}^{\left(2\right)}\left[\omega\right]
\end{array}\right)=\left(\frac{-i}{2\pi}\right)^{2}\hat{B}\left[-\omega\right]iG_{c_{i}c_{i}}^{T}\left[-\omega\right]\vec{\psi}^{\left(1\right)}\label{eq:psi^2_nohp}\\
\text{where } & \vec{\psi}^{\left(1\right)}\equiv\left(\begin{array}{c}
\mathring{T}_{12}\left[0\right]\\
\mathring{T}_{23}\left[0\right]\\
\mathring{T}_{13}\left[0\right]
\end{array}\right)\nonumber 
\end{align}
$G_{c_{i}c_{i}}$ is the atomic Green's function in real space under
the no-hopping assumption (since all atoms are equivalent, these functions
are equal), and we defined
\[
\hat{B}\left[\omega\right]\equiv\left(\begin{array}{ccc}
0 & \mathring{T}_{12}\left[\omega\right] & \mathring{T}_{12}\left[\omega\right]\\
\mathring{T}_{23}\left[\omega\right] & 0 & \mathring{T}_{23}\left[\omega\right]\\
\mathring{T}_{13}\left[\omega\right] & \mathring{T}_{13}\left[\omega\right] & 0
\end{array}\right)
\]
It is important to note, that $\vec{\psi}^{\left(1\right)}$ is not
present in the expansion of Eq. (\ref{eq:aaa_int}), as it is a part
of disconnected diagram series. The expansion of the $\psi_{12}$
component to the third order in $T$ is shown on Fig. \ref{Fig_combination}
c) (the integration over each closed loop is implicit). From diagrammatics,
we deduce:

\begin{align*}
\psi_{12}^{\left(3\right)} & =\frac{-i}{2\pi}T_{12}\left[-\omega\right]\int d\xi iG_{c_{i}c_{i}}\left[-\xi-\omega\right]iG_{c_{i}c_{i}}\left[\xi\right]\psi_{23}^{\left(2\right)}\left[\xi\right]\\
 & +\frac{-i}{2\pi}T_{12}\left[-\omega\right]\int d\xi iG_{c_{i}c_{i}}\left[-\xi-\omega\right]iG_{c_{i}c_{i}}\left[\xi\right]\psi_{13}^{\left(2\right)}\left[\xi\right]
\end{align*}
The structure of $\psi_{13}^{\left(3\right)}$ and $\psi_{23}^{\left(3\right)}$
is similar to those of $\psi_{12}^{\left(3\right)}$, and the full
vector $\vec{\psi}^{\left(3\right)}$ can be put under the following
matrix form:

\begin{equation}
\vec{\psi}^{\left(3\right)}\equiv\left(\begin{array}{c}
\psi_{12}^{\left(3\right)}\left[\omega\right]\\
\psi_{23}^{\left(3\right)}\left[\omega\right]\\
\psi_{13}^{\left(3\right)}\left[\omega\right]
\end{array}\right)=\frac{-i}{2\pi}\hat{B}\left[-\omega\right]\int d\xi iG_{c_{i}c_{i}}\left[-\xi-\omega\right]iG_{c_{i}c_{i}}\left[\xi\right]\vec{\psi}^{\left(2\right)}\left[\xi\right]\label{eq:psi_3}
\end{equation}
and, analogously, for the $\left(n+1\right)$-th order we can write:

\begin{equation}
\vec{\psi}^{\left(n+1\right)}=\frac{-i}{2\pi}\hat{B}\left[-\omega\right]\int d\xi iG_{c_{i}c_{i}}\left[-\xi-\omega\right]iG_{c_{i}c_{i}}\left[\xi\right]\vec{\psi}^{\left(n\right)}\left[\xi\right]\label{eq:psi_iterative}
\end{equation}
This iterative equation is equivalent to the following self-consistent
equation (this can be checked directly by iterating the latter):

\begin{equation}
\vec{\psi}\left[\omega\right]=\frac{-i}{2\pi}\hat{B}\left[-\omega\right]\int d\xi iG_{c_{i}c_{i}}\left[-\xi-\omega\right]iG_{c_{i}c_{i}}\left[\xi\right]\vec{\psi}\left[\xi\right]+\vec{\psi}^{\left(2\right)}\left[\omega\right]\label{eq:psi_SC}
\end{equation}

Let us first analyze the expression Eq. (\ref{eq:psi_iterative}).
The Green's function $G_{c_{i}c_{i}}$ has two poles defined by $\omega_{p,1},\omega_{p,2}$
in the lower part of the complex plane \citep{AGD} (poles correspond
to the two possible dressed states). Let us assume for now that the
full solution of Eq. (\ref{eq:psi_SC}) $\vec{\psi}\left[\omega\right]$
has poles only in the upper part; we can then perform the integration
in the right hand side of Eq. (\ref{eq:psi_SC}) and get via the residue
theorem:

\begin{equation}
\vec{\psi}\left[\omega\right]=\hat{B}\left[-\omega\right]\eta_{1}\left[-\omega\right]\vec{\psi}\left[\omega_{p_{1}}\right]+\hat{B}\left[-\omega\right]\eta_{2}\left[-\omega\right]\vec{\psi}\left[\omega_{p_{2}}\right]+\vec{\psi}^{\left(2\right)}\left[\omega\right]\label{eq:psi_Integrated}
\end{equation}
where $\eta_{j}\left[\omega\right]\equiv G_{c_{i}c_{i}}\left[\omega-\omega_{p_{j}}\right]\times\text{Res}_{\xi\rightarrow\omega_{p_{j}}}G_{c_{i}c_{i}}\left[\xi\right]$.
In order to resolve the self-dependence in Eq. (\ref{eq:psi_Integrated}),
we consider equations for $\vec{\psi}\left[\omega_{p_{1}}\right]$
and $\vec{\psi}\left[\omega_{p_{2}}\right]$ respectively:

\begin{align*}
\vec{\psi}\left[\omega_{p_{1}}\right]= & \hat{B}\left[-\omega_{p_{1}}\right]\eta_{1}\left[-\omega_{p_{1}}\right]\vec{\psi}\left[\omega_{p_{1}}\right]+\hat{B}\left[-\omega_{p_{1}}\right]\eta_{2}\left[-\omega_{p_{1}}\right]\vec{\psi}\left[\omega_{p_{2}}\right]+\vec{\psi}^{\left(2\right)}\left[\omega_{p_{1}}\right]\\
\vec{\psi}\left[\omega_{p_{2}}\right]= & \hat{B}\left[-\omega_{p_{2}}\right]\eta_{1}\left[-\omega_{p_{2}}\right]\vec{\psi}\left[\omega_{p_{1}}\right]+\hat{B}\left[-\omega_{p_{2}}\right]\eta_{2}\left[-\omega_{p_{2}}\right]\vec{\psi}\left[\omega_{p_{2}}\right]+\vec{\psi}^{\left(2\right)}\left[\omega_{p_{2}}\right]
\end{align*}
which form a closed set of linear equations yielding:

\begin{equation}
\left(\begin{array}{c}
\vec{\psi}\left[\omega_{p_{1}}\right]\\
\vec{\psi}\left[\omega_{p_{2}}\right]
\end{array}\right)=\left(\begin{array}{cc}
1-\hat{B}\left[-\omega_{p_{1}}\right]\eta_{1}\left[-\omega_{p_{1}}\right] & -\hat{B}\left[-\omega_{p_{1}}\right]\eta_{2}\left[-\omega_{p_{1}}\right]\\
-\hat{B}\left[-\omega_{p_{2}}\right]\eta_{1}\left[-\omega_{p_{2}}\right] & 1-\hat{B}\left[-\omega_{p_{2}}\right]\eta_{2}\left[-\omega_{p_{2}}\right]
\end{array}\right)^{-1}\left(\begin{array}{c}
\vec{\psi}^{\left(2\right)}\left[\omega_{p_{1}}\right]\\
\vec{\psi}^{\left(2\right)}\left[\omega_{p_{2}}\right]
\end{array}\right)\label{eq:psi_vectorEq}
\end{equation}
The analytical solution of Eq. (\ref{eq:psi_vectorEq}) which can
be obtained in e.g. Mathematica is cumbersome even in many limiting
cases. It can, however, be dealt with numerically. Using\textbf{ }Eq.
(\ref{eq:psi_vectorEq}) we can write the general solution of the
self-consistent equation:

\begin{equation}
\vec{\psi}\left[\omega\right]=\left(\hat{B}\left[-\omega\right]\eta_{1}\left[-\omega\right],\hat{B}\left[-\omega\right]\eta_{2}\left[-\omega\right]\right)\times\left(\begin{array}{c}
\vec{\psi}\left[\omega_{p_{1}}\right]\\
\vec{\psi}\left[\omega_{p_{2}}\right]
\end{array}\right)+\vec{\psi}^{\left(2\right)}\left[\omega\right]\label{eq:psi_final}
\end{equation}
Finally, we check \emph{a posteriori }our initial assumption, \emph{i.e.}
that Eq. (\ref{eq:psi_final}) indeed has poles only in the upper
part of the complex plane.

To conclude this subsection, we now briefly discuss how the reintroduction
of hopping between atoms modifies the previous results at lowest order
in $\nicefrac{V_{b}}{V}$. We first note that it is possible to derive
the exact expression of the term $\vec{\psi}^{\left(2\right)}\left[\omega\right]$
standing on the right handside of Eq. (\ref{eq:psi_final}) without
assuming $g=0$, \emph{i.e.} including hopping. Here we provide the
final expression only, as the derivation is the same as above, but
in the spatial Fourier space:

\begin{align*}
\sum_{r_{1},r_{2},r_{3}}\vec{\psi}_{g\neq0}^{\left(2\right)}\left[\omega\right] & =\left(\frac{-i}{2\pi}\right)^{2}iG_{c_{0}c_{0}}\left[-\omega\right]T_{0}\left[0\right]T_{0}\left[-\omega\right]\left(\begin{array}{c}
1\\
1\\
1
\end{array}\right)\\
 & \approx\left(\frac{-i}{2\pi}\right)^{2}iG_{c_{0}c_{0}}\left[-\omega\right]\mathring{T}_{0}\left[0\right]\mathring{T}_{0}\left[-\omega\right]\left(\begin{array}{c}
1\\
1\\
1
\end{array}\right)
\end{align*}
where $T_{0}$ is given by Eq. (\ref{eq:CTP_T_0_Final}), and in the
last line we kept only the leading terms in $V_{b}/V$. It can also
be shown that the contribution of hopping between atoms in the first
term on the right handside of Eq. (\ref{eq:psi_final}) is at least
of the third order in $\nicefrac{V_{b}}{V}$, and can therefore be
neglected in our approximation framework. Combining these remarks
we can now numerically derive the three body correlation function.
We note that this method can be extended to higher-order correlation
functions.

\subsection{Numerical results and discussion}

In this subsection we present the numerical results for the frequency
distribution of correlation function Eq. (\ref{eq:a_out^3}). 

\begin{figure}
\begin{centering}
\includegraphics[scale=0.6]{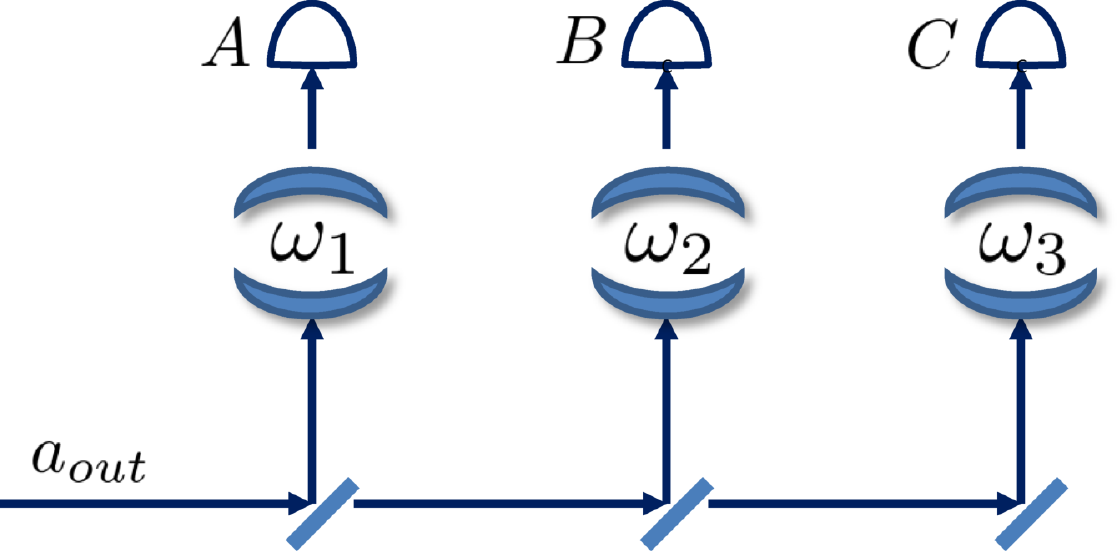}
\par\end{centering}
\caption{An experimental scheme for measurement of the three-photon correlation
function $\left|\left\langle \delta a_{out}\left(\omega_{1}\right)\delta a_{out}\left(\omega_{2}\right)\delta a_{out}\left(\omega_{3}\right)\right\rangle ^{\left(3\right)}\right|^{2}$.}

\label{FIG_detection}
\end{figure}

A possible experimental scheme to measure the three-body (frequency-resolved)
correlation function is suggested on Fig. \ref{FIG_detection}. We
assume that the light transmitted through the cavity is split into
three paths by means of standard beam-splitters and sent to three
different detectors, combined with narrow-band filters represented
by cavities. In this case, it can be shown, that the total number
of photons, jointly detected by all three detectors, is proportional
to:

\begin{equation}
\int_{-\infty}^{\infty}dt_{1}dt_{2}dt_{3}\left\langle A^{\dagger}\left(t_{1}\right)B^{\dagger}\left(t_{2}\right)C^{\dagger}\left(t_{3}\right)C\left(t_{3}\right)B\left(t_{2}\right)A\left(t_{1}\right)\right\rangle \sim\left\langle a_{out}^{\dagger}\left(\omega_{1}\right)a_{out}^{\dagger}\left(\omega_{2}\right)a_{out}^{\dagger}\left(\omega_{3}\right)a_{out}\left(\omega_{3}\right)a_{out}\left(\omega_{2}\right)a_{out}\left(\omega_{1}\right)\right\rangle \label{eq:ABC}
\end{equation}
where $A$, $B$ and $C$ are the respective annihilation operators
of the input fields impinging on the three corresponding detectors
(see Fig. \ref{FIG_detection}). $\omega_{1},\omega_{2}$ and $\omega_{3}$
are the frequencies (in the frame, rotating at $\omega_{p}$) of filters
(cavities). As can be seen from Eq. (\ref{eq:ABC}), it gives not
only the connected three-body contribution to the correlation function
but also the disconnected ones. However, if none of $\omega_{k}$'s
in Eq. Eq. (\ref{eq:ABC}) is equal to zero, the result will contain
only the desired part. 

\begin{figure}
\begin{centering}
\includegraphics[width=0.7\paperwidth]{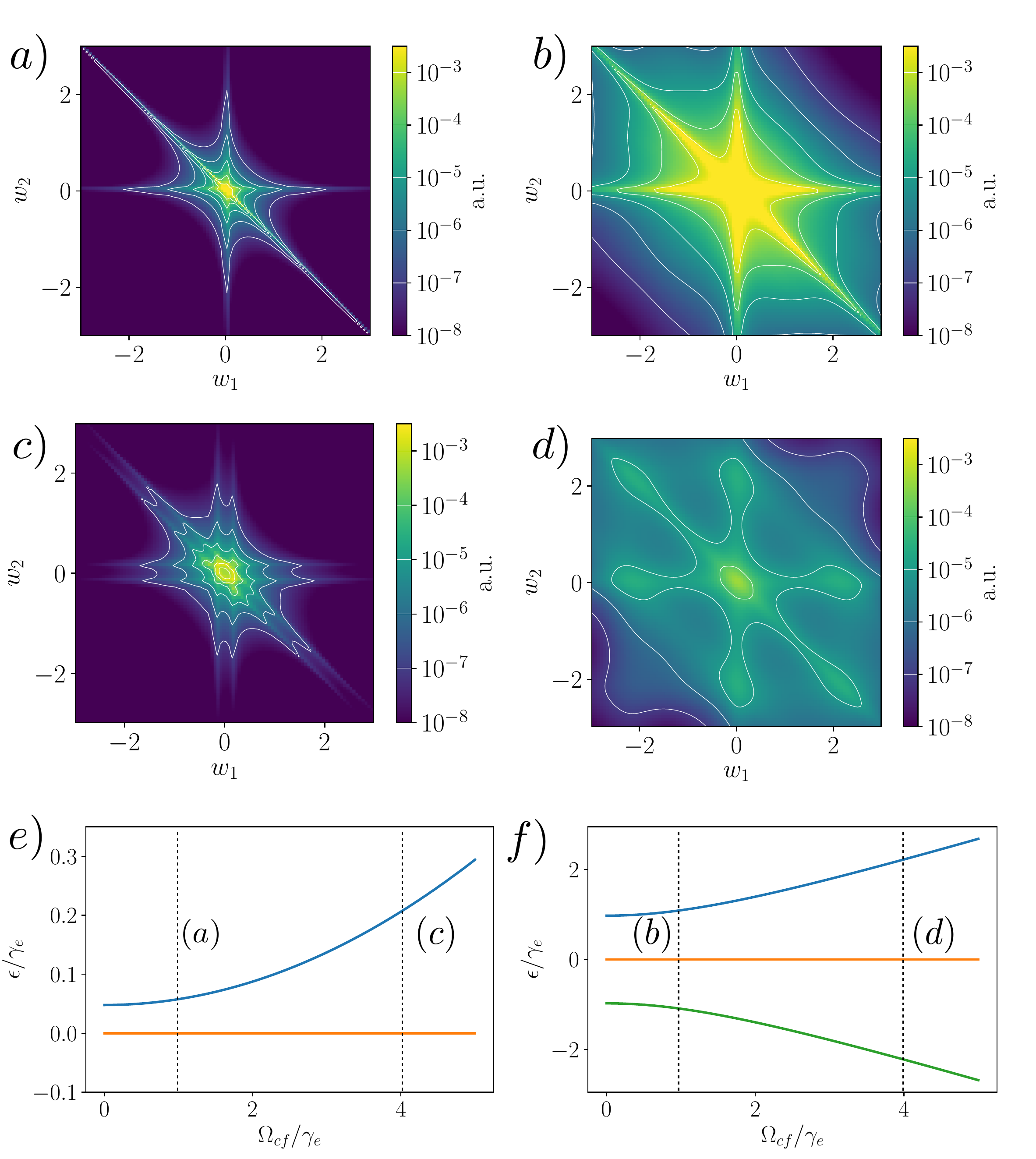}
\par\end{centering}
\caption{Three-photon correlation function $\left|\left\langle \delta a\left(\omega_{1}\right)\delta a\left(\omega_{2}\right)\delta a\left(-\omega_{1}-\omega_{2}\right)\right\rangle ^{\left(3\right)}\right|$
as a function of $\omega_{1}$ and $\omega_{2}$: for the parameters
a) $\Delta_{e}=-25\gamma_{e}$, $\Omega_{cf}=\gamma_{e}$, b) $\Delta_{e}=0$,
$\Omega_{cf}=\gamma_{e}$, c) $\Delta_{e}=-25\gamma_{e}$, $,\Omega_{cf}=4\gamma_{e}$,
d) $\Delta_{e}=0$, $\Omega_{cf}=4\gamma_{e}$. The polariton eigenstate
energy for: e) detuned regime $\Delta_{e}=-25\gamma_{e}$, f) resonant
case $\Delta_{e}=0$, $\epsilon_{1}$, $\epsilon_{2}$, $\epsilon_{3}$
energies are given by the blue, orange and green curves respectively.
In the far detuned regime one of the polaritons eigenenergy is equal
to $\epsilon_{3}$$\approx-25\gamma_{e}$ (not shown).}

\label{FIG_resum}
\end{figure}

Fig. \ref{FIG_resum} displays numerical simulations of the third-order
correlation function $\left\langle \delta a\left(\omega_{1}\right)\delta a\left(\omega_{2}\right)\delta a\left(-\omega_{1}-\omega_{2}\right)\right\rangle $
with respect to the frequencies $\left(\omega_{1},\omega_{2}\right)$
of two of the emitted photons \textendash{} the third frequency is
automatically set to $-\left(\omega_{1}+\omega_{2}\right)$ by conservation
of energy in the frame rotating at $\omega_{p}$ \textendash{} for
different values of the control field $\Omega_{cf}$ and intermediate
state detuning $\Delta_{e}$. 

As for the second-order correlation function, we may interpret the
structures observed on these plots by resorting to the three-photon
cascade picture of Fig. \ref{FIG2} (b) and the polaritonic structure
on Fig. \ref{FIG_resum} (e,f). When the coupled atom-cavity system
absorbs three incoming probe photons, it is promoted to the three-excitation
subspace, schematically represented as a quasi-continuum. When deexciting
to its ground-state, the system reemits three photons of respective
frequencies $\omega_{a}$ (upper transition), $\omega_{b}$ (intermediate
transition) and $\omega_{c}$ (lower transition). Note that the triplet
of frequencies $\left(\omega_{1},\omega_{2},-\omega_{1}-\omega_{2}\right)$
used in Fig. \ref{FIG_resum} alternatively play the role of all $6$
permutations of $\left(\omega_{a},\omega_{b},\omega_{c}\right)$.
The frequency $\omega_{c}$ of the lower transition is constrained
to three values given by the polaritonic eigenenergies $\epsilon_{1,2,3}$
represented on Fig. \ref{FIG_resum} (e,f). The frequency of the upper
transition frequency which couples the three-excitation and two-excitation
subspaces is not fixed but is rather constrained to belong to a window
of width $\Delta\omega$ that we shall assume symmetric around origin
(i.e. $\omega_{p}$ in the rotating frame). The intermediate transition
frequency is related to the other two by the energy conservation $\omega_{a}+\omega_{b}+\omega_{c}=0$.
Finally we get
\begin{align*}
-\frac{1}{2}\Delta\omega & \leq\omega_{a}\leq\frac{1}{2}\Delta\omega\\
\omega_{b} & =-\omega_{a}-\omega_{c}\\
\omega_{c} & =\epsilon_{k=1,2,3}
\end{align*}

In the far detuned case, diagonalizing the Hamiltonian of the atom-cavity-system
in the single-excitation subspace yields the polaritonic structure
of Fig. \ref{FIG_resum} (e) with two polaritons of close energies,
the third being too far away to be represented on the plot. When these
two energies are two close, i.e. when $\Omega_{cf}$ is too weak,
the two cases $\omega_{c}=\epsilon_{1,2}$ cannot be distinguished
: this case is represented on Fig. 8 (a) where three main lines can
be identified, one vertical corresponding to $\omega_{1}=\omega_{c}=\epsilon_{1,2}$,
one horizontal corresponding to $\omega_{2}=\omega_{c}=\epsilon_{1,2}$,
one antidiagonal corresponding to $\omega_{3}=-\omega_{1}-\omega_{2}=\omega_{c}=\epsilon_{1,2}$.
Note that the width $\Delta\omega$ limits the ``visibility window''
where the correlation function takes substantial values. By contrast,
when the control field is larger, the two cases $\omega_{c}=\epsilon_{1,2}$
can be distinguished, which yields a double structure of six lines
\textendash{} two vertical, two horizontal, two antidiagonal \textendash{}
as can be seen on Fig. 8 (c).

In the resonant case, the symmetry of the polaritonic energy scheme
of Fig. \ref{FIG_resum} (f) leads to a structure with $3\times3=9$
lines \textendash{} three vertical, three horizontal, three antidiagonal
\textendash{} which can be distinguished when $\Omega_{cf}$ is large
enough, as in Fig. \ref{FIG_resum} (d), but merge when $\Omega_{cf}$
is too weak, as in Fig. \ref{FIG_resum} (b).

For sake of completeness, we underline that, although the above interpretation
seems to agree well with our observations, there exists a slight discrepancy,
for instance on Fig. \ref{FIG_resum} (c). According to the values
of the polaritonic energies, we indeed expect to observe a vertical
(horizontal) line centered on $\omega_{1}=0$ ($\omega_{2}=0$) :
by contrast, the line we obtain is slightly shifted from the vertical
(horizontal) axis. The reason for this effect could be an underlying
structure within the two-excitation semi-continuum, that may be the
object of further studies. 

\section{Conclusion and perspectives}

In this article, we theoretically investigated the nonlinear optical
response of an atomic medium placed in a cavity and excited towards
a Rydberg level in an EIT-configuration by a weak quantum probe and
a strong control field. To this end, we made use of the so-called
Schwinger-Keldysh contour technique, commonly employed in condensed
matter non-equilibrium physics, which allows for the systematic perturbation
expansion of expectation values of interest. As expected, our analytic
calculations show that the strong dipole-dipole interactions between
Rydberg polaritons lead to quantum nonlinearities, i.e. nonlinearities
noticeable in the few-photon regime. In particular, here, we presented
the detailed calculation of the $G^{(1)}$ correlation function up
to the fourth order in the probe amplitude : this allowed us to derive
the shape of the spectrum of the light transmitted through the cavity
and reveal the existence of an inelastic component that we interpreted
physically \citep{GBB16}. Using the Fadeev approach, we also investigated
three-body effects by computing the three-photon correlation function
of the transmitted light up to third order in the probe amplitude
and suggested an experimental setup to measure this quantity. We moreover
identified a strong correlation of the frequencies of the photons
transmitted, reminiscent of the time correlation recently observed
in MIT experiment in a free-space setup \citep{LVC17}. The results
presented here show the power and versatility of the Schwinger-Keldysh
approach for the treatment strongly interacting atomic media for quantum
optics. Though we used it to analyze a single-mode cavity setup, we
think it should be profitable in the treatment of several-mode cavity
systems for the investigation of quantum fluids of light and exotic
states that can be designed in such experiments \citep{CC13}, as
well as in free-space configurations where it could be an alternative
to effective field theory. 
\begin{acknowledgments}
This work is supported by the European Union project RySQ ( FET \#
640378), by the « Chaire SAFRAN - IOGS Photonique Ultime » and \foreignlanguage{american}{by
the Army Research Laboratory Center for Distributed Quantum Information
via the project SciNet, the ERC Synergy Grant UQUAM and the SFB FoQuS
(FWF Project No. F4016-N23)}.
\end{acknowledgments}

\appendix

\section{Pair correlation function $\left\langle {\cal T}\left(a\left(t\right)a\left(t^{\prime}\right)\right)\right\rangle $\label{sec:Pair-correlation-function}}

In this appendix we provide the technical details of the derivation
of Eq. (\ref{eq:CTP_aa_p>1}) omitted in the main text. Keeping only
the ${\cal C}_{+}$ part of the contour (for shortness we will omit
$"+"$ indices in this section) Eq (\ref{eq:CTP_aa}) writes:

\begin{equation}
\left\langle {\cal T}\left(a\left(t\right)a\left(t^{\prime}\right)\right)\right\rangle ^{\left(2\right)}=\frac{\left(-\mathrm{i}\alpha\right)^{2}}{2!}\left\langle {\cal T}\left\{ e^{-\mathrm{i}\frac{1}{2}\sum_{m,n}\kappa_{mn}\int dsc_{n}^{\dagger}c_{m}^{\dagger}c_{n}c_{m}}\left(\int dsa^{\dagger}\left(s\right)\right)^{2}a\left(t\right)a\left(t^{\prime}\right)\right\} \right\rangle \label{eq:CTP_aa_1-1}
\end{equation}

To evaluate this expression we now perform the perturbative expansion
with respect to $H_{dd}$, to expressed in the spinwave basis derived
in App. \ref{sec:A_bos_spinwave}:

\[
H_{dd}=\frac{1}{2}\sum_{\vec{q},\vec{k}_{1},\vec{k}_{2}}U_{\vec{q}}c_{\vec{k}_{2}-\vec{q}}^{\dagger}c_{\vec{q}+\vec{k}_{1}}^{\dagger}c_{\vec{k}_{1}}c_{\vec{k}_{2}}
\]

The zeroth order of the expansion of Eq. (\ref{eq:CTP_aa_1-1}) in
$H_{dd}$ yields

\begin{eqnarray*}
\left\langle {\cal T}\left(a\left(t\right)a\left(t^{\prime}\right)\right)\right\rangle ^{\left(2,0\right)} & = & \frac{\left(-\mathrm{i}\alpha\right)^{2}}{2!}\left\langle {\cal T}\left\{ \left(\int dsa^{\dagger}\left(s\right)\right)^{2}a\left(t\right)a\left(t^{\prime}\right)\right\} \right\rangle \\
 & = & \left(-\mathrm{i}\alpha\right)^{2}\left(\int ds_{1}{\rm i}G_{aa}\left[t-s_{1}\right]\int ds_{2}{\rm i}G_{aa}\left[t^{\prime}-s_{2}\right]\right)\\
 & = & \left(-\mathrm{i}\alpha\right)^{2}{\rm i}G_{aa}\left[\omega=0\right]{\rm i}G_{aa}\left[\omega=0\right]\\
 & \equiv & \left\langle a\left(t\right)\right\rangle ^{\left(1\right)}\left\langle a\left(t^{\prime}\right)\right\rangle ^{\left(1\right)}
\end{eqnarray*}
where the superscript $^{\left(p,q\right)}$ denotes the $p-th$ order
in expansion in $\alpha$ and $q-th$ in $H_{dd}$. The factorization
of $\left\langle {\cal T}\left(a\left(t\right)a\left(t^{\prime}\right)\right)\right\rangle ^{\left(2,0\right)}$
constitutes an obvious consequence of the fact that, at zeroth order
in $H_{dd}$, the system is completely linear.

The first order of the expansion in $H_{dd}$ writes:

\begin{eqnarray}
\left\langle {\cal T}\left\{ a\left(t\right)a\left(t^{\prime}\right)\right\} \right\rangle ^{\left(2,1\right)} & = & \sum_{\vec{q},\vec{k}_{1},\vec{k}_{2}}\frac{-{\rm i}U_{\vec{q}}}{2}\frac{\left(-\mathrm{i}\alpha\right)^{2}}{2!}\left\langle {\cal T}\left\{ \int dsc_{\vec{k}_{2}-\vec{q}}^{\dagger}c_{\vec{q}+\vec{k}_{1}}^{\dagger}c_{\vec{k}_{1}}c_{\vec{k}_{2}}\left(\int ds^{\prime}a^{\dagger}\left(s^{\prime}\right)\right)^{2}a\left(t\right)a\left(t^{\prime}\right)\right\} \right\rangle \label{eq:CTP_aa_21-1}
\end{eqnarray}

According to Wick's theorem, we now have to review all possible ways
to pair creation and annihilation operators in (\ref{eq:CTP_aa_21-1}).
As shown in Subsection \ref{subsec:CTP_Green}, the matrix representation
$\hat{G}^{T}\left[\omega\right]$ of the time-ordered Green's function
shows a block-diagonal structure in the basis $\left\{ a_{0},b_{0},c_{0},\right.$$\left.\left\{ b_{\vec{k}\neq0},c_{\vec{k}\neq0}\right\} \right\} $
which implies that the Green's functions $G_{xy}^{T}\left(t-t^{\prime}\right)=-{\rm i}\left\langle {\cal T}\left\{ x\left(t\right)y^{\dagger}\left(t^{\prime}\right)\right\} \right\rangle $
vanishes unless $x$ and $y$ simultaneously belong to the same set,
either $\left\{ a_{0},b_{0},c_{0}\right\} $ or $\left\{ b_{\vec{k}\neq0},c_{\vec{k}\neq0}\right\} $.
Therefore\textbf{ }only contractions of operators all picked either
in the set $\left\{ a,b_{0},c_{0}\right\} $ or in the set $\left\{ b_{\vec{k}\neq0},c_{\vec{k}\neq0}\right\} $
give non-vanishing contractions, whence

\begin{equation}
\left\langle {\cal T}\left\{ a\left(t\right)a\left(t^{\prime}\right)\right\} \right\rangle ^{\left(2,1\right)}=-{\rm i}\times U_{0}\left(-\mathrm{i}\alpha\right)^{2}\int ds_{1}ds_{2}ds_{3}G_{ac_{0}}\left[t,s_{1}\right]G_{ac_{0}}\left[t^{\prime},s_{1}\right]G_{c_{0}a}\left[s,s_{2}\right]G_{c_{0}a}\left[s,s_{3}\right]\label{eq:CTP_aa_21-final-1}
\end{equation}

Fourier transforming of Eq. (\ref{eq:CTP_aa_21-final-1}) with respect
to both $t$ and $t^{\prime}$ we get:

\begin{eqnarray}
\left\langle {\cal T}\left\{ a\left(\omega_{out,1}\right)a\left(\omega_{out,2}\right)\right\} \right\rangle ^{\left(2,1\right)} & = & \frac{1}{2\pi}\int dtdt^{\prime}e^{{\rm i}\omega_{out,1}t}e^{{\rm i}\omega_{out,2}t^{\prime}}\left\langle {\cal T}\left\{ a\left(t\right)a\left(t^{\prime}\right)\right\} \right\rangle ^{\left(2,1\right)}\nonumber \\
 & = & \left(-\mathrm{i}\sqrt{2\pi}\alpha\right)^{2}\left(-{\rm i}\times\frac{U_{0}}{2\pi}\right)\delta\left(\omega_{out,1}+\omega_{out,2}\right)G_{ac_{0}}^{T}\left[\omega\right]G_{ac_{0}}^{T}\left[\omega^{\prime}\right]\left(G_{c_{0}a}^{T}\left[0\right]\right)^{2}\label{eq:CTP_aa_21_f-1}
\end{eqnarray}
Note that the operator ${\cal T}$ appearing in $\left\langle {\cal T}\left\{ a\left(\omega_{out,1}\right)a\left(\omega_{out,2}\right)\right\} \right\rangle ^{\left(2,1\right)}$
does not refer to any hypothetical ordering in the frequency space;
it is a mere notation meant to remind that this quantity was obtained
by Fourier transforming the average of a time-ordered product in real
time space. 

Consider now the second order in expansion in $H_{dd}$:

\begin{eqnarray*}
\left\langle {\cal T}\left\{ a\left(t\right)a\left(t^{\prime}\right)\right\} \right\rangle ^{\left(2,2\right)} & = & \frac{\left(-\mathrm{i}\sqrt{2\pi}\alpha\right)^{2}}{2^{3}}\sum_{\vec{q},\vec{k}_{1},\vec{k}_{2},\vec{k}_{1}^{\prime},\vec{k}_{2}^{\prime}\vec{q}^{\prime},}\left(\frac{-{\rm i}U_{-\vec{q}}}{2\pi}\right)\left(\frac{-{\rm i}U_{\vec{q}^{\prime}}}{2\pi}\right)\\
 & \times & \left\langle {\cal T}\left\{ \begin{array}{c}
a\left(t\right)a\left(t^{\prime}\right)\int ds_{1}c_{\vec{k}_{2}-\vec{q}}^{\dagger}\left(s\right)c_{\vec{q}+\vec{k}_{1}}^{\dagger}\left(s\right)c_{\vec{k}_{1}}\left(s\right)c_{\vec{k}_{2}}\left(s\right)\\
\times\int ds_{2}c_{\vec{k}_{2}^{\prime}-\vec{q}^{\prime}}^{\dagger}\left(s_{2}\right)c_{\vec{q}^{\prime}+\vec{k}_{1}^{\prime}}^{\dagger}\left(s_{2}\right)c_{\vec{k}_{1}^{\prime}}\left(s_{2}\right)c_{\vec{k}_{2}^{\prime}}\left(s_{2}\right)\left(\frac{1}{\sqrt{2\pi}}\int dsa^{\dagger}\left(s\right)\right)^{2}
\end{array}\right\} \right\rangle 
\end{eqnarray*}

Using the same remark as made above for the first order and using
that, the contribution of ``disconnected diagrams'' (\emph{i.e.}
the contraction arrangement in which the $\left(4p\right)$ atomic
operators of $\left(-{\rm i}\int_{{\cal C}_{+}}H_{dd}\right)^{p}$are
paired with each other and therefore are disconnected from the other
terms) vanishes\citep{AGD}, we get in the temporal Fourier space:

\begin{align}
 & \left\langle {\cal T}\left\{ a\left(\omega_{out,1}\right)a\left(\omega_{out,2}\right)\right\} \right\rangle ^{\left(2,2\right)}\label{eq:CTP_aa^22-1}\\
= & \left(-\mathrm{i}\sqrt{2\pi}\alpha G_{c_{0}a}^{T}\left[0\right]\right)^{2}\delta\left(\omega_{out,1}+\omega_{out,2}\right)G_{ac_{0}}^{T}\left[\omega_{out,1}\right]G_{ac_{0}}^{T}\left[\omega_{out,2}\right]\sum_{\vec{q}}\left(\frac{-{\rm i}U_{-\vec{q}}}{2\pi}\right)\left(\frac{{\rm i}U_{\vec{q}}}{2\pi}\right)\left(\int d\omega G_{c_{\vec{q}},c_{\vec{q}}}^{T}\left[\omega\right]G_{c_{-\vec{q}},c_{-\vec{q}}}^{T}\left[-\omega\right]\right)\nonumber 
\end{align}

The further expansion in $H_{dd}$ reveals a self-similar form which
can be conveniently expressed using a diagrammatic representation.
According to the latter, the Green's functions of different kinds
are represented by different arrows (see Fig. \ref{Fig_feyn}), while
the interaction potential is represented by a vertical dashed line;
it is moreover implicit that, for each loop in a diagram, integration
(summation) should be performed over internal variables (indices)
and that the overall expression obtained should be multiplied by the
factor $\left(-\mathrm{i}\sqrt{2\pi}\alpha\right)^{2}\delta\left(\omega_{out,1}+\omega_{out,2}\right)\left(\frac{-{\rm i}}{2\pi}\right)^{p}$
where $p$ is the order in $H_{dd}$\emph{, i.e.} the number of dashed
vertical lines. Note that we do not distinguish $G_{ac_{0}}\left[\omega\right]$
and $G_{c_{0}a}\left[\omega\right]$ graphically since their expressions
coincide (see Eq. (\ref{eq:CTP_G_k})). 

It is easy to see that diagrams (a), (b), (c) in Fig. \ref{Fig_feyn},
which represent $\left\langle {\cal T}\left\{ a\left(\omega_{out,1}\right)a\left(\omega_{out,2}\right)\right\} \right\rangle ^{\left(2,p\right)}$
for $p=1,2,3$, have four thick lines in common. These thick lines
represent the conversion of a photon from the cavity mode to the symmetric
Rydberg polariton and back. As there is no integration over the arguments
of the corresponding Green's function we can factorize them (Fig.
\ref{Fig_feyn} d). The remaining part of the correlation function
is denoted by $T_{0}$; its perturbative expansion is diagrammatically
represented in Fig. \ref{Fig_feyn} (e).

From Fig. \ref{Fig_feyn} (d) we finally get Eq. (\ref{eq:CTP_aa_p>1})
of the main text. 

\section{Calculation of the 2 body T-matrix\label{subsec:CTP_T-matrix}}
\begin{center}
\textbf{\emph{}}
\begin{figure}
\begin{centering}
\emph{a})\includegraphics[height=2cm]{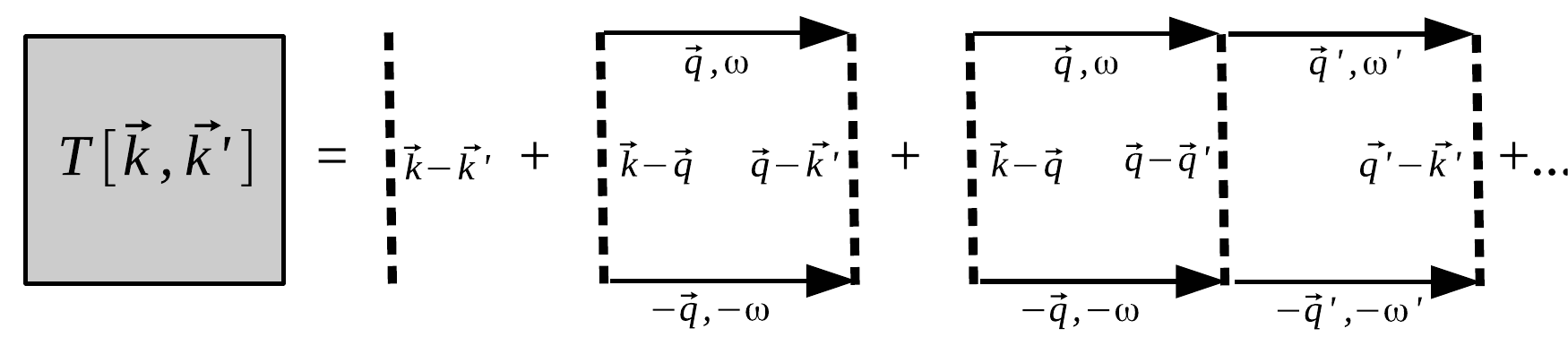}
\par\end{centering}
\begin{centering}
\emph{b})\includegraphics[height=2cm]{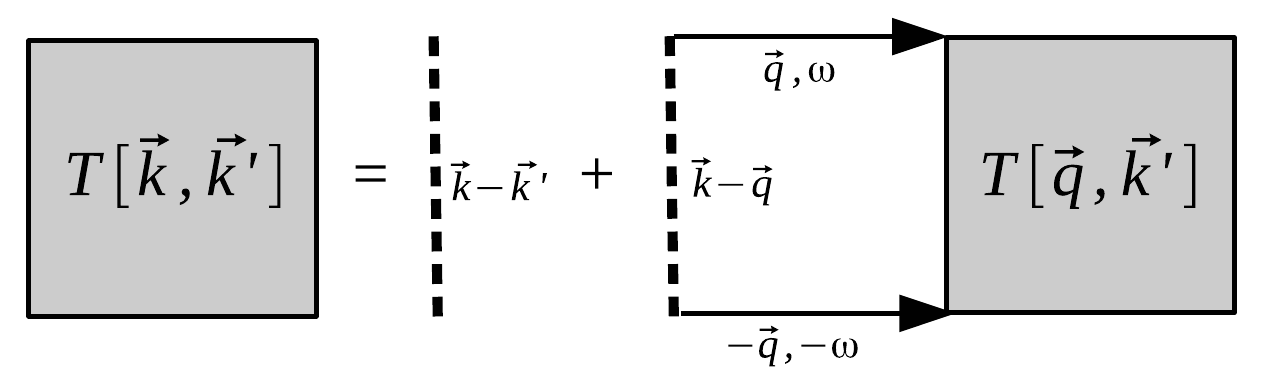}
\par\end{centering}
\caption{Diagrammatic representation of $T\left[\vec{k},\vec{k}^{\prime}\right]$
in \emph{a}) perturbative form, \emph{b}) self-consistent form.}

\label{FigTkk}
\end{figure}
\par\end{center}

As discussed in App. \ref{sec:Pair-correlation-function}, $T_{0}$
describes the combination of all possible interaction-induced scattering
processes in which two incoming Rydberg polaritons are converted back
to the same symmetric spinwaves. This quantity actually appears as
a specific value of a more general function which describes the scattering
of two arbitrary (\emph{i.e.} not necessarily symmetric) Rydberg polaritons
only constrained by the conservation of the sum of the wavevectors.
This latter function is denoted by $T\left[\vec{k},\vec{k}^{\prime},\vec{P}\right]$
where $k\equiv\frac{\vec{k}_{in,1}-\vec{k}_{in,2}}{2}$ and $\vec{k}^{\prime}\equiv\frac{\vec{k}_{out,1}-\vec{k}_{out,2}}{2}$
are the differences of the incoming/outgoing spinwave's wavevectors
$\vec{k}_{\left(in/out\right),1},\vec{k}_{\left(in/out\right),2}$
respectively, and $\vec{P}=\vec{k}_{in,1}+\vec{k}_{in,2}$ is their
(necessarily conserved) sum. The symmetry of the coupling between
the cavity mode and the atoms restricts the possible values of the
wavevectors: we are therefore entitled to consider only the $\vec{P}=0$
component, and denote $T\left[\vec{k},\vec{k}^{\prime},\vec{P}=0\right]\equiv T\left[\vec{k},\vec{k}^{\prime}\right]$. 

The diagrammatic representation of $T\left[\vec{k},\vec{k}^{\prime}\right]$
( given in Fig.\ref{FigTkk} a) is similar to the one obtained for
$T_{0}$. From the diagrammatic structure it is easy to infer its
self-consistent definition shown in Fig.\ref{FigTkk} b). The corresponding
equation is readily obtained using the correspondence rules specified
in Fig. \ref{Fig_feyn}:

\begin{eqnarray}
T\left[\vec{k},\vec{k}^{\prime}\right] & = & U_{\vec{k}-\vec{k}^{\prime}}+{\rm i}\sum_{\vec{q}}U_{\vec{k}-\vec{q}}S_{\vec{q}}T\left[\vec{q},\vec{k}^{\prime}\right]\label{eq:CTP_Tkk}
\end{eqnarray}
where $S_{\vec{q}}$ is defined in Eq. \ref{eq:S_q}. As shown in
Sec. \ref{subsec:CTP_Green}, all Green's functions $G_{c_{\vec{q}}c_{\vec{q}}}$
have the same expression for $\vec{q}\neq0$; we therefore define
$S_{\vec{q}\neq0}\equiv S$ and Eq. \ref{eq:CTP_Tkk} writes:

\begin{eqnarray}
T\left[\vec{k},\vec{k}^{\prime}\right] & = & U_{\vec{k}-\vec{k}^{\prime}}+{\rm i}\sum_{\vec{q}}U_{\vec{k}-\vec{q}}ST\left[\vec{q},\vec{k^{\prime}}\right]+{\rm i}U_{\vec{k}}\left(S_{0}-S\right)T\left[0,\vec{k}^{\prime}\right]\label{eq:CTP_Tkk_mod}
\end{eqnarray}
It is convenient to represent this equation in the matrix form 
\[
\widehat{T}=\widehat{{\cal U}}+{\rm i}S\,\widehat{{\cal U}}\cdotp\widehat{T}+{\rm i}\left(S_{0}-S\right)\widehat{{\cal U}}\cdotp{\cal P}^{\left(0\right)}\cdotp\widehat{T}
\]
where $\widehat{T}_{\vec{k},\vec{k}^{\prime}}\equiv T\left[\vec{k},\vec{k}^{\prime}\right]$,
$\widehat{{\cal U}}_{\vec{k},\vec{k}^{\prime}}\equiv U_{\vec{k}-\vec{k}^{\prime}}$
and ${\cal P}_{\vec{k},\vec{k}^{\prime}}^{\left(0\right)}\equiv\delta\left(\vec{k}\right)\delta\left(\vec{k}^{\prime}\right)$
is the projector onto the zeroth spinwave. Finally,

\begin{equation}
\widehat{T}=\left(1-{\rm i}S\,\widehat{{\cal U}}\right)^{-1}\cdotp\widehat{{\cal U}}\cdotp\left(\mathbb{I}+{\rm i}\left(S_{0}-S\right)\,{\cal P}^{\left(0\right)}\cdotp\widehat{T}\right)\label{eq:CTP_Tkk_matr}
\end{equation}
There is no obvious straightforward way to extract $\widehat{T}$
from Eq. (\ref{eq:CTP_Tkk_matr}) in the general case. We may however
relate $\widehat{T}$ to its value in the \emph{hypothetical }configuration
when the atoms decouple from the cavity, \emph{i.e. }when\emph{ }atom-cavity
coupling coefficient vanishes\emph{ i.e. $g=0$.}

In the latter condition from Eq. (\ref{eq:CTP_G_k}) we infer that
$G_{c_{\vec{q}}c_{\vec{q}}}^{T}\left[\omega\right]=G_{c_{0}c_{0}}^{T}\left[\omega\right]$
whence $S=S_{0}$. In this specific configuration, the matrix $\widehat{T}$,
that we shall denote $\widehat{\mathring{T}}$, to distinguish it
from the general case, obeys $\widehat{\mathring{T}}=\widehat{{\cal U}}+{\rm i}S\,\widehat{{\cal U}}\cdotp\widehat{\mathring{T}}$
, whence $\widehat{\mathring{T}}=\left(1-{\rm i}S\widehat{{\cal U}}\right)^{-1}\cdotp\widehat{{\cal U}}$.
Eq. (\ref{eq:CTP_Tkk_matr}) yields: 

\begin{equation}
\widehat{T}=\widehat{\mathring{T}}\cdotp\left(\mathbb{I}+{\rm i}\left(S_{0}-S\right)\,{\cal P}^{0}\cdotp\widehat{T}\right)\label{eq:CTP_Tkk_matr-1}
\end{equation}
Multiplying both sides of Eq. (\ref{eq:CTP_Tkk_matr-1}) by ${\cal P}^{\left(0\right)}\cdotp$
and solving for ${\cal P}^{\left(0\right)}\cdotp\widehat{T}$ we get:

\[
{\cal P}^{\left(0\right)}\cdotp\widehat{T}=\frac{{\cal P}^{\left(0\right)}\cdotp\widehat{\mathring{T}}}{1-{\rm i}\left(S_{0}-S\right)\,\mathring{T}_{0}}
\]
where ${\cal P}^{\left(0\right)}\cdotp\widehat{\mathring{T}}\cdotp{\cal P}^{\left(0\right)}\equiv\mathring{T}_{0}{\cal P}^{\left(0\right)}$.
Substituting this expression into Eq. (\ref{eq:CTP_Tkk_matr-1}) we
finally get the expression for the $T$ matrix:

\begin{equation}
\widehat{T}=\widehat{\mathring{T}}+{\rm i}\left(S_{0}-S\right)\,\frac{\widehat{\mathring{T}}\cdotp{\cal P}^{\left(0\right)}\cdotp\widehat{\mathring{T}}}{1-{\rm i}\left(S_{0}-S\right)\,\mathring{T}_{0}}\label{eq:CTP_Tkk_final}
\end{equation}
Since we are interested in determining $T_{0}=T\left[0,0\right]={\rm Tr}\left\{ {\cal P}^{\left(0\right)}\cdotp\widehat{T}\cdotp{\cal P}^{\left(0\right)}\right\} $
we multiply Eq. (\ref{eq:CTP_Tkk_final}) by ${\cal P}^{\left(0\right)}$
on the left and right sides and get:

\begin{equation}
T_{0}=\frac{\mathring{T_{0}}}{1-{\rm i}\mathring{T_{0}}\left(S_{0}-S\right)}\label{eq:CTP_T0_Eq}
\end{equation}

The advantage of this relation is that $\mathring{T_{0}}$ can be
evaluated exactly. From Eq. (\ref{eq:CTP_Tkk}) with $S=S_{0}$ we
get:

\begin{eqnarray*}
\mathring{T}\left[\vec{k},\vec{k}^{\prime}\right] & = & U_{\vec{k}-\vec{k}^{\prime}}+{\rm i}S\sum_{\vec{q}}U_{\vec{k}-\vec{q}}\mathring{T}\left[\vec{q},\vec{k}^{\prime}\right]
\end{eqnarray*}
This equation can be easily solved in the real space using $U_{\vec{K}}=\frac{1}{N}\sum_{m}^{N}\kappa\left(\vec{r}_{m}\right)e^{i\vec{K}\vec{r_{m}}}$
(see App. \ref{sec:A_bos_spinwave}) :

\begin{eqnarray*}
\mathring{T}\left[\vec{r},\vec{r}^{\prime}\right] & \equiv & \frac{1}{N}\sum_{\vec{k},\vec{k}^{\prime}}e^{-{\rm i}\vec{k}\vec{r}}e^{{\rm i}\vec{k}^{\prime}\vec{r}^{\prime}}\mathring{T}\left[\vec{k},\vec{k}^{\prime}\right]=\kappa\left(\vec{r}\right)\delta_{\vec{r},\vec{r}^{\prime}}+{\rm i}S\kappa\left(\vec{r}\right)\mathring{T}\left[\vec{r},\vec{r}^{\prime}\right]
\end{eqnarray*}
where $\vec{r}$ and $\vec{r}^{\prime}$ denote the real space conjugate
coordinates to $\vec{k}$ and $\vec{k}^{\prime}$, respectively, and
$\kappa\left(R\right)=\frac{C_{6}}{R^{6}}$. Finally we get:

\begin{eqnarray}
\mathring{T}\left[\vec{r},\vec{r}^{\prime}\right] & = & \frac{\kappa\left(\vec{r}\right)}{1-{\rm i}S\kappa\left(\vec{r}\right)}\delta_{\vec{r},\vec{r}^{\prime}}\label{eq:T=00005Br=00005D}
\end{eqnarray}
Using this expression and transforming back to the spinwave space
we get $\mathring{T}\left[0,0\right]=\frac{1}{N}\sum_{i}\frac{\kappa\left(\vec{r}_{i}\right)}{1-{\rm i}S\kappa\left(\vec{r}_{i}\right)}$.
We may finally approximate $\mathring{T}_{0}$ by an integral assuming
the size of the sample to be sufficiently big, and in the case $C_{6}<0$
we get:
\begin{eqnarray}
\mathring{T}_{0} & \approx & \frac{1}{V}\int_{V}d^{3}R\frac{\kappa\left(R\right)}{1-{\rm i}SV\left(R\right)}\approx-\frac{2\pi^{2}}{3V}\sqrt{\frac{-{\rm i}\left|C_{6}\right|}{S}}\label{eq:CTP_T_ring}
\end{eqnarray}

Note, that the Eq. (\ref{eq:T=00005Br=00005D}) characterizes the
atomic density-density correlation function: it is maximal if two
atoms are blocking each other and zero if atoms are not interacting.
Consequently, $\mathring{T}_{0}$ is proportional to the ratio of
the volume of a blockade sphere $\left(V_{b}\right)$ to the total
volume of the sample. The expression for $V_{b}$ is consistent with
the previously obtained results \citep{SPB13,GBB15}. 

Now combining Eqs. (\ref{eq:CTP_T0_Eq}, \ref{eq:CTP_T_ring}) we
finally have: 
\begin{equation}
T_{0}=\frac{-\frac{2\pi^{2}}{3V}\sqrt{\frac{-{\rm i}\left|C_{6}\right|}{S}}}{1+{\rm i}\left(S_{0}-S\right)\frac{2\pi^{2}}{3V}\sqrt{\frac{-{\rm i}\left|C_{6}\right|}{S}}}\label{eq:CTP_T_0_Final}
\end{equation}

Substituting this expression into Eq. (\ref{eq:CTP_aa_p>1}) we can
get an expression for the non-linear part of the pair correlation
function in Fourier space. We recover exactly the same expression
as in \citep{GBB15}, though in a much more concise form.

\section{Spinwave basis\label{sec:A_bos_spinwave}}

In this appendix, we introduce the collective atomic modes known as
``spinwaves'', which play a crucial role due to the symmetries of
the problem and lead to a simpler expression of the full Hamiltonian.

First, we assume that atoms occupy the vertices of a 3D square lattice
of step $\delta$. In all calculations we will eventually set the
limit $\delta\rightarrow0$ and therefore consider a continuous medium
but we will keep the discrete sums in all expressions for the sake
of convenience. The discrete Fourier transform allows us to relate
the direct space bosonic operators $b_{j}$ and $c_{j}$ to the reciprocal
space collective (so-called) spinwave operators $b_{\vec{k}},c_{\vec{k}}$:

\begin{eqnarray}
c_{\vec{k}} & = & \frac{1}{\sqrt{N}}\sum_{j}e^{i\vec{k}\vec{r_{j}}}c_{j}\leftrightarrow c_{j}=\frac{1}{\sqrt{N}}\sum_{\vec{k}}e^{-i\vec{k}\vec{r_{j}}}c_{\vec{k}}\label{eq:A_bos_collective}\\
b_{\vec{k}} & = & \frac{1}{\sqrt{N}}\sum_{j}e^{i\vec{k}\vec{r_{j}}}b_{j}\leftrightarrow b_{j}=\frac{1}{\sqrt{N}}\sum_{\vec{k}}e^{-i\vec{k}\vec{r_{j}}}b_{\vec{k}}\label{eq:A_bos_collective1}
\end{eqnarray}
where $\vec{r}_{i}$ is the position of the $i$-th atom and $k_{x,y,z}=-\nicefrac{\pi}{\delta},-\nicefrac{\pi}{\delta}+\frac{2\pi}{L_{x,y,z}}.\ldots,\nicefrac{\pi}{\delta}$
are the components of the $\vec{k}$ vector, where $L_{x,y,z}$ is
the lattice dimension in the $\left\{ x,y,z\right\} $ direction .
One readily shows

\begin{eqnarray*}
\left[c_{\vec{k}},c_{\vec{k}^{\prime}}^{\dagger}\right] & = & \frac{1}{N}\sum_{m,n}e^{i\vec{k}\vec{r_{m}}}e^{-i\vec{k}^{\prime}\vec{r_{n}}}\left[c_{m},c_{n}^{\dagger}\right]=\frac{1}{N}\sum_{m}e^{i\vec{k}\vec{r}_{m}}e^{-i\vec{k}^{\prime}\vec{r}_{m}}=\delta_{\vec{k},\vec{k}^{\prime}}
\end{eqnarray*}
and similarly $\left[b_{\vec{k}},b_{\vec{k}^{\prime}}^{\dagger}\right]=\delta_{\vec{k},\vec{k}^{\prime}}$.
We now rewrite the dipole-dipole Hamiltonian in terms of the spinwaves
operators defined above

\begin{align}
H_{dd} & =\frac{1}{2N^{2}}\sum_{m,n}^{N}\kappa_{mn}\sum_{\vec{k}^{\prime\prime\prime},\vec{k}^{\prime\prime},\vec{k}^{\prime},\vec{k}}e^{i\left(\vec{k}^{\prime\prime\prime}-\vec{k}\right)\vec{r_{n}}}e^{i\left(\vec{k}^{\prime\prime}-\vec{k}^{\prime}\right)\vec{r_{m}}}c_{\vec{k}^{\prime\prime\prime}}c_{\vec{k}^{\prime\prime}}c_{\vec{k}^{\prime}}c_{\vec{k}}\label{eq:A_bos_H_dd}
\end{align}
Imposing periodic boundary conditions on $\kappa_{mn}$, we obtain
\begin{eqnarray}
\frac{1}{N^{2}}\sum_{m,n}^{N}\kappa_{mn}e^{i\left(\vec{k}^{\prime\prime\prime}-\vec{k}\right)\vec{r_{n}}}e^{i\left(\vec{k}^{\prime\prime}-\vec{k}^{\prime}\right)\vec{r_{m}}} & = & U_{\vec{k}^{\prime\prime}-\vec{k}^{\prime}}\frac{1}{N}\sum_{n}^{N}e^{i\left(\vec{k}^{\prime\prime}-\vec{k}^{\prime}+\vec{k}^{\prime\prime\prime}-\vec{k}\right)\vec{r_{n}}}\nonumber \\
 & = & U_{\vec{k}^{\prime\prime}-\vec{k}^{\prime}}\delta\left(\vec{k}^{\prime\prime}-\vec{k}^{\prime}+\vec{k}^{\prime\prime\prime}-\vec{k}\right)\label{eq:A_bos_periodis}
\end{eqnarray}
where we defined the Fourier transform of the interaction matrix $\kappa_{mn}$
as $\frac{1}{N}\sum_{m}^{N}\kappa_{nm}e^{i\vec{K}\vec{r_{m}}}\equiv U_{\vec{K}}e^{i\vec{K}\vec{r_{n}}}$.
Substituting Eq. (\ref{eq:A_bos_periodis}) into Eq. (\ref{eq:A_bos_H_dd})
we get

\begin{align}
H_{dd} & =\sum_{\vec{k}^{\prime},\vec{k},\vec{q}}U_{\vec{q}}c_{\vec{k}^{\prime}-\vec{q}}c_{\vec{k}^{\prime}+\vec{q}}c_{\vec{k}^{\prime}}c_{\vec{k}}\label{eq:A_bos_H_dd-1}
\end{align}
where summations can be taken within any period of the lattice and
will be omitted for the sake of conciseness. 

\section{Transmission Spectrum technical details\label{sec:Transmission-Spectrum-technical} }

In this appendix we provide technical details of the derivation of
the elastic and inelastic contributions of the transmission spectrum,
omitted in Sec. \ref{sec:CTP_Intensity-correlation}.

\subsection{Elastic contribution }

We define the elastic contribtion to the spectrum as the partial resummation
${\cal E}\left(t,t^{\prime}\right)=\sum_{p=0,q>0}+\sum_{p>0,q=0}$
of Eq. (\ref{eq:CTP_pq_expansion}). We first consider the partial
sum of Eq. (\ref{eq:CTP_pq_expansion}) including the terms $p\neq0,q=0$,
we get:

\begin{equation}
\frac{\left(-\mathrm{i}\sqrt{2\pi}\alpha\right)^{4}}{4}\sum_{p}\frac{\left(-{\rm i}\right)^{p}}{p!}\left\langle {\cal T}_{{\cal C}}\left\{ \left(\frac{1}{2}\sum_{m,n}\kappa_{mn}\int_{\mathcal{C}_{+}}dsc_{n}^{\dagger}c_{m}^{\dagger}c_{n}c_{m}\right)^{p}A_{q}^{2}A_{q}^{\dagger2}a_{-}^{\dagger}\left(t\right)a_{+}\left(t^{\prime}\right)\right\} \right\rangle \label{eq:CTP_aa^4_q=00003D0-1}
\end{equation}
Using the results of Subsec. \ref{sec:CTP_First-order}\textbf{ }we
see that the operator $a_{-}^{\dagger}\left(t\right)$ does not have
other candidates for contraction than $A_{q}$. Since $\left\langle a^{\dagger}\left(t\right)\right\rangle ^{\left(1\right)}=\left(-\sqrt{2\pi}{\rm i}\alpha\right)\left\langle {\cal T}_{{\cal C}}\left\{ a_{-}^{\dagger}\left(t\right)A_{q}\right\} \right\rangle $
(see Sec. \ref{sec:CTP_First-order}),

\begin{equation}
\frac{\left(-\mathrm{i}\sqrt{2\pi}\alpha\right)^{3}}{2}\sum_{p}\frac{\left(-{\rm i}\right)^{p}}{p!}\left\langle {\cal T}_{{\cal C}}\left\{ \left(\frac{1}{2}\sum_{m,n}\kappa_{mn}\int_{\mathcal{C}_{+}}dsc_{n}^{\dagger}c_{m}^{\dagger}c_{n}c_{m}\right)^{p}A_{q}A_{q}^{\dagger2}a_{+}\left(t^{\prime}\right)\right\} \right\rangle \times\left\langle a^{\dagger}\left(t\right)\right\rangle ^{\left(1\right)}\label{eq:CTP_aa^4_q=00003D0_2-1}
\end{equation}
From the general formula Eq. (\ref{eq:CTP_the_most_general}) we have:

\begin{eqnarray*}
\left\langle a\left(t\right)\right\rangle ^{\left(3\right)} & \equiv & \frac{\left(-\mathrm{i}\sqrt{2\pi}\alpha\right)^{3}}{2!}\sum_{p,q}\frac{\left(-{\rm i}\right)^{p+q}}{p!q!}\left\langle \begin{array}{c}
{\cal T}_{{\cal C}}\left\{ \left(\frac{1}{2}\sum_{m,n}\kappa_{mn}\int_{{\cal C}_{+}}c_{n}^{\dagger}c_{m}^{\dagger}c_{n}c_{m}\right)^{p}\right.\\
\times\left.\left(\frac{1}{2}\sum_{m,n}\kappa_{mn}\int_{{\cal C}_{-}}c_{n}^{\dagger}c_{m}^{\dagger}c_{n}c_{m}\right)^{q}A_{q}A_{q}^{\dagger2}a_{+}\left(t\right)\right\} 
\end{array}\right\rangle 
\end{eqnarray*}
In the expression above one of the operators $c_{n}^{\dagger},c_{m}^{\dagger}$
belonging the ${\cal C}_{-}$ branch does not have any partner for
contraction. Therefore only the $q=0$ term will contribute to the
sum and

\begin{equation}
\left\langle a\left(t\right)\right\rangle ^{\left(3\right)}=\frac{\left(-\mathrm{i}\sqrt{2\pi}\alpha\right)^{3}}{2!}\left\langle {\cal T}_{{\cal C}}\left\{ e^{-\mathrm{i}\frac{1}{2}\sum_{m,n}\kappa_{mn}\int_{{\cal C}_{+}}dsc_{n}^{\dagger}c_{m}^{\dagger}c_{n}c_{m}}A_{q}A_{q}^{\dagger2}a_{+}\left(t\right)\right\} \right\rangle \label{eq:CTP_a^3_def-1}
\end{equation}
Eq. (\ref{eq:CTP_aa^4_q=00003D0_2-1}) therefore writes $\left\langle a^{\dagger}\left(t\right)\right\rangle ^{\left(1\right)}\left\langle a\left(t^{\prime}\right)\right\rangle ^{\left(3\right)}$. 

Analogously, the partial sum of the terms $\left(p=0,q\neq0\right)$
yields $\left\langle a^{\dagger}\left(t\right)\right\rangle ^{\left(3\right)}\left\langle a\left(t^{\prime}\right)\right\rangle ^{\left(1\right)}$.
Finally we get 
\begin{equation}
{\cal E}=\left\langle a^{\dagger}\left(t\right)\right\rangle ^{\left(3\right)}\left\langle a\left(t^{\prime}\right)\right\rangle ^{\left(1\right)}+\left\langle a^{\dagger}\left(t\right)\right\rangle ^{\left(1\right)}\left\langle a\left(t^{\prime}\right)\right\rangle ^{\left(3\right)}\label{eq:fact}
\end{equation}

We shall now determine the expression for $\left\langle a\left(t\right)\right\rangle ^{\left(3\right)}$.
Let us perform an expansion of Eq. (\ref{eq:CTP_a^3_def-1}) with
respect to $H_{dd}$. Due to the symmetry properties of the system,
it is more convenient to work in the spatial Fourier space. We therefore
get:

\begin{eqnarray*}
\left\langle a\left(t\right)\right\rangle ^{\left(3\right)} & = & \frac{\left(-\mathrm{i}\sqrt{2\pi}\alpha\right)^{3}}{2!}\sum_{p}\frac{1}{p!}\left\langle {\cal T}_{{\cal C}}\left\{ \left(-\frac{{\rm i}}{2}\sum_{\vec{q},\vec{k}_{1},\vec{k}_{2}}U_{\vec{q}}\int_{{\cal C}_{+}}c_{\vec{k}_{2}-\vec{q}}^{\dagger}c_{\vec{q}+\vec{k}_{1}}^{\dagger}c_{\vec{k}_{1}}c_{\vec{k}_{2}}\right)^{p}A_{q}A_{q}^{\dagger2}a_{+}\left(t\right)\right\} \right\rangle 
\end{eqnarray*}
As expected, the zeroth order $\left(p=0\right)$ of this expansion
is zero since it necessarily involves the vanishing contraction of
two ``quantum'' operators $A_{q}$ and $A_{q}^{\dagger}$ (see Sec.
\ref{subsec:CTP_Green}). Apart from the external lines, the terms
in the expansion with $p>1$ form the same ladder series as derived
in App. \ref{subsec:CTP_T-matrix}. The difference in the external
lines is given by the replacement of one of $a_{+}$ operators by
$A_{q}$: the corresponding contraction is given by:
\begin{eqnarray*}
\left\langle {\cal T}_{{\cal C}}\left\{ A_{q}c_{+,\vec{k}}^{\dagger}\left(t\right)\right\} \right\rangle  & = & \frac{1}{\sqrt{2\pi}}\int ds\left\langle {\cal T}_{{\cal C}}\left\{ \left(a_{+}\left(s\right)-a_{-}\left(s\right)\right)c_{+,\vec{k}}^{\dagger}\left(t\right)\right\} \right\rangle =\frac{1}{\sqrt{2\pi}}\int ds\left\{ -iG_{ac_{0}}^{\tilde{T}}\left[s,t\right]\right\} \delta_{\vec{k},0}
\end{eqnarray*}
Combining these remarks, we get in the frequency space:

\begin{equation}
\left\langle a\left(\omega\right)\right\rangle ^{\left(3\right)}=\left(-\mathrm{i}\sqrt{2\pi}\alpha\right)^{3}\delta\left(\omega\right)\left(-{\rm i}G_{ac_{0}}^{T}\left[0\right]{\rm i}G_{ac_{0}}^{\tilde{T}}\left[0\right]\right)\left(\frac{-{\rm i}T_{0}}{2\pi}\right)\left({\rm i}G_{c_{0}a}^{T}\left[0\right]\right)^{2}\label{eq:CTP_a^3_final-1}
\end{equation}
where $T_{0}$ was defined in Eq. (\ref{eq:CTP_t_0_vague}), and finally,
using Eq. (\ref{eq:fact}) we recover Eq. (\ref{eq:Eww}) of the main
text. 

\subsection{Inelastic contribution to $G_{out}^{\left(1\right)}$\label{subsec:CTP_inelastic-1}}

Here we provide the partial resummation in Eq. (\ref{eq:CTP_pq_expansion})
${\cal I}\left(t,t^{\prime}\right)\equiv\sum_{p>0,q>0}=\left\langle a^{\dagger}\left(t\right)a\left(t^{\prime}\right)\right\rangle ^{\left(4\right)}-{\cal E}\left(t,t^{\prime}\right)$
which stands for the inelastic part of the spectrum.

We have ${\cal I}\left(t,t^{\prime}\right)=\sum_{p>0,q>0}{\cal I}_{p,q}$
, where

\[
{\cal I}_{p,q}\equiv\frac{\left(-\mathrm{i}\sqrt{2\pi}\alpha\right)^{4}}{4}\left\langle {\cal T}_{{\cal C}}\left\{ \frac{\left(-{\rm i}\int_{{\cal C}_{+}}dsH_{dd}\left(s\right)\right)^{p}\left(-{\rm i}\int_{{\cal C}_{-}}dsH_{dd}\left(s\right)\right)^{q}}{p!q!}A_{q}^{2}A_{q}^{\dagger2}a_{-}^{\dagger}\left(t\right)a_{+}\left(t^{\prime}\right)\right\} \right\rangle 
\]
Let us first consider the term $\left(p=1,q=1\right)$ still writing
$H_{dd}$ in the spinwave basis:

\begin{eqnarray}
{\cal I}_{1,1}\left(t,t^{\prime}\right) & = & \frac{\left(-\mathrm{i}\sqrt{2\pi}\alpha\right)^{4}}{4}\left\langle {\cal T}_{{\cal C}}\left\{ \left(-{\rm i}\frac{1}{2}\sum_{\vec{q},\vec{k}_{1},\vec{k}_{2}}\int dsU_{\vec{q}}c_{\vec{k}_{2}-\vec{q},+}^{\dagger}c_{\vec{q}+\vec{k}_{1},+}^{\dagger}c_{\vec{k}_{1},+}c_{\vec{k}_{2},+}\right)\right.\right.\nonumber \\
 & \times & \left.\left.\left(-{\rm i}\frac{1}{2}\sum_{\vec{q}^{\prime},\vec{k}_{1}^{\prime},\vec{k}_{2}^{\prime}}U_{\vec{q}^{\prime}}\int dsc_{\vec{k}_{2}^{\prime}-\vec{q}^{\prime},-}^{\dagger}c_{\vec{q}^{\prime}+\vec{k}_{1}^{\prime},-}^{\dagger}c_{\vec{k}_{1}^{\prime},-}c_{\vec{k}_{2}^{\prime},-}\right)A_{q}^{2}A_{q}^{\dagger2}a_{-}^{\dagger}\left(t\right)a_{+}\left(t^{\prime}\right)\right\} \right\rangle \label{eq:CTP_aa_inelast-4}
\end{eqnarray}
In Eq. (\ref{eq:CTP_aa_inelast-4}) operators $c_{\vec{k}_{1},+}c_{\vec{k}_{2},+}$
and $c_{\vec{k}_{2}^{\prime}-\vec{q}^{\prime},-}^{\dagger}c_{\vec{q}^{\prime}+\vec{k}_{1}^{\prime},-}^{\dagger}$
can only be contracted with $A_{q}^{\dagger2}$ and $A_{q}^{2}$,
respectively with $\left\langle {\cal T}_{{\cal C}}\left\{ c_{\vec{k},+}\left(t\right)A_{q}^{\dagger}\right\} \right\rangle $
$={\rm \frac{i}{\sqrt{2\pi}}}\int dsG_{c_{0}a}^{T}\left(t-s\right)\delta_{\vec{k},0}$
and $\left\langle {\cal T}_{{\cal C}}\left\{ c_{\vec{k},-}^{\dagger}\left(t\right)A_{q}\right\} \right\rangle $
$={\rm \frac{-i}{\sqrt{2\pi}}}\int dsG_{c_{0}a}^{\tilde{T}}\left(s-t\right)\delta_{\vec{k},0}$
we get

\begin{eqnarray}
{\cal I}_{1,1}\left(t,t^{\prime}\right) & = & \frac{\left(-\mathrm{i}\sqrt{2\pi}\alpha\right)^{4}}{4}\left\langle {\cal T}_{{\cal C}}\left\{ \left(-{\rm i}\int_{{\cal C}_{+}}dz_{1}\sum_{\vec{q}}U_{\vec{q}}c_{-\vec{q}}^{\dagger}c_{\vec{q}}^{\dagger}\right)\left({\rm i}\int_{-{\cal C}_{-}}dz_{2}\sum_{\vec{q}^{\prime}}U_{\vec{q}^{\prime}}c_{-\vec{q}^{\prime}}c_{\vec{q}^{\prime}}\right)\right.\right.\nonumber \\
 & \times & \left.\left.a_{-}^{\dagger}\left(t\right)a_{+}\left(t^{\prime}\right)\right\} \right\rangle \left(\int\frac{ds}{\sqrt{2\pi}}{\rm i}G_{c_{0}a}^{T}\left[z_{1},s\right]\right)^{2}\left(-\int\frac{ds}{\sqrt{2\pi}}{\rm i}G_{ac_{0}}^{\tilde{T}}\left[s,z_{2}\right]\right)^{2}\label{eq:CTP_aa_inelast-1-1}
\end{eqnarray}
In this expression, $a_{-}^{\dagger}\left(t\right)$ and $a_{+}\left(t^{\prime}\right)$
can only be contracted with one of $c_{\vec{q},-}$ and $c_{\vec{q},+}^{\dagger}$
operators respectively. Whence:

\begin{eqnarray}
{\cal I}_{1,1}\left(t,t^{\prime}\right) & = & \left(-\mathrm{i}\sqrt{2\pi}\alpha\right)^{4}\left\langle {\cal T}_{{\cal C}}\left\{ \left(-{\rm i}\int_{{\cal C}_{+}}dz_{1}U_{0}c_{0}^{\dagger}\right)\left({\rm i}\int_{-{\cal C}_{-}}dz_{2}U_{0}c_{0}\right)\right\} \right\rangle \label{eq:CTP_aa_inelast-2-1}\\
 & \times & {\rm i}G_{c_{0}a}^{\tilde{T}}\left[z_{2},t\right]{\rm i}G_{ac_{0}}^{T}\left[t^{\prime},z_{1}\right]\left(\int\frac{ds}{\sqrt{2\pi}}{\rm i}G_{c_{0}a}^{T}\left[z_{1},s\right]\right)^{2}\left(-\int\frac{ds}{\sqrt{2\pi}}{\rm i}G_{ac_{0}}^{\tilde{T}}\left[s,z_{2}\right]\right)^{2}\nonumber 
\end{eqnarray}
and finally

\begin{eqnarray}
{\cal I}_{1,1}\left(t,t^{\prime}\right) & = & \left(-\mathrm{i}\sqrt{2\pi}\alpha\right)^{4}\left(-{\rm i}U_{0}\right)\left({\rm i}U_{0}\right)\int_{-\infty}^{\infty}dz_{1}dz_{2}{\rm i}G_{c_{0}c_{0}}^{>}\left[z_{2},z_{1}\right]{\rm i}G_{c_{0}a}^{\tilde{T}}\left[z_{2},t\right]\nonumber \\
 & \times & {\rm i}G_{ac_{0}}^{T}\left[t^{\prime},z_{1}\right]\left(\int\frac{ds}{\sqrt{2\pi}}{\rm i}G_{c_{0}a}^{T}\left[z_{1},s\right]\right)^{2}\left(-\int\frac{ds}{\sqrt{2\pi}}{\rm i}G_{ac_{0}}^{\tilde{T}}\left[s,z_{2}\right]\right)^{2}\label{eq:CTP_aa_inelast-3-2}
\end{eqnarray}
or equivalently in frequency domain

\begin{equation}
\left(-\mathrm{i}\sqrt{2\pi}\alpha\right)^{4}\delta\left(\omega-\omega^{\prime}\right)\left(-{\rm i}\frac{U_{0}}{2\pi}\right)\left({\rm i}\frac{U_{0}}{2\pi}\right){\rm i}G_{c_{0}c_{0}}^{>}\left[-\omega\right]{\rm i}G_{c_{0}a}^{\tilde{T}}\left[\omega^{\prime}\right]{\rm i}G_{ac_{0}}^{T}\left[\omega\right]\left({\rm i}G_{c_{0}a}^{T}\left[0\right]\right)^{2}\left({\rm i}G_{ac_{0}}^{\tilde{T}}\left[0\right]\right)^{2}\label{eq:CTP_aa_inelast-3-1-2}
\end{equation}
Computing higher-order terms we find again the same ladder structure
of diagrams as in Sec. \ref{sec:CTP_Intensity-correlation}, whose
resummation for $p,q>0$ yields Eq. (\ref{eq:CTP_aa_inelast}) of
the main text.

\bibliographystyle{apsrev4-1}
\bibliography{biblio}

\begin{thebibliography}{32}%
\makeatletter
\providecommand \@ifxundefined [1]{%
 \@ifx{#1\undefined}
}%
\providecommand \@ifnum [1]{%
 \ifnum #1\expandafter \@firstoftwo
 \else \expandafter \@secondoftwo
 \fi
}%
\providecommand \@ifx [1]{%
 \ifx #1\expandafter \@firstoftwo
 \else \expandafter \@secondoftwo
 \fi
}%
\providecommand \natexlab [1]{#1}%
\providecommand \enquote  [1]{``#1''}%
\providecommand \bibnamefont  [1]{#1}%
\providecommand \bibfnamefont [1]{#1}%
\providecommand \citenamefont [1]{#1}%
\providecommand \href@noop [0]{\@secondoftwo}%
\providecommand \href [0]{\begingroup \@sanitize@url \@href}%
\providecommand \@href[1]{\@@startlink{#1}\@@href}%
\providecommand \@@href[1]{\endgroup#1\@@endlink}%
\providecommand \@sanitize@url [0]{\catcode `\\12\catcode `\$12\catcode
  `\&12\catcode `\#12\catcode `\^12\catcode `\_12\catcode `\%12\relax}%
\providecommand \@@startlink[1]{}%
\providecommand \@@endlink[0]{}%
\providecommand \url  [0]{\begingroup\@sanitize@url \@url }%
\providecommand \@url [1]{\endgroup\@href {#1}{\urlprefix }}%
\providecommand \urlprefix  [0]{URL }%
\providecommand \Eprint [0]{\href }%
\providecommand \doibase [0]{http://dx.doi.org/}%
\providecommand \selectlanguage [0]{\@gobble}%
\providecommand \bibinfo  [0]{\@secondoftwo}%
\providecommand \bibfield  [0]{\@secondoftwo}%
\providecommand \translation [1]{[#1]}%
\providecommand \BibitemOpen [0]{}%
\providecommand \bibitemStop [0]{}%
\providecommand \bibitemNoStop [0]{.\EOS\space}%
\providecommand \EOS [0]{\spacefactor3000\relax}%
\providecommand \BibitemShut  [1]{\csname bibitem#1\endcsname}%
\let\auto@bib@innerbib\@empty
\bibitem [{\citenamefont {Grankin}\ \emph {et~al.}(2016)\citenamefont
  {Grankin}, \citenamefont {Brion}, \citenamefont {Boddeda}, \citenamefont
  {{\'C}uk}, \citenamefont {Usmani}, \citenamefont {Ourjoumtsev},\ and\
  \citenamefont {Grangier}}]{GBB16}%
  \BibitemOpen
  \bibfield  {author} {\bibinfo {author} {\bibfnamefont {A.}~\bibnamefont
  {Grankin}}, \bibinfo {author} {\bibfnamefont {E.}~\bibnamefont {Brion}},
  \bibinfo {author} {\bibfnamefont {R.}~\bibnamefont {Boddeda}}, \bibinfo
  {author} {\bibfnamefont {S.}~\bibnamefont {{\'C}uk}}, \bibinfo {author}
  {\bibfnamefont {I.}~\bibnamefont {Usmani}}, \bibinfo {author} {\bibfnamefont
  {A.}~\bibnamefont {Ourjoumtsev}}, \ and\ \bibinfo {author} {\bibfnamefont
  {P.}~\bibnamefont {Grangier}},\ }\href@noop {} {\bibfield  {journal}
  {\bibinfo  {journal} {Physical Review Letters}\ }\textbf {\bibinfo {volume}
  {117}},\ \bibinfo {pages} {253602} (\bibinfo {year} {2016})}\BibitemShut
  {NoStop}%
\bibitem [{\citenamefont {Nielsen}\ and\ \citenamefont {Chuang}(2010)}]{NC10}%
  \BibitemOpen
  \bibfield  {author} {\bibinfo {author} {\bibfnamefont {M.~A.}\ \bibnamefont
  {Nielsen}}\ and\ \bibinfo {author} {\bibfnamefont {I.~L.}\ \bibnamefont
  {Chuang}},\ }\href@noop {} {\emph {\bibinfo {title} {Quantum computation and
  quantum information}}}\ (\bibinfo  {publisher} {Cambridge university press},\
  \bibinfo {year} {2010})\BibitemShut {NoStop}%
\bibitem [{\citenamefont {Firstenberg}\ \emph {et~al.}(2013)\citenamefont
  {Firstenberg}, \citenamefont {Peyronel}, \citenamefont {Liang}, \citenamefont
  {Gorshkov}, \citenamefont {Lukin},\ and\ \citenamefont
  {Vuleti{\'c}}}]{FPL13}%
  \BibitemOpen
  \bibfield  {author} {\bibinfo {author} {\bibfnamefont {O.}~\bibnamefont
  {Firstenberg}}, \bibinfo {author} {\bibfnamefont {T.}~\bibnamefont
  {Peyronel}}, \bibinfo {author} {\bibfnamefont {Q.-Y.}\ \bibnamefont {Liang}},
  \bibinfo {author} {\bibfnamefont {A.~V.}\ \bibnamefont {Gorshkov}}, \bibinfo
  {author} {\bibfnamefont {M.~D.}\ \bibnamefont {Lukin}}, \ and\ \bibinfo
  {author} {\bibfnamefont {V.}~\bibnamefont {Vuleti{\'c}}},\ }\href@noop {}
  {\bibfield  {journal} {\bibinfo  {journal} {Nature}\ }\textbf {\bibinfo
  {volume} {502}},\ \bibinfo {pages} {71} (\bibinfo {year} {2013})}\BibitemShut
  {NoStop}%
\bibitem [{\citenamefont {Peyronel}\ \emph {et~al.}(2012)\citenamefont
  {Peyronel}, \citenamefont {Firstenberg}, \citenamefont {Liang}, \citenamefont
  {Hofferberth}, \citenamefont {Gorshkov}, \citenamefont {Pohl}, \citenamefont
  {Lukin},\ and\ \citenamefont {Vuleti{\'c}}}]{PFL12}%
  \BibitemOpen
  \bibfield  {author} {\bibinfo {author} {\bibfnamefont {T.}~\bibnamefont
  {Peyronel}}, \bibinfo {author} {\bibfnamefont {O.}~\bibnamefont
  {Firstenberg}}, \bibinfo {author} {\bibfnamefont {Q.-Y.}\ \bibnamefont
  {Liang}}, \bibinfo {author} {\bibfnamefont {S.}~\bibnamefont {Hofferberth}},
  \bibinfo {author} {\bibfnamefont {A.~V.}\ \bibnamefont {Gorshkov}}, \bibinfo
  {author} {\bibfnamefont {T.}~\bibnamefont {Pohl}}, \bibinfo {author}
  {\bibfnamefont {M.~D.}\ \bibnamefont {Lukin}}, \ and\ \bibinfo {author}
  {\bibfnamefont {V.}~\bibnamefont {Vuleti{\'c}}},\ }\href@noop {} {\bibfield
  {journal} {\bibinfo  {journal} {Nature}\ }\textbf {\bibinfo {volume} {488}},\
  \bibinfo {pages} {57} (\bibinfo {year} {2012})}\BibitemShut {NoStop}%
\bibitem [{\citenamefont {Gorniaczyk}\ \emph {et~al.}(2014)\citenamefont
  {Gorniaczyk}, \citenamefont {Tresp}, \citenamefont {Schmidt}, \citenamefont
  {Fedder},\ and\ \citenamefont {Hofferberth}}]{GTS14}%
  \BibitemOpen
  \bibfield  {author} {\bibinfo {author} {\bibfnamefont {H.}~\bibnamefont
  {Gorniaczyk}}, \bibinfo {author} {\bibfnamefont {C.}~\bibnamefont {Tresp}},
  \bibinfo {author} {\bibfnamefont {J.}~\bibnamefont {Schmidt}}, \bibinfo
  {author} {\bibfnamefont {H.}~\bibnamefont {Fedder}}, \ and\ \bibinfo {author}
  {\bibfnamefont {S.}~\bibnamefont {Hofferberth}},\ }\href@noop {} {\bibfield
  {journal} {\bibinfo  {journal} {Phys. Rev. Lett.}\ }\textbf {\bibinfo
  {volume} {113}},\ \bibinfo {pages} {053601} (\bibinfo {year}
  {2014})}\BibitemShut {NoStop}%
\bibitem [{\citenamefont {Tiarks}\ \emph {et~al.}(2014)\citenamefont {Tiarks},
  \citenamefont {Baur}, \citenamefont {Schneider}, \citenamefont {D\"urr},\
  and\ \citenamefont {Rempe}}]{TBS14}%
  \BibitemOpen
  \bibfield  {author} {\bibinfo {author} {\bibfnamefont {D.}~\bibnamefont
  {Tiarks}}, \bibinfo {author} {\bibfnamefont {S.}~\bibnamefont {Baur}},
  \bibinfo {author} {\bibfnamefont {K.}~\bibnamefont {Schneider}}, \bibinfo
  {author} {\bibfnamefont {S.}~\bibnamefont {D\"urr}}, \ and\ \bibinfo {author}
  {\bibfnamefont {G.}~\bibnamefont {Rempe}},\ }\href@noop {} {\bibfield
  {journal} {\bibinfo  {journal} {Phys. Rev. Lett.}\ }\textbf {\bibinfo
  {volume} {113}},\ \bibinfo {pages} {053602} (\bibinfo {year}
  {2014})}\BibitemShut {NoStop}%
\bibitem [{\citenamefont {Dudin}\ and\ \citenamefont {Kuzmich}(2012)}]{DK12}%
  \BibitemOpen
  \bibfield  {author} {\bibinfo {author} {\bibfnamefont {Y.}~\bibnamefont
  {Dudin}}\ and\ \bibinfo {author} {\bibfnamefont {A.}~\bibnamefont
  {Kuzmich}},\ }\href@noop {} {\bibfield  {journal} {\bibinfo  {journal}
  {Science}\ }\textbf {\bibinfo {volume} {336}},\ \bibinfo {pages} {887}
  (\bibinfo {year} {2012})}\BibitemShut {NoStop}%
\bibitem [{\citenamefont {Maxwell}\ \emph {et~al.}(2013)\citenamefont
  {Maxwell}, \citenamefont {Szwer}, \citenamefont {Paredes-Barato},
  \citenamefont {Busche}, \citenamefont {Pritchard}, \citenamefont {Gauguet},
  \citenamefont {Weatherill}, \citenamefont {Jones},\ and\ \citenamefont
  {Adams}}]{MSB13}%
  \BibitemOpen
  \bibfield  {author} {\bibinfo {author} {\bibfnamefont {D.}~\bibnamefont
  {Maxwell}}, \bibinfo {author} {\bibfnamefont {D.~J.}\ \bibnamefont {Szwer}},
  \bibinfo {author} {\bibfnamefont {D.}~\bibnamefont {Paredes-Barato}},
  \bibinfo {author} {\bibfnamefont {H.}~\bibnamefont {Busche}}, \bibinfo
  {author} {\bibfnamefont {J.~D.}\ \bibnamefont {Pritchard}}, \bibinfo {author}
  {\bibfnamefont {A.}~\bibnamefont {Gauguet}}, \bibinfo {author} {\bibfnamefont
  {K.~J.}\ \bibnamefont {Weatherill}}, \bibinfo {author} {\bibfnamefont
  {M.~P.~A.}\ \bibnamefont {Jones}}, \ and\ \bibinfo {author} {\bibfnamefont
  {C.~S.}\ \bibnamefont {Adams}},\ }\href@noop {} {\bibfield  {journal}
  {\bibinfo  {journal} {Phys. Rev. Lett.}\ }\textbf {\bibinfo {volume} {110}},\
  \bibinfo {pages} {103001} (\bibinfo {year} {2013})}\BibitemShut {NoStop}%
\bibitem [{\citenamefont {Grankin}\ \emph {et~al.}(2014)\citenamefont
  {Grankin}, \citenamefont {Brion}, \citenamefont {Bimbard}, \citenamefont
  {Boddeda}, \citenamefont {Usmani}, \citenamefont {Ourjoumtsev},\ and\
  \citenamefont {Grangier}}]{GBB14}%
  \BibitemOpen
  \bibfield  {author} {\bibinfo {author} {\bibfnamefont {A.}~\bibnamefont
  {Grankin}}, \bibinfo {author} {\bibfnamefont {E.}~\bibnamefont {Brion}},
  \bibinfo {author} {\bibfnamefont {E.}~\bibnamefont {Bimbard}}, \bibinfo
  {author} {\bibfnamefont {R.}~\bibnamefont {Boddeda}}, \bibinfo {author}
  {\bibfnamefont {I.}~\bibnamefont {Usmani}}, \bibinfo {author} {\bibfnamefont
  {A.}~\bibnamefont {Ourjoumtsev}}, \ and\ \bibinfo {author} {\bibfnamefont
  {P.}~\bibnamefont {Grangier}},\ }\href@noop {} {\bibfield  {journal}
  {\bibinfo  {journal} {New Journal of Physics}\ }\textbf {\bibinfo {volume}
  {16}},\ \bibinfo {pages} {043020} (\bibinfo {year} {2014})}\BibitemShut
  {NoStop}%
\bibitem [{\citenamefont {Grankin}\ \emph {et~al.}(2015)\citenamefont
  {Grankin}, \citenamefont {Brion}, \citenamefont {Bimbard}, \citenamefont
  {Boddeda}, \citenamefont {Usmani}, \citenamefont {Ourjoumtsev},\ and\
  \citenamefont {Grangier}}]{GBB15}%
  \BibitemOpen
  \bibfield  {author} {\bibinfo {author} {\bibfnamefont {A.}~\bibnamefont
  {Grankin}}, \bibinfo {author} {\bibfnamefont {E.}~\bibnamefont {Brion}},
  \bibinfo {author} {\bibfnamefont {E.}~\bibnamefont {Bimbard}}, \bibinfo
  {author} {\bibfnamefont {R.}~\bibnamefont {Boddeda}}, \bibinfo {author}
  {\bibfnamefont {I.}~\bibnamefont {Usmani}}, \bibinfo {author} {\bibfnamefont
  {A.}~\bibnamefont {Ourjoumtsev}}, \ and\ \bibinfo {author} {\bibfnamefont
  {P.}~\bibnamefont {Grangier}},\ }\href@noop {} {\bibfield  {journal}
  {\bibinfo  {journal} {Phys. Rev. A}\ }\textbf {\bibinfo {volume} {92}},\
  \bibinfo {pages} {043841} (\bibinfo {year} {2015})}\BibitemShut {NoStop}%
\bibitem [{\citenamefont {Boddeda}\ \emph {et~al.}(2016)\citenamefont
  {Boddeda}, \citenamefont {Usmani}, \citenamefont {Bimbard}, \citenamefont
  {Grankin}, \citenamefont {Ourjoumtsev}, \citenamefont {Brion},\ and\
  \citenamefont {Grangier}}]{BUB16}%
  \BibitemOpen
  \bibfield  {author} {\bibinfo {author} {\bibfnamefont {R.}~\bibnamefont
  {Boddeda}}, \bibinfo {author} {\bibfnamefont {I.}~\bibnamefont {Usmani}},
  \bibinfo {author} {\bibfnamefont {E.}~\bibnamefont {Bimbard}}, \bibinfo
  {author} {\bibfnamefont {A.}~\bibnamefont {Grankin}}, \bibinfo {author}
  {\bibfnamefont {A.}~\bibnamefont {Ourjoumtsev}}, \bibinfo {author}
  {\bibfnamefont {E.}~\bibnamefont {Brion}}, \ and\ \bibinfo {author}
  {\bibfnamefont {P.}~\bibnamefont {Grangier}},\ }\href@noop {} {\bibfield
  {journal} {\bibinfo  {journal} {Journal of Physics B: Atomic, Molecular and
  Optical Physics}\ }\textbf {\bibinfo {volume} {49}},\ \bibinfo {pages}
  {084005} (\bibinfo {year} {2016})}\BibitemShut {NoStop}%
\bibitem [{\citenamefont {Parigi}\ \emph {et~al.}(2012)\citenamefont {Parigi},
  \citenamefont {Bimbard}, \citenamefont {Stanojevic}, \citenamefont
  {Hilliard}, \citenamefont {Nogrette}, \citenamefont {Tualle-Brouri},
  \citenamefont {Ourjoumtsev},\ and\ \citenamefont {Grangier}}]{PBS12}%
  \BibitemOpen
  \bibfield  {author} {\bibinfo {author} {\bibfnamefont {V.}~\bibnamefont
  {Parigi}}, \bibinfo {author} {\bibfnamefont {E.}~\bibnamefont {Bimbard}},
  \bibinfo {author} {\bibfnamefont {J.}~\bibnamefont {Stanojevic}}, \bibinfo
  {author} {\bibfnamefont {A.~J.}\ \bibnamefont {Hilliard}}, \bibinfo {author}
  {\bibfnamefont {F.}~\bibnamefont {Nogrette}}, \bibinfo {author}
  {\bibfnamefont {R.}~\bibnamefont {Tualle-Brouri}}, \bibinfo {author}
  {\bibfnamefont {A.}~\bibnamefont {Ourjoumtsev}}, \ and\ \bibinfo {author}
  {\bibfnamefont {P.}~\bibnamefont {Grangier}},\ }\href@noop {} {\bibfield
  {journal} {\bibinfo  {journal} {Phys. Rev. Lett.}\ }\textbf {\bibinfo
  {volume} {109}},\ \bibinfo {pages} {233602} (\bibinfo {year}
  {2012})}\BibitemShut {NoStop}%
\bibitem [{\citenamefont {Stanojevic}\ \emph {et~al.}(2013)\citenamefont
  {Stanojevic}, \citenamefont {Parigi}, \citenamefont {Bimbard}, \citenamefont
  {Ourjoumtsev},\ and\ \citenamefont {Grangier}}]{SPB13}%
  \BibitemOpen
  \bibfield  {author} {\bibinfo {author} {\bibfnamefont {J.}~\bibnamefont
  {Stanojevic}}, \bibinfo {author} {\bibfnamefont {V.}~\bibnamefont {Parigi}},
  \bibinfo {author} {\bibfnamefont {E.}~\bibnamefont {Bimbard}}, \bibinfo
  {author} {\bibfnamefont {A.}~\bibnamefont {Ourjoumtsev}}, \ and\ \bibinfo
  {author} {\bibfnamefont {P.}~\bibnamefont {Grangier}},\ }\href@noop {}
  {\bibfield  {journal} {\bibinfo  {journal} {Phys. Rev. A}\ }\textbf {\bibinfo
  {volume} {88}},\ \bibinfo {pages} {053845} (\bibinfo {year}
  {2013})}\BibitemShut {NoStop}%
\bibitem [{\citenamefont {Jia}\ \emph {et~al.}(2017)\citenamefont {Jia},
  \citenamefont {Schine}, \citenamefont {Georgakopoulos}, \citenamefont {Ryou},
  \citenamefont {Sommer},\ and\ \citenamefont {Simon}}]{JSG17}%
  \BibitemOpen
  \bibfield  {author} {\bibinfo {author} {\bibfnamefont {N.}~\bibnamefont
  {Jia}}, \bibinfo {author} {\bibfnamefont {N.}~\bibnamefont {Schine}},
  \bibinfo {author} {\bibfnamefont {A.}~\bibnamefont {Georgakopoulos}},
  \bibinfo {author} {\bibfnamefont {A.}~\bibnamefont {Ryou}}, \bibinfo {author}
  {\bibfnamefont {A.}~\bibnamefont {Sommer}}, \ and\ \bibinfo {author}
  {\bibfnamefont {J.}~\bibnamefont {Simon}},\ }\href@noop {} {\bibfield
  {journal} {\bibinfo  {journal} {arXiv preprint arXiv:1705.07475}\ } (\bibinfo
  {year} {2017})}\BibitemShut {NoStop}%
\bibitem [{\citenamefont {Petrosyan}\ \emph {et~al.}(2011)\citenamefont
  {Petrosyan}, \citenamefont {Otterbach},\ and\ \citenamefont
  {Fleischhauer}}]{POF11}%
  \BibitemOpen
  \bibfield  {author} {\bibinfo {author} {\bibfnamefont {D.}~\bibnamefont
  {Petrosyan}}, \bibinfo {author} {\bibfnamefont {J.}~\bibnamefont
  {Otterbach}}, \ and\ \bibinfo {author} {\bibfnamefont {M.}~\bibnamefont
  {Fleischhauer}},\ }\href@noop {} {\bibfield  {journal} {\bibinfo  {journal}
  {Physical review letters}\ }\textbf {\bibinfo {volume} {107}},\ \bibinfo
  {pages} {213601} (\bibinfo {year} {2011})}\BibitemShut {NoStop}%
\bibitem [{\citenamefont {Gorshkov}\ \emph {et~al.}(2013)\citenamefont
  {Gorshkov}, \citenamefont {Nath},\ and\ \citenamefont {Pohl}}]{GNP13}%
  \BibitemOpen
  \bibfield  {author} {\bibinfo {author} {\bibfnamefont {A.~V.}\ \bibnamefont
  {Gorshkov}}, \bibinfo {author} {\bibfnamefont {R.}~\bibnamefont {Nath}}, \
  and\ \bibinfo {author} {\bibfnamefont {T.}~\bibnamefont {Pohl}},\ }\href@noop
  {} {\bibfield  {journal} {\bibinfo  {journal} {Phys. Rev. Lett.}\ }\textbf
  {\bibinfo {volume} {110}},\ \bibinfo {pages} {153601} (\bibinfo {year}
  {2013})}\BibitemShut {NoStop}%
\bibitem [{\citenamefont {Gorshkov}\ \emph {et~al.}(2011)\citenamefont
  {Gorshkov}, \citenamefont {Otterbach}, \citenamefont {Fleischhauer},
  \citenamefont {Pohl},\ and\ \citenamefont {Lukin}}]{GOF11}%
  \BibitemOpen
  \bibfield  {author} {\bibinfo {author} {\bibfnamefont {A.~V.}\ \bibnamefont
  {Gorshkov}}, \bibinfo {author} {\bibfnamefont {J.}~\bibnamefont {Otterbach}},
  \bibinfo {author} {\bibfnamefont {M.}~\bibnamefont {Fleischhauer}}, \bibinfo
  {author} {\bibfnamefont {T.}~\bibnamefont {Pohl}}, \ and\ \bibinfo {author}
  {\bibfnamefont {M.~D.}\ \bibnamefont {Lukin}},\ }\href@noop {} {\bibfield
  {journal} {\bibinfo  {journal} {Phys. Rev. Lett.}\ }\textbf {\bibinfo
  {volume} {107}},\ \bibinfo {pages} {133602} (\bibinfo {year}
  {2011})}\BibitemShut {NoStop}%
\bibitem [{\citenamefont {Schwinger}(1961)}]{S61}%
  \BibitemOpen
  \bibfield  {author} {\bibinfo {author} {\bibfnamefont {J.}~\bibnamefont
  {Schwinger}},\ }\href@noop {} {\bibfield  {journal} {\bibinfo  {journal}
  {Journal of Mathematical Physics}\ }\textbf {\bibinfo {volume} {2}},\
  \bibinfo {pages} {407} (\bibinfo {year} {1961})}\BibitemShut {NoStop}%
\bibitem [{\citenamefont {Rammer}(2007)}]{R07}%
  \BibitemOpen
  \bibfield  {author} {\bibinfo {author} {\bibfnamefont {J.}~\bibnamefont
  {Rammer}},\ }\href@noop {} {\emph {\bibinfo {title} {Quantum field theory of
  non-equilibrium states}}}\ (\bibinfo  {publisher} {Cambridge University
  Press},\ \bibinfo {year} {2007})\BibitemShut {NoStop}%
\bibitem [{\citenamefont {Stefanucci}\ and\ \citenamefont {van
  Leeuwen}(2013)}]{SL13}%
  \BibitemOpen
  \bibfield  {author} {\bibinfo {author} {\bibfnamefont {G.}~\bibnamefont
  {Stefanucci}}\ and\ \bibinfo {author} {\bibfnamefont {R.}~\bibnamefont {van
  Leeuwen}},\ }\href@noop {} {\emph {\bibinfo {title} {Nonequilibrium Many-Body
  Theory of Quantum Systems: A Modern Introduction}}}\ (\bibinfo  {publisher}
  {Cambridge University Press},\ \bibinfo {year} {2013})\BibitemShut {NoStop}%
\bibitem [{\citenamefont {Fleischhauer}\ and\ \citenamefont
  {Lukin}(2000)}]{FL00}%
  \BibitemOpen
  \bibfield  {author} {\bibinfo {author} {\bibfnamefont {M.}~\bibnamefont
  {Fleischhauer}}\ and\ \bibinfo {author} {\bibfnamefont {M.~D.}\ \bibnamefont
  {Lukin}},\ }\href@noop {} {\bibfield  {journal} {\bibinfo  {journal} {Phys.
  Rev. Lett.}\ }\textbf {\bibinfo {volume} {84}},\ \bibinfo {pages} {5094}
  (\bibinfo {year} {2000})}\BibitemShut {NoStop}%
\bibitem [{\citenamefont {Saffman}\ \emph {et~al.}(2010)\citenamefont
  {Saffman}, \citenamefont {Walker},\ and\ \citenamefont {M\o{}lmer}}]{SWM10}%
  \BibitemOpen
  \bibfield  {author} {\bibinfo {author} {\bibfnamefont {M.}~\bibnamefont
  {Saffman}}, \bibinfo {author} {\bibfnamefont {T.~G.}\ \bibnamefont {Walker}},
  \ and\ \bibinfo {author} {\bibfnamefont {K.}~\bibnamefont {M\o{}lmer}},\
  }\href@noop {} {\bibfield  {journal} {\bibinfo  {journal} {Rev. Mod. Phys.}\
  }\textbf {\bibinfo {volume} {82}},\ \bibinfo {pages} {2313} (\bibinfo {year}
  {2010})}\BibitemShut {NoStop}%
\bibitem [{\citenamefont {Lukin}\ \emph {et~al.}(2001)\citenamefont {Lukin},
  \citenamefont {Fleischhauer}, \citenamefont {Cote}, \citenamefont {Duan},
  \citenamefont {Jaksch}, \citenamefont {Cirac},\ and\ \citenamefont
  {Zoller}}]{LFC01}%
  \BibitemOpen
  \bibfield  {author} {\bibinfo {author} {\bibfnamefont {M.~D.}\ \bibnamefont
  {Lukin}}, \bibinfo {author} {\bibfnamefont {M.}~\bibnamefont {Fleischhauer}},
  \bibinfo {author} {\bibfnamefont {R.}~\bibnamefont {Cote}}, \bibinfo {author}
  {\bibfnamefont {L.~M.}\ \bibnamefont {Duan}}, \bibinfo {author}
  {\bibfnamefont {D.}~\bibnamefont {Jaksch}}, \bibinfo {author} {\bibfnamefont
  {J.~I.}\ \bibnamefont {Cirac}}, \ and\ \bibinfo {author} {\bibfnamefont
  {P.}~\bibnamefont {Zoller}},\ }\href@noop {} {\bibfield  {journal} {\bibinfo
  {journal} {Phys. Rev. Lett.}\ }\textbf {\bibinfo {volume} {87}},\ \bibinfo
  {pages} {037901} (\bibinfo {year} {2001})}\BibitemShut {NoStop}%
\bibitem [{\citenamefont {Walls}\ and\ \citenamefont {Milburn}(2007)}]{Walls}%
  \BibitemOpen
  \bibfield  {author} {\bibinfo {author} {\bibfnamefont {D.~F.}\ \bibnamefont
  {Walls}}\ and\ \bibinfo {author} {\bibfnamefont {G.~J.}\ \bibnamefont
  {Milburn}},\ }\href@noop {} {\emph {\bibinfo {title} {Quantum optics}}}\
  (\bibinfo  {publisher} {Springer Science \& Business Media},\ \bibinfo {year}
  {2007})\BibitemShut {NoStop}%
\bibitem [{\citenamefont {Fleischhauer}\ and\ \citenamefont
  {Yelin}(1999)}]{FY99}%
  \BibitemOpen
  \bibfield  {author} {\bibinfo {author} {\bibfnamefont {M.}~\bibnamefont
  {Fleischhauer}}\ and\ \bibinfo {author} {\bibfnamefont {S.~F.}\ \bibnamefont
  {Yelin}},\ }\href@noop {} {\bibfield  {journal} {\bibinfo  {journal} {Phys.
  Rev. A}\ }\textbf {\bibinfo {volume} {59}},\ \bibinfo {pages} {2427}
  (\bibinfo {year} {1999})}\BibitemShut {NoStop}%
\bibitem [{\citenamefont {Bienias}\ \emph {et~al.}(2014)\citenamefont
  {Bienias}, \citenamefont {Choi}, \citenamefont {Firstenberg}, \citenamefont
  {Maghrebi}, \citenamefont {Gullans}, \citenamefont {Lukin}, \citenamefont
  {Gorshkov},\ and\ \citenamefont {B\"uchler}}]{BCF14}%
  \BibitemOpen
  \bibfield  {author} {\bibinfo {author} {\bibfnamefont {P.}~\bibnamefont
  {Bienias}}, \bibinfo {author} {\bibfnamefont {S.}~\bibnamefont {Choi}},
  \bibinfo {author} {\bibfnamefont {O.}~\bibnamefont {Firstenberg}}, \bibinfo
  {author} {\bibfnamefont {M.~F.}\ \bibnamefont {Maghrebi}}, \bibinfo {author}
  {\bibfnamefont {M.}~\bibnamefont {Gullans}}, \bibinfo {author} {\bibfnamefont
  {M.~D.}\ \bibnamefont {Lukin}}, \bibinfo {author} {\bibfnamefont {A.~V.}\
  \bibnamefont {Gorshkov}}, \ and\ \bibinfo {author} {\bibfnamefont {H.~P.}\
  \bibnamefont {B\"uchler}},\ }\href@noop {} {\bibfield  {journal} {\bibinfo
  {journal} {Phys. Rev. A}\ }\textbf {\bibinfo {volume} {90}},\ \bibinfo
  {pages} {053804} (\bibinfo {year} {2014})}\BibitemShut {NoStop}%
\bibitem [{\citenamefont {Kamenev}(2011)}]{K11}%
  \BibitemOpen
  \bibfield  {author} {\bibinfo {author} {\bibfnamefont {A.}~\bibnamefont
  {Kamenev}},\ }\href@noop {} {\emph {\bibinfo {title} {Field theory of
  non-equilibrium systems}}}\ (\bibinfo  {publisher} {Cambridge University
  Press},\ \bibinfo {year} {2011})\BibitemShut {NoStop}%
\bibitem [{\citenamefont {Abrikosov}\ \emph {et~al.}(1975)\citenamefont
  {Abrikosov}, \citenamefont {Gorkov},\ and\ \citenamefont
  {Dzyaloshinski}}]{AGD}%
  \BibitemOpen
  \bibfield  {author} {\bibinfo {author} {\bibfnamefont {A.}~\bibnamefont
  {Abrikosov}}, \bibinfo {author} {\bibfnamefont {L.}~\bibnamefont {Gorkov}}, \
  and\ \bibinfo {author} {\bibfnamefont {I.}~\bibnamefont {Dzyaloshinski}},\
  }\href@noop {} {\emph {\bibinfo {title} {Methods of Quantum Field Theory in
  Statistical Physics}}},\ Dover Books on Physics Series\ (\bibinfo
  {publisher} {Dover Publications},\ \bibinfo {year} {1975})\BibitemShut
  {NoStop}%
\bibitem [{\citenamefont {Ourjoumtsev}\ \emph {et~al.}(2011)\citenamefont
  {Ourjoumtsev}, \citenamefont {Kubanek}, \citenamefont {Koch}, \citenamefont
  {Sames}, \citenamefont {Pinkse}, \citenamefont {Rempe},\ and\ \citenamefont
  {Murr}}]{OKK11}%
  \BibitemOpen
  \bibfield  {author} {\bibinfo {author} {\bibfnamefont {A.}~\bibnamefont
  {Ourjoumtsev}}, \bibinfo {author} {\bibfnamefont {A.}~\bibnamefont
  {Kubanek}}, \bibinfo {author} {\bibfnamefont {M.}~\bibnamefont {Koch}},
  \bibinfo {author} {\bibfnamefont {C.}~\bibnamefont {Sames}}, \bibinfo
  {author} {\bibfnamefont {P.~W.}\ \bibnamefont {Pinkse}}, \bibinfo {author}
  {\bibfnamefont {G.}~\bibnamefont {Rempe}}, \ and\ \bibinfo {author}
  {\bibfnamefont {K.}~\bibnamefont {Murr}},\ }\href@noop {} {\bibfield
  {journal} {\bibinfo  {journal} {Nature}\ }\textbf {\bibinfo {volume} {474}},\
  \bibinfo {pages} {623} (\bibinfo {year} {2011})}\BibitemShut {NoStop}%
\bibitem [{\citenamefont {Faddeev}(1961)}]{F60}%
  \BibitemOpen
  \bibfield  {author} {\bibinfo {author} {\bibfnamefont {L.}~\bibnamefont
  {Faddeev}},\ }\href@noop {} {\bibfield  {journal} {\bibinfo  {journal} {Sov.
  Phys. JETP}\ }\textbf {\bibinfo {volume} {12}},\ \bibinfo {pages} {1014}
  (\bibinfo {year} {1961})}\BibitemShut {NoStop}%
\bibitem [{\citenamefont {Liang}\ \emph {et~al.}(2017)\citenamefont {Liang},
  \citenamefont {Venkatramani}, \citenamefont {Cantu}, \citenamefont
  {Nicholson}, \citenamefont {Gullans}, \citenamefont {Gorshkov}, \citenamefont
  {Thompson}, \citenamefont {Chin}, \citenamefont {Lukin},\ and\ \citenamefont
  {Vuletic}}]{LVC17}%
  \BibitemOpen
  \bibfield  {author} {\bibinfo {author} {\bibfnamefont {Q.-Y.}\ \bibnamefont
  {Liang}}, \bibinfo {author} {\bibfnamefont {A.~V.}\ \bibnamefont
  {Venkatramani}}, \bibinfo {author} {\bibfnamefont {S.~H.}\ \bibnamefont
  {Cantu}}, \bibinfo {author} {\bibfnamefont {T.~L.}\ \bibnamefont
  {Nicholson}}, \bibinfo {author} {\bibfnamefont {M.~J.}\ \bibnamefont
  {Gullans}}, \bibinfo {author} {\bibfnamefont {A.~V.}\ \bibnamefont
  {Gorshkov}}, \bibinfo {author} {\bibfnamefont {J.~D.}\ \bibnamefont
  {Thompson}}, \bibinfo {author} {\bibfnamefont {C.}~\bibnamefont {Chin}},
  \bibinfo {author} {\bibfnamefont {M.~D.}\ \bibnamefont {Lukin}}, \ and\
  \bibinfo {author} {\bibfnamefont {V.}~\bibnamefont {Vuletic}},\ }\href@noop
  {} {\bibfield  {journal} {\bibinfo  {journal} {arXiv preprint
  arXiv:1709.01478}\ } (\bibinfo {year} {2017})}\BibitemShut {NoStop}%
\bibitem [{\citenamefont {Carusotto}\ and\ \citenamefont {Ciuti}(2013)}]{CC13}%
  \BibitemOpen
  \bibfield  {author} {\bibinfo {author} {\bibfnamefont {I.}~\bibnamefont
  {Carusotto}}\ and\ \bibinfo {author} {\bibfnamefont {C.}~\bibnamefont
  {Ciuti}},\ }\href@noop {} {\bibfield  {journal} {\bibinfo  {journal} {Rev.
  Mod. Phys.}\ }\textbf {\bibinfo {volume} {85}},\ \bibinfo {pages} {299}
  (\bibinfo {year} {2013})}\BibitemShut {NoStop}%
\end{thebibliography}%

\end{document}